\documentclass[%
 reprint,
superscriptaddress,
longbibliography,
 amsmath,amssymb,
 nobalancelastpage,
prb,
]{revtex4-1}
\usepackage[utf8]{inputenc}
\usepackage[romanian,english]{babel}
\usepackage{combelow}
\usepackage{color}
\usepackage{multirow}
\usepackage{gensymb}

\useshorthands{'}
\defineshorthand{'t}{\cb{t}}
\defineshorthand{'S}{\cb{S}}
\defineshorthand{'T}{\cb{T}}

\usepackage{graphicx}
\usepackage{dcolumn}

\usepackage{bm}
\usepackage{amsmath}
\usepackage{mhchem}
\usepackage{graphicx}
\usepackage{caption}
\usepackage{subcaption}
\usepackage{cleveref}
\newcommand{\NCCO}{\ce{Nd_{1.85}Ce_{0.15}CuO_{4-\delta}}}
\newcommand{\PLCCO}{\ce{Pr_{0.88}LaCe_{0.12}CuO_{4-\delta}}}
\newcommand{\NCCOg}{\ce{Nd_{2-x}Ce_{x}CuO_{4-\delta}}}
\newcommand{\LSCOg}{\ce{La_{2-x}Sr_{x}CuO_{4}}}

\begin{document}

\preprint{APS/123-QED}

\title{Emergence of low-energy spin waves in superconducting electron-doped cuprates}

\author{Kristine M. L. Krighaar}
 \email{kristine.krighaar@nbi.ku.dk}
 \affiliation{Nanoscience Center, Niels Bohr Institute, University of Copenhagen, 2100 Copenhagen, Denmark}

\author{Jeppe J. Cederholm}
\affiliation{Nanoscience Center, Niels Bohr Institute, University of Copenhagen, 2100 Copenhagen, Denmark}
\affiliation{Institute Laue-Langevin (ILL), 71 Avenue des Martyrs, CS20156, 38042 Grenoble, France}
\affiliation{Laboratory for Quantum Magnetism, Institute of Physics, \'{E}cole Polytechnique F\'{e}d\'{e}rale de Lausanne (EPFL), CH-1015 Lausanne, Switzerland}

\author{Ellen M. S. Schriver}
\affiliation{Nanoscience Center, Niels Bohr Institute, University of Copenhagen, 2100 Copenhagen, Denmark}

\author{Henrik Jacobsen}
\affiliation{Nanoscience Center, Niels Bohr Institute, University of Copenhagen, 2100 Copenhagen, Denmark}
\affiliation{Data Management and Software Centre, Asmussens Allé 305, 2800 Kongens Lyngby, Denmark}

\author{Christine P. Lauritzen}
\affiliation{Nanoscience Center, Niels Bohr Institute, University of Copenhagen, 2100 Copenhagen, Denmark}
\affiliation{Condensed Matter and Interfaces, Debye Institute for Nanomaterials Science, Utrecht University, 3508 TA Utrecht, The Netherlands}

\author{Igor Zaliznyak}
\affiliation{Condensed Matter Physics and Materials Science Division, Brookhaven National Laboratory, Upton, NY 11973, USA}

\author{C\'edric H. Qvistgaard}
\affiliation{Nanoscience Center, Niels Bohr Institute, University of Copenhagen, 2100 Copenhagen, Denmark}
\affiliation{Department of Energy Conversion and Storage, Technical University of Denmark, 2800 Kgs. Lyngby, Denmark}

\author{Ursula B. Hansen}
\affiliation{Institute Laue-Langevin (ILL), 71 Avenue des Martyrs, CS20156, 38042 Grenoble, France}

\author{Ahmed Alshemi}
\affiliation{Division of Synchrotron Radiation Research, Department of Physics, Lund University, SE-22100 Lund, Sweden}

\author{Anton P. J. Stampfl}
\affiliation{Australian Nuclear Science and Technology Organisation, Lucas Heights, NSW 2234, Australia}%

\author{Jean-Claude Grivel}
\affiliation{Department of Energy Conversion and Storage, Technical University of Denmark, 2800 Kgs. Lyngby, Denmark}

\author{Dongjoon Song}
\affiliation{Quantum Matter Institute, University of British Columbia, Vancouver, British Columbia, Canada, V6T 1Z4}

\author{Kim Lefmann}%
\affiliation{%
Nanoscience Center, Niels Bohr Institute, University of Copenhagen, 2100 Copenhagen, Denmark}%

\author{Machteld E. Kamminga }
\email{m.e.kamminga@uu.nl}
\affiliation{Condensed Matter and Interfaces, Debye Institute for Nanomaterials Science, Utrecht University, 3508 TA Utrecht, The Netherlands}

\date{\today}

\begin{abstract}
In order to fully utilize the technological potential of unconventional superconductors, an enhanced understanding of the superconducting mechanism is necessary. In the best performing superconductors, the cuprates, superconductivity is intimately linked with magnetism, although the details of this coupling remain elusive. In search of clarity in the magnetism-superconductivity relationship, we focus on the electron-doped cuprate \NCCO\ (NCCO). NCCO has an antiferromagnetic ground state when synthesized, and only becomes superconducting after a reductive annealing process. This makes NCCO an ideal template to study how the magnetism differs in the superconducting and non-superconducting state, while keeping the material template as constant as possible. Using neutron spectroscopy, we reveal that the as-grown crystal exhibits a large spin pseudogap in the magnetic fluctuation spectrum. Upon annealing, defects that are introduced by the commonly employed synthesis method are removed and the spin pseudogap is significantly reduced. While the spin pseudogap in the annealed sample is likely an effect of superconductivity, we argue that the spin pseudogap in the as-grown sample is caused by the absence of long-wavelength spin waves. The defects in as-grown NCCO thus play the dual role of suppressing both superconductivity and low-energy spin waves, highlighting a potential connection between these two phenomena. 
\end{abstract}

\maketitle

\section{Introduction}
tranSuperconducting cuprates have been a focal point of condensed matter research for decades, due to their high critical temperatures and complex interplay of charge, spin, and lattice degrees of freedom.~\cite{bednorz_possible_1986,tranquada_evidence_1995,scalapino_common_2012} The undoped parent compounds of these materials are antiferromagnetic Mott insulators. By doping the \ce{CuO2} planes with charge carriers, the antiferromagnetic phase subsides and superconductivity emerges below a critical temperature $T_c$. \cite{Kofu09} Based on extensive neutron scattering studies, it is now understood that there is a subtle connection between superconductivity and antiferromagnetism in these materials.\cite{lake_spin_1999, keimer_quantum_2015} 
In particular, the vastly asymmetric phase diagram between electron-doping (\textit{n}-type) and hole-doping (\textit{p}-type) has been considered as an archetypal example of that connection; more  robust antiferromagnetic order with lower T$_c$ and a narrower  superconducting dome  are encountered for \textit{n}-type cuprates, whereas less robust antiferromagnetic order with higher T$_c$ and a wider superconducting dome are observed for p-type cuprates. Therefore, investigating similarities and differences between \textit{n}-type and \textit{p}-type cuprates seems crucial to understand the physics behind high-T$_c$ superconductivity.~\cite{Armitage_SC_gap_anisotropy_2001,Armitage_anomolous_2001, Damascelli_phase_diagram_2001}

The vastly studied hole-doped cuprates, such as \LSCOg\ (LSCO), commonly exhibit incommensurate antiferromagnetic fluctuations.~\cite{tranquada_evidence_1995} In contrast, electron-doped cuprates such as \NCCOg\ (NCCO), show commensurate antiferromagnetic fluctuations, both above and below $T_c$, at the antiferromagnetic ($\pi$, $\pi$) points (tetragonal ($0.5$, $0.5$, 0)).~\cite{yamada_commensurate_2003} Similar to LSCO, neutron scattering studies on NCCO reveal the presence of a spin pseudogap of these antiferromagnetic excitations in the superconducting state.~\cite{zhao_neutron-spin_2007, yamada_neutron_1999, motoyama_magnetic_2006,yu_two_2010,yamada_commensurate_2003, Matsuda1992} However, the temperature dependence of the spin pseudogap energy contrasts with the temperature-independent behavior observed in LSCO.

\begin{figure}
    \includegraphics[width=\linewidth]{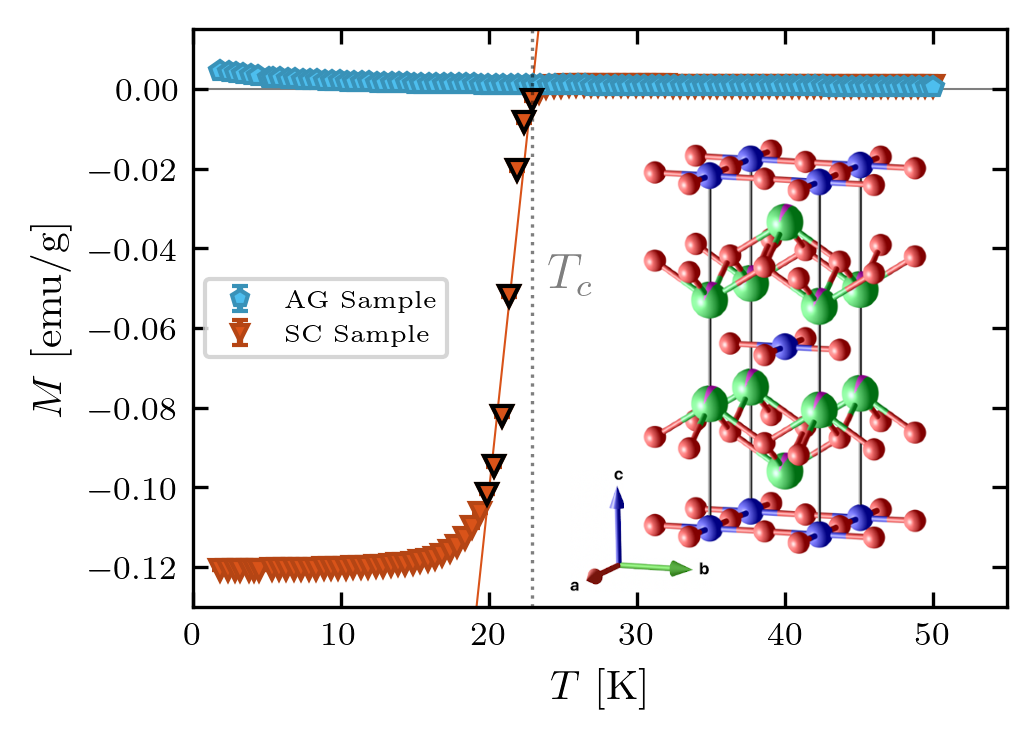}
    \caption{Magnetization as a function of temperature (ZFC: zero-field cooled), measured at 10~Oe field, for the as-grown (AG) and reductively annealed, superconducting (SC) NCCO single crystals, depicted in blue pentagons and orange triangles, respectively. $T_c$ is defined as the onset temperature of superconductivity. Insert: crystal structure of NCCO,\cite{Momma_VESTA_software,belokoneva1991preparation} with Cu, O and Nd depicted in blue, red and green, respectively. The 15\% Ce doping on the Nd site is denoted as a pink slice on the green Nd atoms.}
    \label{fig:squid}
\end{figure}

In this work, we dive deeper into the origin of the spin pseudogap in NCCO and its relation to superconductivity, by focusing on yet another crucial difference between \textit{n}-type and \textit{p}-type cuprates: the effect of reductive annealing. From a materials chemistry perspective, the main peculiarity of \textit{n}-type cuprates is that they, directly after synthesis, exhibit antiferromagnetic order throughout the accessible doping range, and an oxygen reduction treatment is required to induce superconductivity,~\cite{Takagi1989,Tokura_superconducting_1989} leaving the as-grown samples always non-superconducting. This is clearly shown by our susceptibility data for as-grown and annealed crystals of NCCO, depicted in Fig.~\ref{fig:squid}. 

Having two doping degrees of freedom in \NCCOg, \textit{i.e.}\ Ce doping and O reduction, makes \textit{n}-type cuprates chemically more complex than their \textit{p}-type counterparts and the requisite of this reductive annealing immediately diminishes the electron-hole symmetry for cuprate superconductors. Most notably, the exact effect of this reductive annealing step on the materials structure has been a longstanding debate and to date there still appears to be no consensus on the matter. We will elaborate on the common hypotheses and their chemical consequences in more detail in the discussion section. 
 
Yamada \textit{et al.} showed that the static magnetic signal in annealed NCCO is drastically suppressed as compared to the as-grown samples.\cite{yamada_commensurate_2003} This matches the current understanding of competing magnetic and superconducting order in cuprate materials in general.\cite{scalapino_common_2012} Additionally, Yamada \textit{et al.} showed that a spin pseudogap of around 4 meV opens below $T_c$ in the superconducting phase.\cite{yamada_commensurate_2003} This is in agreement with Zhao \textit{et al}, \cite{zhao_neutron-spin_2007} who, among others,~\cite{yamada_neutron_1999, motoyama_magnetic_2006,yu_two_2010,yamada_commensurate_2003, Matsuda1992} used inelastic neutron scattering to probe the magnetic excitations in reductively annealed NCCO  above and below $T_c$. They found that superconductivity opens a spin pseudogap at the antiferromagnetic ordering wave vector \textbf{Q} = (0.5, 0.5, 0). Concomitantly, they state the observation of a resonance peak at the same \textbf{Q} at 9.5 meV above this spin pseudogap, similar to those of \textit{p}-type cuprates and the related \textit{n}-type cuprate \PLCCO \ (PLCCO).\cite{zhao_neutron-spin_2007} Such a resonance is generally understood to draw its intensity from the spin pseudogap in the low-energy part of the spin excitation spectrum in \textit{p}-type cuprates.\cite{rossat-mignod_neutron_1991,Mook_1993_YBCO_resonance,Dai_2001_YBCO_resonance,tranquada_quantum_2004,Hayden_2004_high-energy-spin-excitations} This is consistent with the case of the annealed superconducting NCCO that has a spin pseudogap. However, the fact that superconducting PLCCO is gapless remains puzzling in this respect.\cite{zhao_neutron-spin_2007,dai_evolution_2007} The spin fluctuations in as-grown NCCO have not received a dedicated investigation, leaving the open question how reductive annealing affects the low-energy spin dynamics in this electron-doped cuprate.

In this study, we focus on the effect of reductive annealing on the magnetic excitations in NCCO and determine which part of the excitation spectrum is affected by the annealing process. We show that reductive annealing, which induces superconductivity, indeed opens up a spin pseudogap. Using a thermal neutron triple-axis spectrometer, we measure the low-energy spin fluctuations in NCCO both before and after reductive annealing. To ensure sample consistency necessary for a straight comparison, we used a single, optimally doped \NCCO \ crystal split into two parts. We reductively annealed one half to obtain the superconducting crystal, and left the other half untouched, representing the as-grown crystal. Investigating the effect of reductive annealing on the spin fluctuations and superconductivity in this prototypical \textit{n}-type cuprate directly adds to the understanding of the origin of high-$T_c$ superconductivity and its interplay with magnetic correlations. 

\section{Results}

Fig.~\ref{fig:squid} shows the magnetization measurements of the two crystal pieces from the same growth, of which one has been reductively annealed. Note that here the susceptibility is simply given as magnetization per gram of crystal, as the large crystals needed for our neutron scattering experiments are generally too large for the SQUID magnetometer to convert into absolute units. The annealed sample displays a clear negative magnetization at low temperatures, indicative of the Meissner effect, with an onset temperature of the superconducting transition at $T_c = 23$ K. In contrast, the as-grown sample shows a flat magnetization curve, with only a slight increase at low temperatures. This is typical of an antiferromagnetic response and clearly differs from the sharp superconducting transition. The insert shows the tetragonal crystal structure, $I4/mmm$ for both annealed and as-grown, optimally doped NCCO with lattice parameters $a=b=3.957$ Å and $c= 12.075$ Å. \cite{belokoneva1991preparation}

\begin{figure}
    \centering
    \includegraphics[width=\linewidth]{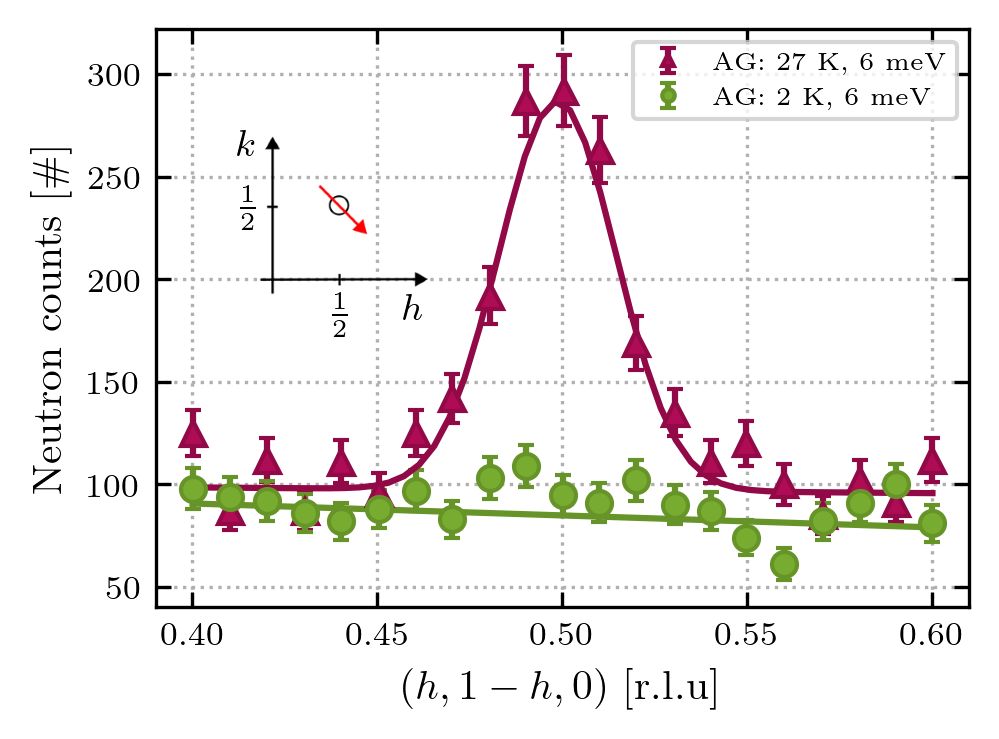}
    \caption{Representative magnetic response of the as-grown (AG) NCCO sample at two temperatures, measured with neutron spectroscopy. At high temperature  (27~K), a peak indicates the presence of magnetic excitations at this temperature. The peak, and therefore the excitations, vanish at low temperature (2~K). The $q$-resolution of the neutron instrument is sufficiently broad to cover the full magnetic response. (Insert) Illustration of the scanning direction (red arrow) across the commensurate magnetic peak (black circle). The solid lines are Gaussian fits to the data. A hypothesis test, using Wilk's theorem, is used to determine whether or not a peak is present, as outlined in the Supplemental Material.}
    \label{fig:raw_scans}
\end{figure}

Fig.~\ref{fig:raw_scans} shows neutron scattering data of two (constant energy) $q$-scans performed on the as-grown sample at the same energy transfer ($\hbar \omega = 6$ meV) but at different temperatures. At 27 K, a clear peak is observed, indicative of a well-defined magnetic response. In contrast, at 2~K, the signal has essentially vanished, indicating a significant reduction in the intensity of the magnetic response. The small insert illustrates the scan direction in the $(h,k)$ plane. The data in Fig.~\ref{fig:raw_scans} are representative for the rest of the $q$-scans, which are given in Fig.~S7-S12 in the Supplemental Material. We normalized the data, as detailed in Ref. [\onlinecite{xu_absolute_2013}], using measurements of a branch of the acoustic phonon at (2, 0, 0), to ensure comparability between the two samples. Details are provided in the Supplemental Material. Subsequently, the integrated intensity is determined by fitting the area of the Gaussian peak centered at (0.5, 0.5, 0), effectively removing the $q$-dependence and yielding the dynamical auto-correlation function $S(\omega)$ which can be seen in the Supplemental Material in Figs.~S13, S14 and S21. The integrated intensities are converted into dynamic susceptibility $\chi''(\omega)$ and we observe no substantial differences between the $S(\omega)$ and $\chi''(\omega)$ results in this range of energies and temperatures.

The onset of the spin pseudogap was determined by fitting $\chi''(\omega)$ with an error function, and defining the onset energy as the $E_{gap}$ parameter further described in the Supplemental Material section II.E. The associated uncertainty was obtained from the fitted parameters. This procedure was applied where the data exhibited a well-defined saturation of magnetic response. For the 2 K as-grown sample, no clear saturation of the signal is observed. We therefore estimate the spin pseudogap as the first data point where the errorbars between 2~K and 27~K overlap. The associated uncertainty is taken as the distance to the nearest adjacent data point. 

\begin{figure}
    \centering
    \includegraphics[width=\linewidth]{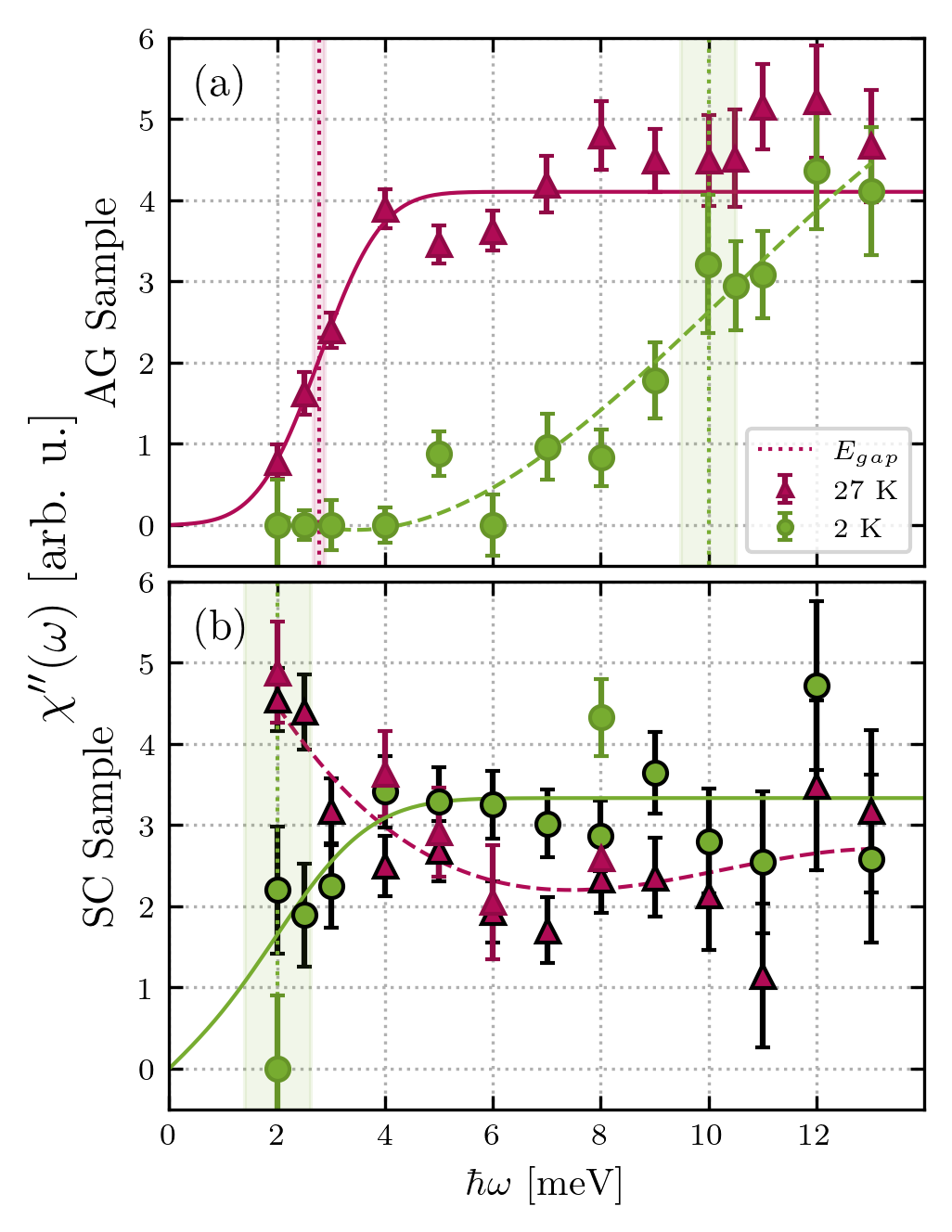}
    \caption{Dynamic susceptibility $\chi''(\omega)$, as a function of energy transfer. (a) as-grown (AG) sample. (b) annealed, superconducting (SC) sample. The black outlined points indicate 3-point scans, while colored outlined points indicate $q$-scans. The solid lines are fits to the response described in the Supplemental Material. The dashed lines are drawn as guide to the eye, while the colored vertical dotted lines are the estimate of the spin pseudogap onset with the faded area representing the uncertainty.}
    \label{fig:E_IN20}
\end{figure}

Fig.~\ref{fig:E_IN20} shows $\chi''(\omega)$ as a function of energy transfer ($\hbar\omega$) for the as-grown sample and the annealed sample, at 2 K and 27 K, extracted from all $q$-scans. In our experiment, we were unable to measure the inelastic signal at energies beyond 14 meV because of interference from a strong signal from the crystal electric field level of Nd at $\sim$ 15 meV. In agreement with Zhao \textit{et al},\cite{zhao_neutron-spin_2007} the annealed, superconducting sample in Fig.~\ref{fig:E_IN20}(b) has no spin pseudogap above $T_c$, and only a small spin pseudogap of $2 \pm0.6$~meV in the superconducting phase at 2~K. This is consistent with data by Motoyama \textit{et al},\cite{motoyama_magnetic_2006} which were acquired using a similar experimental procedure.

The as-grown sample in Fig.~3(a) shows completely different behavior. First, we note the shape of the spin pseudogap at 2 K. Here the spin pseudogap displays a gradual emergence from $\sim$10 meV to 4 meV compared to the more rapid emergence from $\sim$4 meV to 2 meV of the 27 K data. Furthermore, in contrast to the annealed sample, the as-grown sample exhibits a pronounced spin pseudogap that increases from $2.8 \pm 0.1$~meV (onset) at 27 K to $10\pm0.5$ meV (onset) at 2 K. This immediately indicates that the common understanding that superconductivity opens a spin pseudogap,\cite{zhao_neutron-spin_2007} is challenged by the presence of an even larger spin pseudogap in the as-grown, non-superconducting sample. Thus, reductive annealing seems to reduce the spin pseudogap.

\begin{figure}
\centering
\includegraphics[width=\linewidth]{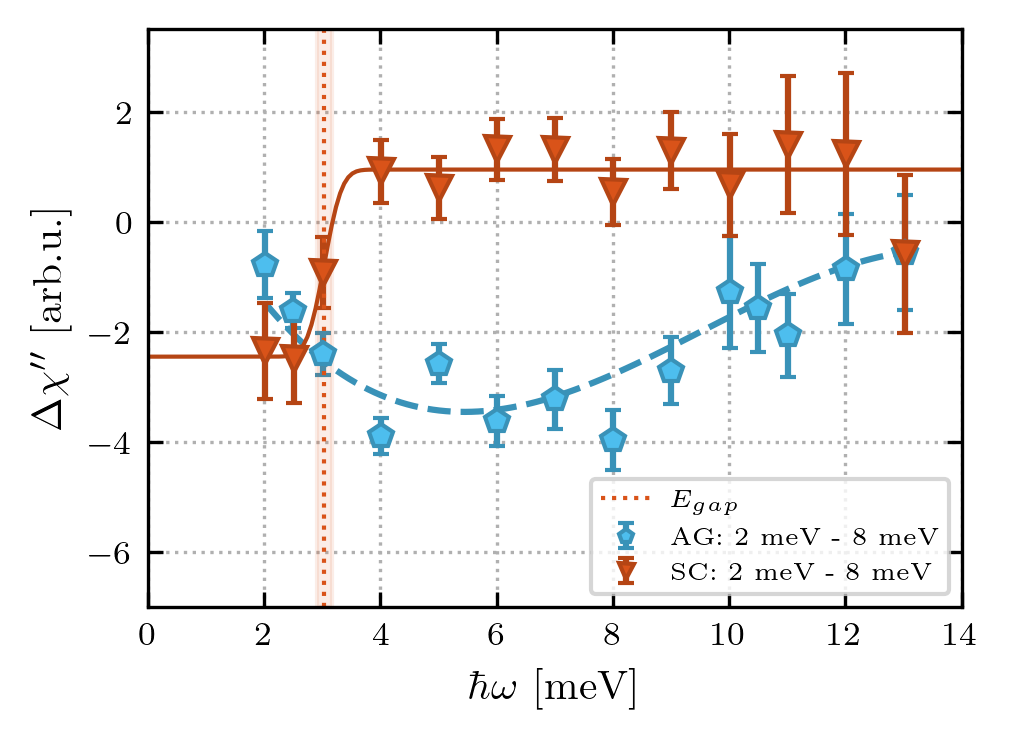}
\caption{Shift in $\chi''(\omega)$ determined by subtraction of the 2 K data from the 27 K data depicted in Fig.~\ref{fig:E_IN20}, for both the as-grown (AG) and annealed, superconducting (SC) samples. The dashed lines are drawn as guide to the eye and The solid lines are fits to the response, the dashed lines are drawn as guide to the eye, while the colored vertical dotted lines are the estimate of the spin pseudogap onset with the faded area representing the uncertainty.}
\label{fig:T_difference}
\end{figure}

To better resolve the onset of the spin pseudogap in Fig.~\ref{fig:E_IN20}, we directly compare the change in spectral weight as a function of energy transfer by subtracting the 27 K data from the 2 K data shown in Fig.~\ref{fig:E_IN20}, following the procedure of Zhao \textit{et al.}\cite{zhao_neutron-spin_2007}. This analysis, shown in Fig.~\ref{fig:T_difference}, illustrates the large spectral weight shift between the two samples. Here it is evident that a large part of the spectral weight moves to lower energies when we anneal the crystal. Moreover, the closing of the spin pseudogap in the annealed, superconducting sample is steep and the spin pseudogap value is more clearly defined at $3.0\pm 0.1$~meV (onset) as seen by the dotted line in Fig. \ref{fig:T_difference}.

\begin{figure}
    \centering
    \includegraphics[width=\linewidth]{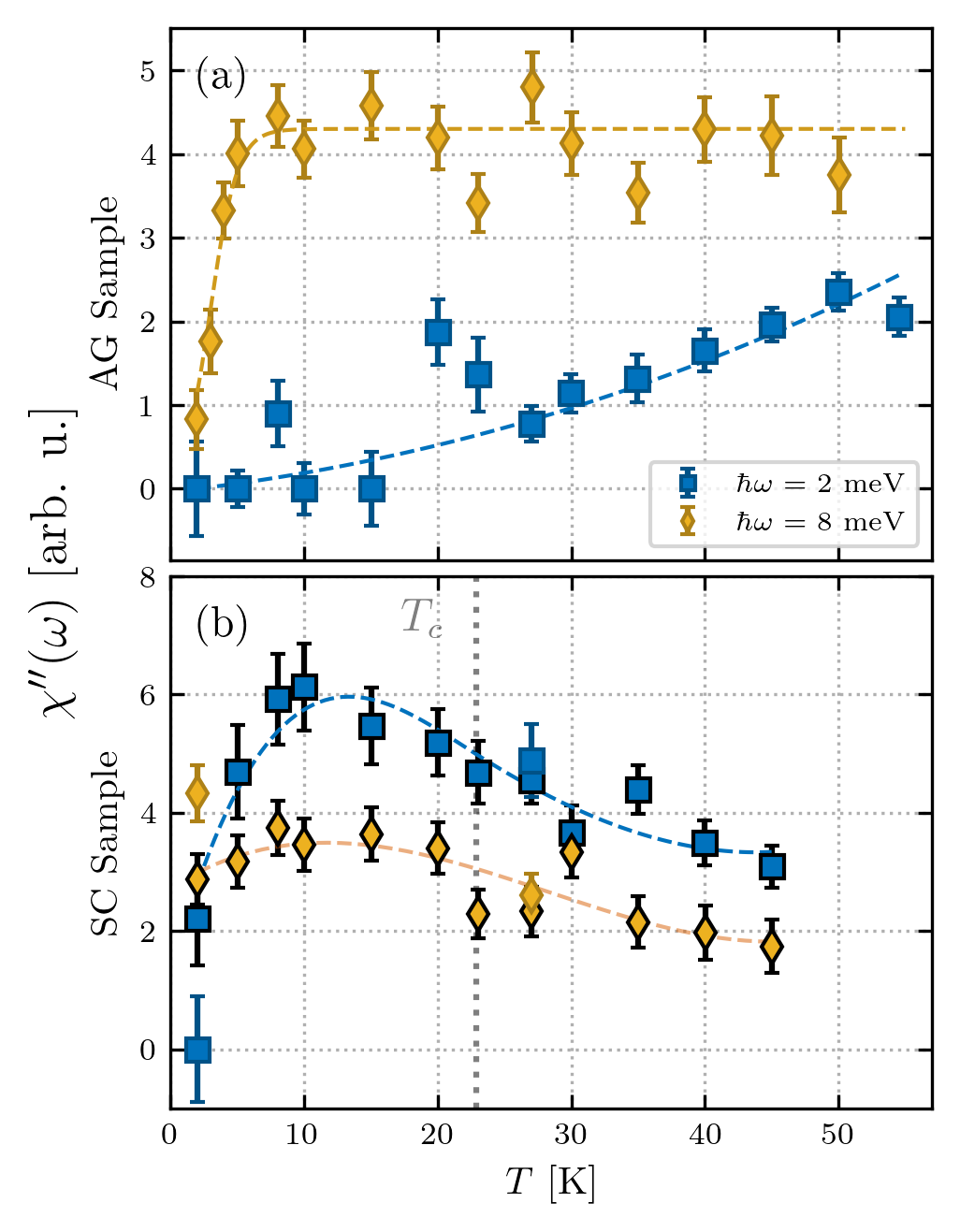}
    \caption{$\chi''(\omega)$ shown as a function of temperature. (a) as-grown sample (AG) and (b) annealed, superconducting (SC) sample. The black outlined points indicate 3-point scans while colored outlined points indicate $q$-scans. Square blue points falling below this line indicate the onset of a spin pseudogap. The dashed blue and yellow lines are guides to the eye.}
    \label{fig:T_IN20}
\end{figure}

To track the temperature dependence of the spin pseudogap, we determine the integrated intensity as a function of temperature measured at energy transfers $\hbar \omega = 2$ meV and $\hbar \omega = 8$ meV as shown in Fig.~\ref{fig:T_IN20}. 
The results from the as-grown sample show that there are stronger energy fluctuations at $8$ meV than at $2$ meV for all temperatures below $\sim 40$ K.  
This is in great contrast to what is observed for the annealed, superconducting sample, where the 2 meV fluctuations dominate until they are gapped out below $\sim 5$~K.

In addition, we measured the antiferromagnetic order at the (3/2, 1/2, 0) peak through the elastic scattering peaks for both samples. The superconducting sample exhibits suppressed antiferromagnetic order compared to the as-grown sample, as previously reported.\cite{yamada_commensurate_2003} Detailed results of these measurements are provided in the Supplemental Material.

\begin{figure}
    \centering
    \includegraphics[width=\linewidth]{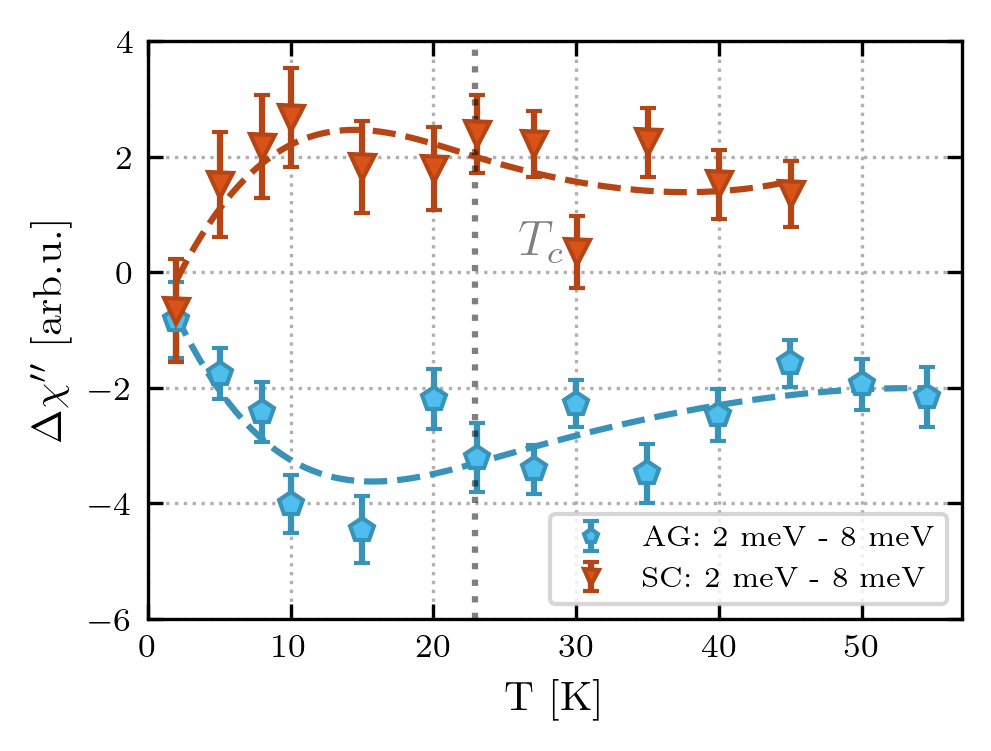}
    \caption{Difference in $\chi''(\omega)$ as a function of temperature, calculated by subtracting the 8 meV data from the 2 meV data. Saturation of the signal for the annealed, superconducting (SC) sample happens close to $T_c$. The dashed lines are guides to the eye.}
    \label{fig:E_difference}
\end{figure}

To emphasize the temperature dependence, we show the difference between the $8$ meV and the $2$ meV data in Fig.~\ref{fig:E_difference}. We observe that the spectral weight at low energies is increasing when the sample is annealed, consistent with the energy dependence reported above. Furthermore, we note that the spectral weight difference in the annealed sample flattens approximately at $T_c$.

\section{Discussion}

We now discuss our results in light of the reductive annealing necessary to obtain superconductivity in NCCO. We first discuss the sample quality and the chemical consequences of reductive annealing. After that we elaborate on how this affects the neutron scattering spectrum and the implications for the relation between reductive annealing and superconductivity.

\subsection{CHEMICAL CONSEQUENCES OF REDUCTIVE ANNEALING}

As stated in the introduction, the exact effect of the reductive annealing step to make \textit{n}-type cuprates superconducting after growth has been a longstanding debate, yet without consensus. 
Many different techniques have shown that the reduction process in fact only removes a small fraction of the O atoms,\cite{Moran_extra_oxygen_1989,Tarascon_growth_1989,Radaelli_apical_oxygen_1994,schultz_single-crystal_1996,Klamut_oxygen_pressure_1997,navarro_oxygen_2001} while having a drastic effect on the conducting and magnetic properties. In fact, the reported amount of oxygen removed generally ranges from 0.1\% to 2\% and decreases with increasing Ce content.\cite{Takayama-Muromachi_oxygen_deficiency_1989,Suzuki_oxygen_nonstoichiometry_1990,Kim_phase_stability_diagram_1993,schultz_single-crystal_1996} However, despite the fact that reduction in principle contributes electrons, this effect cannot be compensated for by simply altering the Ce content in the material. In general, three common hypotheses have been reported to explain the effect of reductive annealing on the materials chemical structure:\cite{armitage_progress_2010}

1) In contrast to LSCO that generally exhibits the so-called \textit{T} crystal structure, \textit{n}-type cuprates such as NCCO have a \textit{T'} crystal structure, characterized by a lack of oxygen atoms in apical positions in the \ce{CuO2} plane (as seen in
Fig.~\ref{fig:squid}). However, some apical O atoms are observed as interstitial defects,\cite{Riou_2004_raman,Richard_2004_raman} even in the undoped parent compound \ce{Nd2CuO4}.\cite{Radaelli_apical_oxygen_1994} These apical oxygen defects are expected to strongly perturb the local ionic potential on the Cu site immediately below it and a systematic study of this effect on resistivity suggests that these extra oxygens introduce defect scattering of conduction electrons without changing the carrier density.\cite{Xu_1996_oxygen_transport_properties} In fact, magnetoresistance data reveal that these defects are spin disordered in nature, potentially counteracting superconductivity in as-grown samples. In turn, reductive annealing is found to lead to a decrease of the apical occupancy,\cite{Radaelli_apical_oxygen_1994,schultz_single-crystal_1996}
 lifting some of the defects present in as-grown NCCO.

2) The argument of apical oxygen reduction is strongly challenged by other reports that found that a local Raman mode associated with the presence of apical oxygens is not at all affected by reductive annealing of cerium-doped cuprates. In fact, new sets of excitations are found to appear upon reductive annealing. These are related to the creation of oxygen vacancies in the \ce{CuO2} plane and in the rare-earth layers, leaving the apical oxygens in place.\cite{Riou_2004_raman,Richard_2004_raman} Note that such in-plane oxygen vacancies would have an effect on the local Cu oxidation state,\cite{navarro_oxygen_2001} and yet another hypothesis argues that the important consequence of reductive annealing is rather related to the Cu sites (see also point 3).

3) Studies on the microstructure of NCCO have reported the appearance and disappearance of an impurity phase associated with annealing and re-oxygenation,\cite{Kurahashi_2002_heat_treatment} later shown to be \ce{(Nd,Ce)2O3}.\cite{Mang_2004_chemical_inhomogeneity} The presence of this small-volume, parasitic phase implied that the additional Cu atoms removed from the affected sheets to form the rare-earth oxide, are now free to cure intrinsic Cu vacancies in the as-grown \ce{CuO2} planes.\cite{Kurahashi_2002_heat_treatment,Kang_2007_microscopic_annealing} Within this scenario, during the reduction process
Cu atoms migrate from these layers to the NCCO
structure to ``repair” defects present in the as-grown materials
resulting in Cu deficient regions with the epitaxial
\ce{(Nd,Ce)2O3} intercalation, and therefore reducing the number of Cu vacancies in the \ce{CuO2} planes. This would in turn remove pair-breaking sites and favor superconductivity.

As implied by these conflicting hypotheses, the exact consequence of the reductive annealing remains unclear with respect to the chemical structure of the material. However, we argue that independent of all evidence from the different studies, two basic principles remain true for NCCO. That is: 1) the crystals of as-grown NCCO are (given the employed synthesis method) imperfect and contain some form of defects, whether they are O or Cu defects (or both) in form of vacancies and/or interstitial sites; 2) reductive annealing targets these defects by removing oxygens  and/or by ``healing'' \  the Cu defects. Thus, reductive annealing enhances the quality of the \ce{CuO2} plane, which appears to be essential for superconductivity to arise. We will address this further below.

\subsection{NEUTRON SCATTERING SPECTRUM}

From our measurements, we were able to study the evolution of the spin pseudogap by annealing, using two types of samples from a single growth. Our results state that for the as-grown sample, both the spin pseudogap and the magnetic order are significantly more pronounced than in the annealed, superconducting samples. This raises the unresolved question of the precise origin of the spin pseudogap and the mechanism that is so strongly influenced by reductive annealing.

We start our argument by noting that a spectral gap behavior similar to our results has been observed in the $S = 1/2$ Heisenberg antiferromagnetic chain compound \ce{SrCuO2}.\cite{motoyama_magnetic_1996,rice_rich_1997,zaliznyak_anisotropic_1999} Here, an opening of a spin pseudogap in the excitation spectrum was observed in samples with low levels of Ca or Ni doping.\cite{hammerath_spin_2011,simutis_spin_2013,simutis_spin_2017} While replacing Sr with Ca does not directly affect the spin chains, in case of Ni-doped \ce{SrCuO2}, the $S = 1$ \ce{Ni^{2+}} impurities are directly inserted in the chains, replacing the $S = 1/2$ \ce{Cu^{2+}} ions. For this system, an inelastic neutron scattering study by Simutis \textit{et al}, reveal that 1\% Ni-doping leads to a spin pseudogap of roughly 8 meV. \cite{simutis_spin_2013} Using a simple model based on defects, they state that the spin dynamics can be fully explained by an effective fragmentation of the spin chains, where such a spin pseudogap implies a depletion of low-energy magnetic states. Expanding this interpretation to two-dimensional \ce{CuO2} planes, as found in NCCO, it is natural to discuss if similar fragmentation could occur. 

\begin{figure}
    \centering
    \includegraphics[width=\linewidth]{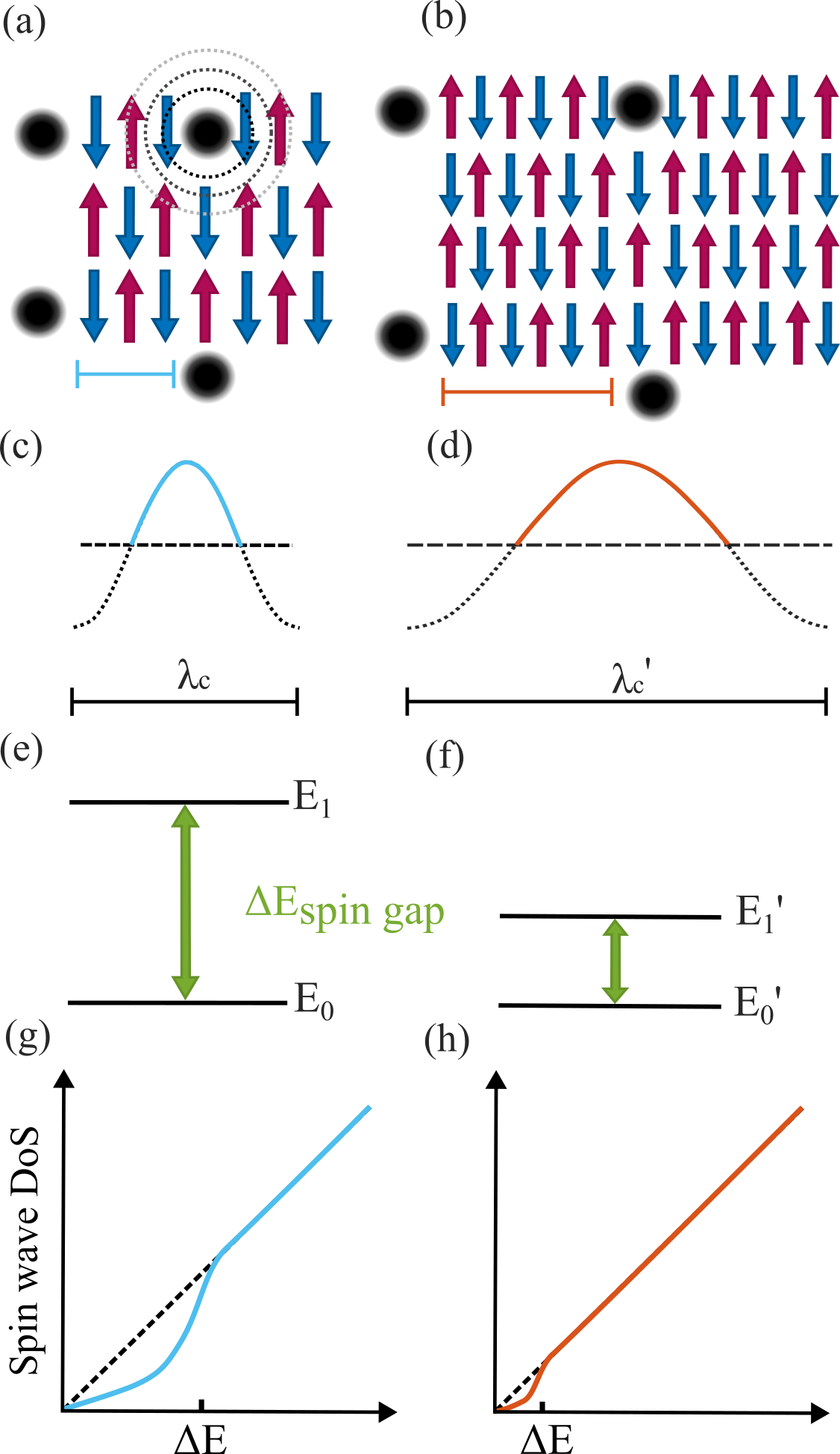}
    \caption{Schematics illustrating how the size of the antiferromagnetic patches influences the spin waves allowed in the system. Left column represents the as-grown sample, while the right is the annealed, superconducting sample. (a) and (b) show the antiferromagnetic structure in the two cases, with the structure composed of smaller patches created by defects (black circles), but still weakly antiferromagnetically interacting. In the annealed sample the undisturbed patches are largest. (c) and (d) show how the patches restrict the spin waves above a certain wavelength. 
    By having larger patches, more low-energy states are occupied, minimizing the energy spin pseudogap, as illustrated in (e) and (f). This is more quantitatively expressed as a (partial) suppression of the spin wave density of states at low energies, seen in (g) and (h).}
    \label{fig:interpretation}
\end{figure}

A previous study has investigated the magnetic correlation length $\xi$ for as-grown NCCO.\cite{mang_spin_2004} Using energy-integrating neutron scattering, they measured the instantaneous correlation function, $S(q)$, and observed that $\xi$ increases with decreasing temperature, eventually creating a weakly long-range ordered magnetic state. This is consistent with what data of as-grown NCCO, both in Ref.~\onlinecite{yamada_commensurate_2003} and in our work (see Fig.~S15 in the Supplemental Material), where the intensity of the (3/2, 1/2, 0) antiferromagnetic peak decreases fast with increasing temperature. In a similar work, Motoyama {\em et al.} studied the behavior of $\xi$ in annealed and superconducting NCCO.\cite{motoyama_spin_2007} For optimal Ce doping they found $\xi$ to be an order of magnitude smaller than for as-grown NCCO. While these studies appear to contradict our model of lattice ``healing''\ by reductive annealing, these properties differ from those examined in our study. For starters, their scattering method integrates only up to the initial neutron energy, i.e. $E_i = 14.7$~meV. Secondly, the shorter correlation length measured in the superconducting state is consistent with spin waves slightly perturbed by superconductivity, while the true elastic signal is strongly suppressed, as also seen in our diffraction measurements. Therefore, we consider these earlier findings to not contradict our physical picture of the system.

Inelastic measurements are distinctly different. Here neutron scattering effectively measures an intensity that is proportional to the density of state of the spin waves.
In the weakly long-range ordered state, these spin waves are strongly affected by the defects, which act as scattering centers. This causes a reduction of the density of states at long wavelengths, corresponding to a reduction in intensity towards low energies. In the ultimate fragmented case of isolated spin clusters, Hendriksen {\em et al.}\ were able to directly relate the size of the spin wave gap to the spatial extend of a cluster.\cite{hendriksen_finite-size_1993} As illustrated in Fig.~\ref{fig:interpretation}, their argument state that the largest wavelength allowed is the one where $\lambda_c/2$ is approximately the same size as the spin cluster. Correspondingly, the inverse relationship between energy and wavelength gives us the lowest allowed energy state.

Returning to our NCCO compound, we now argue that a similar model could explain our results. In terms of spin dynamics, NCCO could be considered a two-dimensional analogue of \ce{SrCuO2} separated by a series of point (or line) defects. As mentioned above, we argue that the defects (regardless of their origin, as discussed in section A) act like scattering centers, working as effective magnetic boundaries. The illustration of Fig.~\ref{fig:interpretation}(a, c, e and g) represents the as-grown sample, where a high concentration of defects results in smaller antiferromagnetic patches. Correspondingly, it sets a limit for the largest allowed wavelength $\lambda_c$ and thus reduces the low-energy density of states of the spin waves. This, in turn, suppresses the low-energy signal leading to a spin pseudogap, consistent with our experimental observations shown in Fig.~\ref{fig:E_IN20}. The effect of reductive annealing is then represented by the illustration of Fig.~\ref{fig:interpretation}(b, d, f and h), where defects are removed and more pristine \ce{CuO2} sheets are created, therefore allowing a larger wavelength $\lambda_c'$, which in turn reduces, or completely eliminates, the spin pseudogap. 
A simple linear spin wave theory calculation correlates the observed spin pseudogap of around 10~meV (onset) in the as-grown sample to an effective patch size of the order 40~nm, vs. 130 nm for a spin pseudogap of 3.0~meV in the annealed sample; for details see the Supplemental Material. 

The above-mentioned reductive annealing behavior in NCCO is consistent with the results for as-grown and annealed electron doped cuprate PLCCO ($x = 0.11$), which show clear commensurate low-energy spin fluctuations in both samples.\cite{fujita_neutron-scattering_2003} Similar to NCCO, PLCCO also needs to be reductively annealed in order to become superconducting. A study on the spin excitation spectrum by Dai \textit{et al},\cite{dai_evolution_2007} reveals a reduction of the spin pseudogap upon annealing. The main difference is the fact that in PLCCO the spin pseudogap upon annealing closes completely, whereas in NCCO a smaller spin pseudogap persists. The authors make no further comments with regard to the spin pseudogap origin.\cite{dai_evolution_2007}

Our model is supported by a previous study using resonant inelastic x-ray scattering on 16\% NCCO.\cite{ishii_post-growth_2021} Herein, the authors observe that the high-energy magnetic excitation spectra are almost identical for as-grown and reductively annealed samples in the energy range from approximately 100~meV to 1~eV. Held up against the results of our neutron data, this signifies that the main effect of annealing lies in the low-energy (\textit{i.e.} long wavelength) spin waves, consistent with the picture provided in Fig.~\ref{fig:interpretation}.

A different cause of the excitation gap could be anisotropy in the magnetic interactions. 
We first note that the pure $S=1/2$ of Cu$^{2+}$ does not support crystal electric field splittings and therefore single-ion anisotropy is to first order ruled out, even in the presence of defects. Instead, one could consider anisotropic exchange interactions, such as the Dzyaloshinskii–Moriya interaction. In the LSCO parent compound La$_2$CuO$_4$, these effect are known to support magnetic gaps up to 5-6~meV.\cite{keimer_magnetic_1992} 
However, for the doped samples this value is reduced. Even in the most magnetic version of LSCO, $x=0.12$, these spin pseudogaps are seen to diminish to below 1~meV.\cite{romer_glassy_2013} 
In light of the similarity between these compounds, we believe that discrepancy between 1~meV and 10~meV (onset) is so large that exchange anisotropy is not a realistic scenario.

\subsection{RELATION TO SUPERCONDUCTIVITY}
We now discuss the potential connection between our observations and the presence/absence of superconductivity in our system. In earlier work, there was a large emphasis on a peak-like structure in the magnetic spectrum that was interpreted as a resonance peak.\cite{zhao_neutron-spin_2007} However, despite the fact that our data in Fig.~\ref{fig:T_difference} shows an increase in $\Delta \chi''$, we do not observe a prominent peak but rather an overall increase of spectral weight between 5 and 12 meV, likely due to the appearance of superconductivity.  
These observations are consistent with more recent work on LSCO \cite{li_low-energy_2018}, where superconductivity near optimal doping is associated with the suppression of low-energy antiferromagnetic spin fluctuations and a spin pseudogap-like depletion, which in turn limits the coherent superconducting gap rather than yielding a sharp resonance feature. 

Another significant feature is the behavior of the spin pseudogap. 
In the superconducting sample this spin pseudogap is likely connected to superconductivity, in analogy with many other cuprates.\cite{lake_spin_1999}
However, our results show that the spin pseudogap is larger in the non-superconducting sample, leaving this larger effect unlikely to be related to superconducting tendencies, such as pre-formed pairs. The shift suggests there is a correlation between the low-energy magnetic excitations and superconductivity. We interpret the spin pseudogap change as the healing of the \ce{CuO}-planes allowing longer wavelength spin waves to form, which in turn can occupy the lower energy states.

A comparison of our work with ARPES data offers an additional perspective on the momentum dependence of the pairing interaction inferred from our neutron measurements. In work by Song \textit{et al}, the electronic excitation spectrum is measured for both as-grown and annealed NCCO.\cite{song_oxygen-content-dependent_2012} Here, they observe sharper excitations in the annealed sample and argue that this is evidence of stronger coherence, which arises from removing defects in the system. This picture is consistent with our picture of longer patches of undisrupted magnetism due to better crystalline order.

It has been suggested by Scalapino that the magnetic excitations act as the superconducting pairing mechanism.\cite{dagotto_superconductivity_1992,scalapino_common_2012} Although our findings align with the general principles, it seems surprising that the appearance of superconductivity correlates with the low-energy parts of the spectrum as high-temperature superconductivity has been believed to originate from the higher energy part of the spectrum\cite{dahm_strength_2009} or a resonance energy.\cite{yu_universal_2009} A comprehensive comparison is limited by the scope of our study and further theoretical insights would be necessary to provide a definitive answer. 

\section{Conclusion}

The electron doped NCCO system is a valuable model system for investigating the emergence of superconductivity, owing to the necessity of reductive annealing. It provides a way to directly compare superconducting and non-superconducting samples originating from the same crystal growth, thereby minimizing uncertainties associated with sample variability.

By studying the effect of reductive annealing using inelastic neutron spectroscopy, we have investigated how annealing alters the magnetic excitation spectrum in NCCO. Although no clear resonance peak is observed in the superconducting sample, we do find a redistribution of spectral weight. This shift reflects the closing of the spin pseudogap upon annealing and coincides with the emergence of superconductivity. We interpret this behavior as annealing removing defects from the system, thereby allowing longer-wavelength spin fluctuations to develop. These enhanced low-energy fluctuations lend support to the spin-fluctuation-mediated pairing mechanism proposed by Scalapino,\cite{scalapino_common_2012} and we believe that the connection between superconductivity and magnetic correlations merits further theoretical and experimental investigation.

 
\section{Experimental Method}
\subsection{Crystal growth and  characterization}
A 39 mm long and 6 mm diameter single crystal of optimally doped \NCCO \ was synthesized using the traveling-solvent floating-zone method. The crystal was cut into three pieces, where one piece was kept as the as-grown sample, while the two other pieces were treated with reductive annealing procedures. The details can be found in the Supplemental Material.

Magnetic susceptibility measurements were conducted using a Quantum Design MPMS-XL superconducting quantum interference device (SQUID) magnetometer. Measurements were performed on warming in a d.c. field of 10 Oe in the temperature range 1.8–50 K, after cooling in zero applied field (ZFC). In a second measurement, the sample was cooled in the applied field of 10 Oe (FC), see Supplemental Material.

\subsection{Neutron Spectroscopy}
Inelastic neutron measurements were conducted at the Australian Nuclear Science and Technology Organisation (ANSTO) using the TAIPAN instrument and at Institut Laue-Langevin (ILL) utilizing the IN20 instrument. Both instruments are thermal triple-axis-spectrometers, and were chosen over cold-neutron triple axis spectrometers due to larger resolution volumes, which enable us to integrate weak signals stemming from small magnetic moments of Cu$^{2+}$ ($S=1/2$), to better verify the existence or absence of magnetic fluctuations. All crystal samples were aligned in the $(h, k, 0)$-plane using a combination of x-ray and neutron Laue diffraction.

Measurements were taken by performing diagonal $q$-scans around the magnetic Bragg point $(h, 1-h, 0)$ for $h=0.5$, see insert of Fig.~\ref{fig:raw_scans}, and with $\Delta E$ ranging from 2 meV to 13 meV. These scans were performed using an Orange cryostat at the base temperature of 1.9 K and at 27 K, which is above the superconducting transition temperature. Furthermore, the temperature dependence (from 2 K up to 55 K) of the magnetic signal was measured at $\Delta E =$ 2 meV to investigate the spin pseudogap, and again at $\Delta E =$ 8 meV.

As shown in more detail in the Supplemental Material, the reductively annealed, superconducting sample showed an additional twin domain 45\degree \ rotated, having almost half the sample mass. Due to the reduced effective sample mass, 3-point scans were utilized to obtain results of significant quality to verify that the behavior was as previously reported\cite{zhao_neutron-spin_2007}. The data treatment method is detailed in section II.C.3 of the Supplemental Material.

To allow a direct comparison of the data from the two samples, the intensities have been normalized to an acoustic phonon scan in units of $S(\mathbf{Q},\omega)$.\cite{xu_absolute_2013} Examples of normalized $q$-scans are shown in the Supplemental Material. 

To assess the presence or absence of a magnetic signal at $h=0.5$, each $q$-scan was fitted to both a linear (slope-only) model and a combined slope-plus-Gaussian model, with the Gaussian area as a free parameter. To determine the best fit model for each scan, in particular whether or not a Gaussian peak is present, Wilks theorem\cite{wilks_large-sample_1938} was used with a confidence interval of $p=0.05$. Figures of the fits as well as an overview of the fitting parameters can be found in the Supplemental Material. We observe for the $q$-scan fits that all Gaussians have approximately the same width, which we then use to estimate the integrated intensity from the 3-point scans. The results of the integrated intensities are then expressed as $\chi''(\omega)$. In addition, we include corresponding versions of Fig.~\ref{fig:E_IN20} - Fig.~\ref{fig:E_difference} expressed for $S(\mathbf{Q},\omega)$, see section II.D and III.B in Supplemental Material. All analysis code is made available via the github repository\cite{NBI_Magnetism_NCCOgithub_2026}.

\section*{Acknowledgments}
We like to thank Andreas Kreisel and Brian M. Andersen for valuable perspectives and discussions.

We are grateful for the access to neutron beamtime at ILL under proposal number TEST-3346 (doi:10.5291/ILL-DATA.TEST-3346) and acknowledge the support of ANSTO in providing access to facilities used in proposal P13914. 

MEK  acknowledges the research program \textit{Materials for the Quantum Age} (QuMat) for financial support. This program (registration number 024.005.006) is part of the Gravitation program financed by the Dutch Ministry of Education, Culture and Science (OCW). The project was supported by the Danish National Committee for Research Infrastructure through DanScatt and the ESS-Lighthouse Q-MAT. HJ was funded by the Carlsberg Foundation Grant No.~cf20-0497. The work at Brookhaven was supported by the Office of Basic Energy Sciences, U.S. Department of Energy (DOE) under Contract No.~DE-SC0012704. Furthermore, this research was undertaken thanks in part to funding from the Max Planck–UBC–UTokyo Centre for Quantum Materials and the Canada First Research Excellence Fund (CFREF), Quantum Materials and Future Technologies.

\bibliographystyle{apsrev4-1}
\bibliography{Main_NCCO}

\begin{thebibliography}{62}%
\makeatletter
\providecommand \@ifxundefined [1]{%
 \@ifx{#1\undefined}
}%
\providecommand \@ifnum [1]{%
 \ifnum #1\expandafter \@firstoftwo
 \else \expandafter \@secondoftwo
 \fi
}%
\providecommand \@ifx [1]{%
 \ifx #1\expandafter \@firstoftwo
 \else \expandafter \@secondoftwo
 \fi
}%
\providecommand \natexlab [1]{#1}%
\providecommand \enquote  [1]{``#1''}%
\providecommand \bibnamefont  [1]{#1}%
\providecommand \bibfnamefont [1]{#1}%
\providecommand \citenamefont [1]{#1}%
\providecommand \href@noop [0]{\@secondoftwo}%
\providecommand \href [0]{\begingroup \@sanitize@url \@href}%
\providecommand \@href[1]{\@@startlink{#1}\@@href}%
\providecommand \@@href[1]{\endgroup#1\@@endlink}%
\providecommand \@sanitize@url [0]{\catcode `\\12\catcode `\$12\catcode
  `\&12\catcode `\#12\catcode `\^12\catcode `\_12\catcode `\%12\relax}%
\providecommand \@@startlink[1]{}%
\providecommand \@@endlink[0]{}%
\providecommand \url  [0]{\begingroup\@sanitize@url \@url }%
\providecommand \@url [1]{\endgroup\@href {#1}{\urlprefix }}%
\providecommand \urlprefix  [0]{URL }%
\providecommand \Eprint [0]{\href }%
\providecommand \doibase [0]{http://dx.doi.org/}%
\providecommand \selectlanguage [0]{\@gobble}%
\providecommand \bibinfo  [0]{\@secondoftwo}%
\providecommand \bibfield  [0]{\@secondoftwo}%
\providecommand \translation [1]{[#1]}%
\providecommand \BibitemOpen [0]{}%
\providecommand \bibitemStop [0]{}%
\providecommand \bibitemNoStop [0]{.\EOS\space}%
\providecommand \EOS [0]{\spacefactor3000\relax}%
\providecommand \BibitemShut  [1]{\csname bibitem#1\endcsname}%
\let\auto@bib@innerbib\@empty
\bibitem [{\citenamefont {Bednorz}\ and\ \citenamefont
  {M{\"u}ller}(1986)}]{bednorz_possible_1986}%
  \BibitemOpen
  \bibfield  {author} {\bibinfo {author} {\bibfnamefont {J.~G.}\ \bibnamefont
  {Bednorz}}\ and\ \bibinfo {author} {\bibfnamefont {K.~A.}\ \bibnamefont
  {M{\"u}ller}},\ }\href {\doibase 10.1007/BF01303701} {\bibfield  {journal}
  {\bibinfo  {journal} {Zeitschrift f{\"u}r Physik B Condensed Matter}\
  }\textbf {\bibinfo {volume} {64}},\ \bibinfo {pages} {189} (\bibinfo {year}
  {1986})}\BibitemShut {NoStop}%
\bibitem [{\citenamefont {Tranquada}\ \emph {et~al.}(1995)\citenamefont
  {Tranquada}, \citenamefont {Sternlieb}, \citenamefont {Axe}, \citenamefont
  {Nakamura},\ and\ \citenamefont {Uchida}}]{tranquada_evidence_1995}%
  \BibitemOpen
  \bibfield  {author} {\bibinfo {author} {\bibfnamefont {J.~M.}\ \bibnamefont
  {Tranquada}}, \bibinfo {author} {\bibfnamefont {B.~J.}\ \bibnamefont
  {Sternlieb}}, \bibinfo {author} {\bibfnamefont {J.~D.}\ \bibnamefont {Axe}},
  \bibinfo {author} {\bibfnamefont {Y.}~\bibnamefont {Nakamura}}, \ and\
  \bibinfo {author} {\bibfnamefont {S.}~\bibnamefont {Uchida}},\ }\href
  {\doibase 10.1038/375561a0} {\bibfield  {journal} {\bibinfo  {journal}
  {Nature}\ }\textbf {\bibinfo {volume} {375}},\ \bibinfo {pages} {561}
  (\bibinfo {year} {1995})}\BibitemShut {NoStop}%
\bibitem [{\citenamefont {Scalapino}(2012)}]{scalapino_common_2012}%
  \BibitemOpen
  \bibfield  {author} {\bibinfo {author} {\bibfnamefont {D.~J.}\ \bibnamefont
  {Scalapino}},\ }\href {\doibase 10.1103/RevModPhys.84.1383} {\bibfield
  {journal} {\bibinfo  {journal} {Reviews of Modern Physics}\ }\textbf
  {\bibinfo {volume} {84}},\ \bibinfo {pages} {1383} (\bibinfo {year}
  {2012})}\BibitemShut {NoStop}%
\bibitem [{\citenamefont {Kofu}\ \emph {et~al.}(2009)\citenamefont {Kofu},
  \citenamefont {Lee}, \citenamefont {Fujita}, \citenamefont {Kang},
  \citenamefont {Eisaki},\ and\ \citenamefont {Yamada}}]{Kofu09}%
  \BibitemOpen
  \bibfield  {author} {\bibinfo {author} {\bibfnamefont {M.}~\bibnamefont
  {Kofu}}, \bibinfo {author} {\bibfnamefont {S.-H.}\ \bibnamefont {Lee}},
  \bibinfo {author} {\bibfnamefont {M.}~\bibnamefont {Fujita}}, \bibinfo
  {author} {\bibfnamefont {H.-J.}\ \bibnamefont {Kang}}, \bibinfo {author}
  {\bibfnamefont {H.}~\bibnamefont {Eisaki}}, \ and\ \bibinfo {author}
  {\bibfnamefont {K.}~\bibnamefont {Yamada}},\ }\href {\doibase
  10.1103/PhysRevLett.102.047001} {\bibfield  {journal} {\bibinfo  {journal}
  {Physical Review Letters}\ }\textbf {\bibinfo {volume} {102}},\ \bibinfo
  {pages} {047001} (\bibinfo {year} {2009})}\BibitemShut {NoStop}%
\bibitem [{\citenamefont {Lake}\ \emph {et~al.}(1999)\citenamefont {Lake},
  \citenamefont {Aeppli}, \citenamefont {Mason}, \citenamefont {Schröder},
  \citenamefont {McMorrow}, \citenamefont {Lefmann}, \citenamefont {Isshiki},
  \citenamefont {Nohara}, \citenamefont {Takagi},\ and\ \citenamefont
  {Hayden}}]{lake_spin_1999}%
  \BibitemOpen
  \bibfield  {author} {\bibinfo {author} {\bibfnamefont {B.}~\bibnamefont
  {Lake}}, \bibinfo {author} {\bibfnamefont {G.}~\bibnamefont {Aeppli}},
  \bibinfo {author} {\bibfnamefont {T.~E.}\ \bibnamefont {Mason}}, \bibinfo
  {author} {\bibfnamefont {A.}~\bibnamefont {Schröder}}, \bibinfo {author}
  {\bibfnamefont {D.~F.}\ \bibnamefont {McMorrow}}, \bibinfo {author}
  {\bibfnamefont {K.}~\bibnamefont {Lefmann}}, \bibinfo {author} {\bibfnamefont
  {M.}~\bibnamefont {Isshiki}}, \bibinfo {author} {\bibfnamefont
  {M.}~\bibnamefont {Nohara}}, \bibinfo {author} {\bibfnamefont
  {H.}~\bibnamefont {Takagi}}, \ and\ \bibinfo {author} {\bibfnamefont {S.~M.}\
  \bibnamefont {Hayden}},\ }\href {\doibase 10.1038/21840} {\bibfield
  {journal} {\bibinfo  {journal} {Nature}\ }\textbf {\bibinfo {volume} {400}},\
  \bibinfo {pages} {43} (\bibinfo {year} {1999})}\BibitemShut {NoStop}%
\bibitem [{\citenamefont {Keimer}\ \emph {et~al.}(2015)\citenamefont {Keimer},
  \citenamefont {Kivelson}, \citenamefont {Norman}, \citenamefont {Uchida},\
  and\ \citenamefont {Zaanen}}]{keimer_quantum_2015}%
  \BibitemOpen
  \bibfield  {author} {\bibinfo {author} {\bibfnamefont {B.}~\bibnamefont
  {Keimer}}, \bibinfo {author} {\bibfnamefont {S.~A.}\ \bibnamefont
  {Kivelson}}, \bibinfo {author} {\bibfnamefont {M.~R.}\ \bibnamefont
  {Norman}}, \bibinfo {author} {\bibfnamefont {S.}~\bibnamefont {Uchida}}, \
  and\ \bibinfo {author} {\bibfnamefont {J.}~\bibnamefont {Zaanen}},\ }\href
  {\doibase 10.1038/nature14165} {\bibfield  {journal} {\bibinfo  {journal}
  {Nature}\ }\textbf {\bibinfo {volume} {518}},\ \bibinfo {pages} {179}
  (\bibinfo {year} {2015})}\BibitemShut {NoStop}%
\bibitem [{\citenamefont {Armitage}\ \emph
  {et~al.}(2001{\natexlab{a}})\citenamefont {Armitage}, \citenamefont {Lu},
  \citenamefont {Feng}, \citenamefont {Kim}, \citenamefont {Damascelli},
  \citenamefont {Shen}, \citenamefont {Ronning}, \citenamefont {Shen},
  \citenamefont {Onose}, \citenamefont {Taguchi},\ and\ \citenamefont
  {Tokura}}]{Armitage_SC_gap_anisotropy_2001}%
  \BibitemOpen
  \bibfield  {author} {\bibinfo {author} {\bibfnamefont {N.~P.}\ \bibnamefont
  {Armitage}}, \bibinfo {author} {\bibfnamefont {D.~H.}\ \bibnamefont {Lu}},
  \bibinfo {author} {\bibfnamefont {D.~L.}\ \bibnamefont {Feng}}, \bibinfo
  {author} {\bibfnamefont {C.}~\bibnamefont {Kim}}, \bibinfo {author}
  {\bibfnamefont {A.}~\bibnamefont {Damascelli}}, \bibinfo {author}
  {\bibfnamefont {K.~M.}\ \bibnamefont {Shen}}, \bibinfo {author}
  {\bibfnamefont {F.}~\bibnamefont {Ronning}}, \bibinfo {author} {\bibfnamefont
  {Z.-X.}\ \bibnamefont {Shen}}, \bibinfo {author} {\bibfnamefont
  {Y.}~\bibnamefont {Onose}}, \bibinfo {author} {\bibfnamefont
  {Y.}~\bibnamefont {Taguchi}}, \ and\ \bibinfo {author} {\bibfnamefont
  {Y.}~\bibnamefont {Tokura}},\ }\href {\doibase 10.1103/PhysRevLett.86.1126}
  {\bibfield  {journal} {\bibinfo  {journal} {Phys. Rev. Lett.}\ }\textbf
  {\bibinfo {volume} {86}},\ \bibinfo {pages} {1126} (\bibinfo {year}
  {2001}{\natexlab{a}})}\BibitemShut {NoStop}%
\bibitem [{\citenamefont {Armitage}\ \emph
  {et~al.}(2001{\natexlab{b}})\citenamefont {Armitage}, \citenamefont {Lu},
  \citenamefont {Kim}, \citenamefont {Damascelli}, \citenamefont {Shen},
  \citenamefont {Ronning}, \citenamefont {Feng}, \citenamefont {Bogdanov},
  \citenamefont {Shen}, \citenamefont {Onose}, \citenamefont {Taguchi},
  \citenamefont {Tokura}, \citenamefont {Mang}, \citenamefont {Kaneko},\ and\
  \citenamefont {Greven}}]{Armitage_anomolous_2001}%
  \BibitemOpen
  \bibfield  {author} {\bibinfo {author} {\bibfnamefont {N.~P.}\ \bibnamefont
  {Armitage}}, \bibinfo {author} {\bibfnamefont {D.~H.}\ \bibnamefont {Lu}},
  \bibinfo {author} {\bibfnamefont {C.}~\bibnamefont {Kim}}, \bibinfo {author}
  {\bibfnamefont {A.}~\bibnamefont {Damascelli}}, \bibinfo {author}
  {\bibfnamefont {K.~M.}\ \bibnamefont {Shen}}, \bibinfo {author}
  {\bibfnamefont {F.}~\bibnamefont {Ronning}}, \bibinfo {author} {\bibfnamefont
  {D.~L.}\ \bibnamefont {Feng}}, \bibinfo {author} {\bibfnamefont
  {P.}~\bibnamefont {Bogdanov}}, \bibinfo {author} {\bibfnamefont {Z.-X.}\
  \bibnamefont {Shen}}, \bibinfo {author} {\bibfnamefont {Y.}~\bibnamefont
  {Onose}}, \bibinfo {author} {\bibfnamefont {Y.}~\bibnamefont {Taguchi}},
  \bibinfo {author} {\bibfnamefont {Y.}~\bibnamefont {Tokura}}, \bibinfo
  {author} {\bibfnamefont {P.~K.}\ \bibnamefont {Mang}}, \bibinfo {author}
  {\bibfnamefont {N.}~\bibnamefont {Kaneko}}, \ and\ \bibinfo {author}
  {\bibfnamefont {M.}~\bibnamefont {Greven}},\ }\href {\doibase
  10.1103/PhysRevLett.87.147003} {\bibfield  {journal} {\bibinfo  {journal}
  {Phys. Rev. Lett.}\ }\textbf {\bibinfo {volume} {87}},\ \bibinfo {pages}
  {147003} (\bibinfo {year} {2001}{\natexlab{b}})}\BibitemShut {NoStop}%
\bibitem [{\citenamefont {Damascelli}\ \emph {et~al.}(2001)\citenamefont
  {Damascelli}, \citenamefont {Lu},\ and\ \citenamefont
  {Shen}}]{Damascelli_phase_diagram_2001}%
  \BibitemOpen
  \bibfield  {author} {\bibinfo {author} {\bibfnamefont {A.}~\bibnamefont
  {Damascelli}}, \bibinfo {author} {\bibfnamefont {D.}~\bibnamefont {Lu}}, \
  and\ \bibinfo {author} {\bibfnamefont {Z.-X.}\ \bibnamefont {Shen}},\ }\href
  {\doibase https://doi.org/10.1016/S0368-2048(01)00264-X} {\bibfield
  {journal} {\bibinfo  {journal} {Journal of Electron Spectroscopy and Related
  Phenomena}\ }\textbf {\bibinfo {volume} {117-118}},\ \bibinfo {pages} {165}
  (\bibinfo {year} {2001})}\BibitemShut {NoStop}%
\bibitem [{\citenamefont {Yamada}\ \emph {et~al.}(2003)\citenamefont {Yamada},
  \citenamefont {Kurahashi}, \citenamefont {Uefuji}, \citenamefont {Fujita},
  \citenamefont {Park}, \citenamefont {Lee},\ and\ \citenamefont
  {Endoh}}]{yamada_commensurate_2003}%
  \BibitemOpen
  \bibfield  {author} {\bibinfo {author} {\bibfnamefont {K.}~\bibnamefont
  {Yamada}}, \bibinfo {author} {\bibfnamefont {K.}~\bibnamefont {Kurahashi}},
  \bibinfo {author} {\bibfnamefont {T.}~\bibnamefont {Uefuji}}, \bibinfo
  {author} {\bibfnamefont {M.}~\bibnamefont {Fujita}}, \bibinfo {author}
  {\bibfnamefont {S.}~\bibnamefont {Park}}, \bibinfo {author} {\bibfnamefont
  {S.-H.}\ \bibnamefont {Lee}}, \ and\ \bibinfo {author} {\bibfnamefont
  {Y.}~\bibnamefont {Endoh}},\ }\href {\doibase 10.1103/PhysRevLett.90.137004}
  {\bibfield  {journal} {\bibinfo  {journal} {Physical Review Letters}\
  }\textbf {\bibinfo {volume} {90}},\ \bibinfo {pages} {137004} (\bibinfo
  {year} {2003})}\BibitemShut {NoStop}%
\bibitem [{\citenamefont {Zhao}\ \emph {et~al.}(2007)\citenamefont {Zhao},
  \citenamefont {Dai}, \citenamefont {Li}, \citenamefont {Freeman},
  \citenamefont {Onose},\ and\ \citenamefont
  {Tokura}}]{zhao_neutron-spin_2007}%
  \BibitemOpen
  \bibfield  {author} {\bibinfo {author} {\bibfnamefont {J.}~\bibnamefont
  {Zhao}}, \bibinfo {author} {\bibfnamefont {P.}~\bibnamefont {Dai}}, \bibinfo
  {author} {\bibfnamefont {S.}~\bibnamefont {Li}}, \bibinfo {author}
  {\bibfnamefont {P.~G.}\ \bibnamefont {Freeman}}, \bibinfo {author}
  {\bibfnamefont {Y.}~\bibnamefont {Onose}}, \ and\ \bibinfo {author}
  {\bibfnamefont {Y.}~\bibnamefont {Tokura}},\ }\href {\doibase
  10.1103/PhysRevLett.99.017001} {\bibfield  {journal} {\bibinfo  {journal}
  {Physical Review Letters}\ }\textbf {\bibinfo {volume} {99}},\ \bibinfo
  {pages} {017001} (\bibinfo {year} {2007})}\BibitemShut {NoStop}%
\bibitem [{\citenamefont {Yamada}\ \emph {et~al.}(1999)\citenamefont {Yamada},
  \citenamefont {Kurahashi}, \citenamefont {Endoh}, \citenamefont {Birgeneau},\
  and\ \citenamefont {Shirane}}]{yamada_neutron_1999}%
  \BibitemOpen
  \bibfield  {author} {\bibinfo {author} {\bibfnamefont {K.}~\bibnamefont
  {Yamada}}, \bibinfo {author} {\bibfnamefont {K.}~\bibnamefont {Kurahashi}},
  \bibinfo {author} {\bibfnamefont {Y.}~\bibnamefont {Endoh}}, \bibinfo
  {author} {\bibfnamefont {R.}~\bibnamefont {Birgeneau}}, \ and\ \bibinfo
  {author} {\bibfnamefont {G.}~\bibnamefont {Shirane}},\ }\href {\doibase
  https://doi.org/10.1016/S0022-3697(99)00042-6} {\bibfield  {journal}
  {\bibinfo  {journal} {Journal of Physics and Chemistry of Solids}\ }\textbf
  {\bibinfo {volume} {60}},\ \bibinfo {pages} {1025} (\bibinfo {year}
  {1999})}\BibitemShut {NoStop}%
\bibitem [{\citenamefont {Motoyama}\ \emph {et~al.}(2006)\citenamefont
  {Motoyama}, \citenamefont {Mang}, \citenamefont {Petitgrand}, \citenamefont
  {Yu}, \citenamefont {Vajk}, \citenamefont {Vishik},\ and\ \citenamefont
  {Greven}}]{motoyama_magnetic_2006}%
  \BibitemOpen
  \bibfield  {author} {\bibinfo {author} {\bibfnamefont {E.~M.}\ \bibnamefont
  {Motoyama}}, \bibinfo {author} {\bibfnamefont {P.~K.}\ \bibnamefont {Mang}},
  \bibinfo {author} {\bibfnamefont {D.}~\bibnamefont {Petitgrand}}, \bibinfo
  {author} {\bibfnamefont {G.}~\bibnamefont {Yu}}, \bibinfo {author}
  {\bibfnamefont {O.~P.}\ \bibnamefont {Vajk}}, \bibinfo {author}
  {\bibfnamefont {I.~M.}\ \bibnamefont {Vishik}}, \ and\ \bibinfo {author}
  {\bibfnamefont {M.}~\bibnamefont {Greven}},\ }\href {\doibase
  10.1103/PhysRevLett.96.137002} {\bibfield  {journal} {\bibinfo  {journal}
  {Physical Review Letters}\ }\textbf {\bibinfo {volume} {96}},\ \bibinfo
  {pages} {137002} (\bibinfo {year} {2006})}\BibitemShut {NoStop}%
\bibitem [{\citenamefont {Yu}\ \emph {et~al.}(2010)\citenamefont {Yu},
  \citenamefont {Li}, \citenamefont {Motoyama}, \citenamefont {Hradil},
  \citenamefont {Mole},\ and\ \citenamefont {Greven}}]{yu_two_2010}%
  \BibitemOpen
  \bibfield  {author} {\bibinfo {author} {\bibfnamefont {G.}~\bibnamefont
  {Yu}}, \bibinfo {author} {\bibfnamefont {Y.}~\bibnamefont {Li}}, \bibinfo
  {author} {\bibfnamefont {E.~M.}\ \bibnamefont {Motoyama}}, \bibinfo {author}
  {\bibfnamefont {K.}~\bibnamefont {Hradil}}, \bibinfo {author} {\bibfnamefont
  {R.~A.}\ \bibnamefont {Mole}}, \ and\ \bibinfo {author} {\bibfnamefont
  {M.}~\bibnamefont {Greven}},\ }\href {\doibase 10.1103/PhysRevB.82.172505}
  {\bibfield  {journal} {\bibinfo  {journal} {Physical Review B}\ }\textbf
  {\bibinfo {volume} {82}},\ \bibinfo {pages} {172505} (\bibinfo {year}
  {2010})}\BibitemShut {NoStop}%
\bibitem [{\citenamefont {Matsuda}\ \emph {et~al.}(1992)\citenamefont
  {Matsuda}, \citenamefont {Endoh}, \citenamefont {Yamada}, \citenamefont
  {Kojima}, \citenamefont {Tanaka}, \citenamefont {Birgeneau}, \citenamefont
  {Kastner},\ and\ \citenamefont {Shirane}}]{Matsuda1992}%
  \BibitemOpen
  \bibfield  {author} {\bibinfo {author} {\bibfnamefont {M.}~\bibnamefont
  {Matsuda}}, \bibinfo {author} {\bibfnamefont {Y.}~\bibnamefont {Endoh}},
  \bibinfo {author} {\bibfnamefont {K.}~\bibnamefont {Yamada}}, \bibinfo
  {author} {\bibfnamefont {H.}~\bibnamefont {Kojima}}, \bibinfo {author}
  {\bibfnamefont {I.}~\bibnamefont {Tanaka}}, \bibinfo {author} {\bibfnamefont
  {R.~J.}\ \bibnamefont {Birgeneau}}, \bibinfo {author} {\bibfnamefont {M.~A.}\
  \bibnamefont {Kastner}}, \ and\ \bibinfo {author} {\bibfnamefont
  {G.}~\bibnamefont {Shirane}},\ }\href@noop {} {\bibfield  {journal} {\bibinfo
   {journal} {Phys. Rev. B}\ }\textbf {\bibinfo {volume} {45}},\ \bibinfo
  {pages} {12548} (\bibinfo {year} {1992})}\BibitemShut {NoStop}%
\bibitem [{\citenamefont {Takagi}\ \emph {et~al.}(1989)\citenamefont {Takagi},
  \citenamefont {Ido}, \citenamefont {Ishibashi}, \citenamefont {Uota},\ and\
  \citenamefont {Uchida}}]{Takagi1989}%
  \BibitemOpen
  \bibfield  {author} {\bibinfo {author} {\bibfnamefont {H.}~\bibnamefont
  {Takagi}}, \bibinfo {author} {\bibfnamefont {T.}~\bibnamefont {Ido}},
  \bibinfo {author} {\bibfnamefont {S.}~\bibnamefont {Ishibashi}}, \bibinfo
  {author} {\bibfnamefont {M.}~\bibnamefont {Uota}}, \ and\ \bibinfo {author}
  {\bibfnamefont {S.}~\bibnamefont {Uchida}},\ }\href@noop {} {\bibfield
  {journal} {\bibinfo  {journal} {Physical Review B}\ }\textbf {\bibinfo
  {volume} {40}},\ \bibinfo {pages} {2254} (\bibinfo {year}
  {1989})}\BibitemShut {NoStop}%
\bibitem [{\citenamefont {Tokura}\ \emph {et~al.}(1989)\citenamefont {Tokura},
  \citenamefont {Takagi},\ and\ \citenamefont
  {Uchida}}]{Tokura_superconducting_1989}%
  \BibitemOpen
  \bibfield  {author} {\bibinfo {author} {\bibfnamefont {Y.}~\bibnamefont
  {Tokura}}, \bibinfo {author} {\bibfnamefont {H.}~\bibnamefont {Takagi}}, \
  and\ \bibinfo {author} {\bibfnamefont {S.}~\bibnamefont {Uchida}},\
  }\href@noop {} {\bibfield  {journal} {\bibinfo  {journal} {Nature (London)}\
  }\textbf {\bibinfo {volume} {337}},\ \bibinfo {pages} {345} (\bibinfo {year}
  {1989})}\BibitemShut {NoStop}%
\bibitem [{\citenamefont {Rossat-Mignod}\ \emph {et~al.}(1991)\citenamefont
  {Rossat-Mignod}, \citenamefont {Regnault}, \citenamefont {Vettier},
  \citenamefont {Bourges}, \citenamefont {Burlet}, \citenamefont {Bossy},
  \citenamefont {Henry},\ and\ \citenamefont
  {Lapertot}}]{rossat-mignod_neutron_1991}%
  \BibitemOpen
  \bibfield  {author} {\bibinfo {author} {\bibfnamefont {J.}~\bibnamefont
  {Rossat-Mignod}}, \bibinfo {author} {\bibfnamefont {L.}~\bibnamefont
  {Regnault}}, \bibinfo {author} {\bibfnamefont {C.}~\bibnamefont {Vettier}},
  \bibinfo {author} {\bibfnamefont {P.}~\bibnamefont {Bourges}}, \bibinfo
  {author} {\bibfnamefont {P.}~\bibnamefont {Burlet}}, \bibinfo {author}
  {\bibfnamefont {J.}~\bibnamefont {Bossy}}, \bibinfo {author} {\bibfnamefont
  {J.}~\bibnamefont {Henry}}, \ and\ \bibinfo {author} {\bibfnamefont
  {G.}~\bibnamefont {Lapertot}},\ }\href {\doibase
  10.1016/0921-4534(91)91955-4} {\bibfield  {journal} {\bibinfo  {journal}
  {Physica C: Superconductivity}\ }\textbf {\bibinfo {volume} {185-189}},\
  \bibinfo {pages} {86} (\bibinfo {year} {1991})}\BibitemShut {NoStop}%
\bibitem [{\citenamefont {Mook}\ \emph {et~al.}(1993)\citenamefont {Mook},
  \citenamefont {Yethiraj}, \citenamefont {Aeppli}, \citenamefont {Mason},\
  and\ \citenamefont {Armstrong}}]{Mook_1993_YBCO_resonance}%
  \BibitemOpen
  \bibfield  {author} {\bibinfo {author} {\bibfnamefont {H.~A.}\ \bibnamefont
  {Mook}}, \bibinfo {author} {\bibfnamefont {M.}~\bibnamefont {Yethiraj}},
  \bibinfo {author} {\bibfnamefont {G.}~\bibnamefont {Aeppli}}, \bibinfo
  {author} {\bibfnamefont {T.~E.}\ \bibnamefont {Mason}}, \ and\ \bibinfo
  {author} {\bibfnamefont {T.}~\bibnamefont {Armstrong}},\ }\href {\doibase
  10.1103/PhysRevLett.70.3490} {\bibfield  {journal} {\bibinfo  {journal}
  {Phys. Rev. Lett.}\ }\textbf {\bibinfo {volume} {70}},\ \bibinfo {pages}
  {3490} (\bibinfo {year} {1993})}\BibitemShut {NoStop}%
\bibitem [{\citenamefont {Dai}\ \emph {et~al.}(2001)\citenamefont {Dai},
  \citenamefont {Mook}, \citenamefont {Hunt},\ and\ \citenamefont
  {Do\ifmmode~\breve{g}\else \u{g}\fi{}an}}]{Dai_2001_YBCO_resonance}%
  \BibitemOpen
  \bibfield  {author} {\bibinfo {author} {\bibfnamefont {P.}~\bibnamefont
  {Dai}}, \bibinfo {author} {\bibfnamefont {H.~A.}\ \bibnamefont {Mook}},
  \bibinfo {author} {\bibfnamefont {R.~D.}\ \bibnamefont {Hunt}}, \ and\
  \bibinfo {author} {\bibfnamefont {F.}~\bibnamefont {Do\ifmmode~\breve{g}\else
  \u{g}\fi{}an}},\ }\href {\doibase 10.1103/PhysRevB.63.054525} {\bibfield
  {journal} {\bibinfo  {journal} {Phys. Rev. B}\ }\textbf {\bibinfo {volume}
  {63}},\ \bibinfo {pages} {054525} (\bibinfo {year} {2001})}\BibitemShut
  {NoStop}%
\bibitem [{\citenamefont {Tranquada}\ \emph {et~al.}(2004)\citenamefont
  {Tranquada}, \citenamefont {Woo}, \citenamefont {Perring}, \citenamefont
  {Goka}, \citenamefont {Gu}, \citenamefont {Xu}, \citenamefont {Fujita},\ and\
  \citenamefont {Yamada}}]{tranquada_quantum_2004}%
  \BibitemOpen
  \bibfield  {author} {\bibinfo {author} {\bibfnamefont {J.~M.}\ \bibnamefont
  {Tranquada}}, \bibinfo {author} {\bibfnamefont {H.}~\bibnamefont {Woo}},
  \bibinfo {author} {\bibfnamefont {T.~G.}\ \bibnamefont {Perring}}, \bibinfo
  {author} {\bibfnamefont {H.}~\bibnamefont {Goka}}, \bibinfo {author}
  {\bibfnamefont {G.~D.}\ \bibnamefont {Gu}}, \bibinfo {author} {\bibfnamefont
  {G.}~\bibnamefont {Xu}}, \bibinfo {author} {\bibfnamefont {M.}~\bibnamefont
  {Fujita}}, \ and\ \bibinfo {author} {\bibfnamefont {K.}~\bibnamefont
  {Yamada}},\ }\href {\doibase 10.1038/nature02574} {\bibfield  {journal}
  {\bibinfo  {journal} {Nature}\ }\textbf {\bibinfo {volume} {429}},\ \bibinfo
  {pages} {534} (\bibinfo {year} {2004})}\BibitemShut {NoStop}%
\bibitem [{\citenamefont {Hayden}\ \emph {et~al.}(2004)\citenamefont {Hayden},
  \citenamefont {Mook}, \citenamefont {Dai}, \citenamefont {Perring},\ and\
  \citenamefont {Doğan}}]{Hayden_2004_high-energy-spin-excitations}%
  \BibitemOpen
  \bibfield  {author} {\bibinfo {author} {\bibfnamefont {S.~M.}\ \bibnamefont
  {Hayden}}, \bibinfo {author} {\bibfnamefont {H.~A.}\ \bibnamefont {Mook}},
  \bibinfo {author} {\bibfnamefont {P.}~\bibnamefont {Dai}}, \bibinfo {author}
  {\bibfnamefont {T.~G.}\ \bibnamefont {Perring}}, \ and\ \bibinfo {author}
  {\bibfnamefont {F.}~\bibnamefont {Doğan}},\ }\href {\doibase
  10.1038/nature02576} {\bibfield  {journal} {\bibinfo  {journal} {Nature}\
  }\textbf {\bibinfo {volume} {429}},\ \bibinfo {pages} {531} (\bibinfo {year}
  {2004})}\BibitemShut {NoStop}%
\bibitem [{\citenamefont {Dai}\ \emph {et~al.}(2007)\citenamefont {Dai},
  \citenamefont {Wilson},\ and\ \citenamefont {Li}}]{dai_evolution_2007}%
  \BibitemOpen
  \bibfield  {author} {\bibinfo {author} {\bibfnamefont {P.}~\bibnamefont
  {Dai}}, \bibinfo {author} {\bibfnamefont {S.~D.}\ \bibnamefont {Wilson}}, \
  and\ \bibinfo {author} {\bibfnamefont {S.}~\bibnamefont {Li}},\ }\href
  {\doibase 10.1016/j.physc.2007.03.090} {\bibfield  {journal} {\bibinfo
  {journal} {Physica C: Superconductivity and its Applications}\ }\textbf
  {\bibinfo {volume} {460-462}},\ \bibinfo {pages} {52} (\bibinfo {year}
  {2007})}\BibitemShut {NoStop}%
\bibitem [{\citenamefont {Momma}\ and\ \citenamefont
  {Izumi}(2011)}]{Momma_VESTA_software}%
  \BibitemOpen
  \bibfield  {author} {\bibinfo {author} {\bibfnamefont {K.}~\bibnamefont
  {Momma}}\ and\ \bibinfo {author} {\bibfnamefont {F.}~\bibnamefont {Izumi}},\
  }\href {\doibase 10.1107/S0021889811038970} {\bibfield  {journal} {\bibinfo
  {journal} {Journal of Applied Crystallography}\ }\textbf {\bibinfo {volume}
  {44}},\ \bibinfo {pages} {1272} (\bibinfo {year} {2011})}\BibitemShut
  {NoStop}%
\bibitem [{\citenamefont {Belokoneva}\ \emph {et~al.}(1991)\citenamefont
  {Belokoneva}, \citenamefont {Leonyuk},\ and\ \citenamefont
  {Leonyuk}}]{belokoneva1991preparation}%
  \BibitemOpen
  \bibfield  {author} {\bibinfo {author} {\bibfnamefont {E.}~\bibnamefont
  {Belokoneva}}, \bibinfo {author} {\bibfnamefont {L.}~\bibnamefont {Leonyuk}},
  \ and\ \bibinfo {author} {\bibfnamefont {N.}~\bibnamefont {Leonyuk}},\
  }\href@noop {} {\bibfield  {journal} {\bibinfo  {journal} {Sverkhprovodimost
  Fiz Khim Tek}\ }\textbf {\bibinfo {volume} {4}},\ \bibinfo {pages} {563}
  (\bibinfo {year} {1991})}\BibitemShut {NoStop}%
\bibitem [{\citenamefont {Xu}\ \emph {et~al.}(2013)\citenamefont {Xu},
  \citenamefont {Xu},\ and\ \citenamefont {Tranquada}}]{xu_absolute_2013}%
  \BibitemOpen
  \bibfield  {author} {\bibinfo {author} {\bibfnamefont {G.}~\bibnamefont
  {Xu}}, \bibinfo {author} {\bibfnamefont {Z.}~\bibnamefont {Xu}}, \ and\
  \bibinfo {author} {\bibfnamefont {J.~M.}\ \bibnamefont {Tranquada}},\ }\href
  {\doibase 10.1063/1.4818323} {\bibfield  {journal} {\bibinfo  {journal}
  {Review of Scientific Instruments}\ }\textbf {\bibinfo {volume} {84}},\
  \bibinfo {pages} {083906} (\bibinfo {year} {2013})}\BibitemShut {NoStop}%
\bibitem [{\citenamefont {Moran}\ \emph {et~al.}(1989)\citenamefont {Moran},
  \citenamefont {Nazzal}, \citenamefont {Huang},\ and\ \citenamefont
  {Torrance}}]{Moran_extra_oxygen_1989}%
  \BibitemOpen
  \bibfield  {author} {\bibinfo {author} {\bibfnamefont {E.}~\bibnamefont
  {Moran}}, \bibinfo {author} {\bibfnamefont {A.}~\bibnamefont {Nazzal}},
  \bibinfo {author} {\bibfnamefont {T.}~\bibnamefont {Huang}}, \ and\ \bibinfo
  {author} {\bibfnamefont {J.}~\bibnamefont {Torrance}},\ }\href {\doibase
  https://doi.org/10.1016/0921-4534(89)90448-6} {\bibfield  {journal} {\bibinfo
   {journal} {Physica C: Superconductivity}\ }\textbf {\bibinfo {volume}
  {160}},\ \bibinfo {pages} {30} (\bibinfo {year} {1989})}\BibitemShut
  {NoStop}%
\bibitem [{\citenamefont {Tarascon}\ \emph {et~al.}(1989)\citenamefont
  {Tarascon}, \citenamefont {Wang}, \citenamefont {Greene}, \citenamefont
  {Bagley}, \citenamefont {Hull}, \citenamefont {D'Egidio}, \citenamefont
  {Miceli}, \citenamefont {Wang}, \citenamefont {Jing}, \citenamefont
  {Clayhold}, \citenamefont {Brawner},\ and\ \citenamefont
  {Ong}}]{Tarascon_growth_1989}%
  \BibitemOpen
  \bibfield  {author} {\bibinfo {author} {\bibfnamefont {J.-M.}\ \bibnamefont
  {Tarascon}}, \bibinfo {author} {\bibfnamefont {E.}~\bibnamefont {Wang}},
  \bibinfo {author} {\bibfnamefont {L.~H.}\ \bibnamefont {Greene}}, \bibinfo
  {author} {\bibfnamefont {B.~G.}\ \bibnamefont {Bagley}}, \bibinfo {author}
  {\bibfnamefont {G.~W.}\ \bibnamefont {Hull}}, \bibinfo {author}
  {\bibfnamefont {S.~M.}\ \bibnamefont {D'Egidio}}, \bibinfo {author}
  {\bibfnamefont {P.~F.}\ \bibnamefont {Miceli}}, \bibinfo {author}
  {\bibfnamefont {Z.~Z.}\ \bibnamefont {Wang}}, \bibinfo {author}
  {\bibfnamefont {T.~W.}\ \bibnamefont {Jing}}, \bibinfo {author}
  {\bibfnamefont {J.}~\bibnamefont {Clayhold}}, \bibinfo {author}
  {\bibfnamefont {D.}~\bibnamefont {Brawner}}, \ and\ \bibinfo {author}
  {\bibfnamefont {N.~P.}\ \bibnamefont {Ong}},\ }\href {\doibase
  10.1103/PhysRevB.40.4494} {\bibfield  {journal} {\bibinfo  {journal} {Phys.
  Rev. B}\ }\textbf {\bibinfo {volume} {40}},\ \bibinfo {pages} {4494}
  (\bibinfo {year} {1989})}\BibitemShut {NoStop}%
\bibitem [{\citenamefont {Radaelli}\ \emph {et~al.}(1994)\citenamefont
  {Radaelli}, \citenamefont {Jorgensen}, \citenamefont {Schultz}, \citenamefont
  {Peng},\ and\ \citenamefont {Greene}}]{Radaelli_apical_oxygen_1994}%
  \BibitemOpen
  \bibfield  {author} {\bibinfo {author} {\bibfnamefont {P.~G.}\ \bibnamefont
  {Radaelli}}, \bibinfo {author} {\bibfnamefont {J.~D.}\ \bibnamefont
  {Jorgensen}}, \bibinfo {author} {\bibfnamefont {A.~J.}\ \bibnamefont
  {Schultz}}, \bibinfo {author} {\bibfnamefont {J.~L.}\ \bibnamefont {Peng}}, \
  and\ \bibinfo {author} {\bibfnamefont {R.~L.}\ \bibnamefont {Greene}},\
  }\href {\doibase 10.1103/PhysRevB.49.15322} {\bibfield  {journal} {\bibinfo
  {journal} {Phys. Rev. B}\ }\textbf {\bibinfo {volume} {49}},\ \bibinfo
  {pages} {15322} (\bibinfo {year} {1994})}\BibitemShut {NoStop}%
\bibitem [{\citenamefont {Schultz}\ \emph {et~al.}(1996)\citenamefont
  {Schultz}, \citenamefont {Jorgensen}, \citenamefont {Peng},\ and\
  \citenamefont {Greene}}]{schultz_single-crystal_1996}%
  \BibitemOpen
  \bibfield  {author} {\bibinfo {author} {\bibfnamefont {A.~J.}\ \bibnamefont
  {Schultz}}, \bibinfo {author} {\bibfnamefont {J.~D.}\ \bibnamefont
  {Jorgensen}}, \bibinfo {author} {\bibfnamefont {J.~L.}\ \bibnamefont {Peng}},
  \ and\ \bibinfo {author} {\bibfnamefont {R.~L.}\ \bibnamefont {Greene}},\
  }\href {\doibase 10.1103/PhysRevB.53.5157} {\bibfield  {journal} {\bibinfo
  {journal} {Physical Review B}\ }\textbf {\bibinfo {volume} {53}},\ \bibinfo
  {pages} {5157} (\bibinfo {year} {1996})}\BibitemShut {NoStop}%
\bibitem [{\citenamefont {Klamut}\ \emph {et~al.}(1997)\citenamefont {Klamut},
  \citenamefont {Sikora}, \citenamefont {Bukowski}, \citenamefont {Dabrowski},\
  and\ \citenamefont {Klamut}}]{Klamut_oxygen_pressure_1997}%
  \BibitemOpen
  \bibfield  {author} {\bibinfo {author} {\bibfnamefont {P.}~\bibnamefont
  {Klamut}}, \bibinfo {author} {\bibfnamefont {A.}~\bibnamefont {Sikora}},
  \bibinfo {author} {\bibfnamefont {Z.}~\bibnamefont {Bukowski}}, \bibinfo
  {author} {\bibfnamefont {B.}~\bibnamefont {Dabrowski}}, \ and\ \bibinfo
  {author} {\bibfnamefont {J.}~\bibnamefont {Klamut}},\ }\href {\doibase
  https://doi.org/10.1016/S0921-4534(97)00340-7} {\bibfield  {journal}
  {\bibinfo  {journal} {Physica C: Superconductivity}\ }\textbf {\bibinfo
  {volume} {282-287}},\ \bibinfo {pages} {541} (\bibinfo {year}
  {1997})}\BibitemShut {NoStop}%
\bibitem [{\citenamefont {Navarro}\ \emph {et~al.}(2001)\citenamefont
  {Navarro}, \citenamefont {Jaque}, \citenamefont {Villegas}, \citenamefont
  {Martin}, \citenamefont {Serquis}, \citenamefont {Prado}, \citenamefont
  {Caneiro},\ and\ \citenamefont {Vicent}}]{navarro_oxygen_2001}%
  \BibitemOpen
  \bibfield  {author} {\bibinfo {author} {\bibfnamefont {E.}~\bibnamefont
  {Navarro}}, \bibinfo {author} {\bibfnamefont {D.}~\bibnamefont {Jaque}},
  \bibinfo {author} {\bibfnamefont {J.}~\bibnamefont {Villegas}}, \bibinfo
  {author} {\bibfnamefont {J.}~\bibnamefont {Martin}}, \bibinfo {author}
  {\bibfnamefont {A.}~\bibnamefont {Serquis}}, \bibinfo {author} {\bibfnamefont
  {F.}~\bibnamefont {Prado}}, \bibinfo {author} {\bibfnamefont
  {A.}~\bibnamefont {Caneiro}}, \ and\ \bibinfo {author} {\bibfnamefont
  {J.}~\bibnamefont {Vicent}},\ }\href {\doibase 10.1016/S0925-8388(01)01198-7}
  {\bibfield  {journal} {\bibinfo  {journal} {Journal of Alloys and Compounds}\
  }\textbf {\bibinfo {volume} {323-324}},\ \bibinfo {pages} {580} (\bibinfo
  {year} {2001})}\BibitemShut {NoStop}%
\bibitem [{\citenamefont {Takayama-Muromachi}\ \emph
  {et~al.}(1989)\citenamefont {Takayama-Muromachi}, \citenamefont {Izumi},
  \citenamefont {Uchida}, \citenamefont {Kato},\ and\ \citenamefont
  {Asano}}]{Takayama-Muromachi_oxygen_deficiency_1989}%
  \BibitemOpen
  \bibfield  {author} {\bibinfo {author} {\bibfnamefont {E.}~\bibnamefont
  {Takayama-Muromachi}}, \bibinfo {author} {\bibfnamefont {F.}~\bibnamefont
  {Izumi}}, \bibinfo {author} {\bibfnamefont {Y.}~\bibnamefont {Uchida}},
  \bibinfo {author} {\bibfnamefont {K.}~\bibnamefont {Kato}}, \ and\ \bibinfo
  {author} {\bibfnamefont {H.}~\bibnamefont {Asano}},\ }\href {\doibase
  https://doi.org/10.1016/0921-4534(89)91296-3} {\bibfield  {journal} {\bibinfo
   {journal} {Physica C: Superconductivity}\ }\textbf {\bibinfo {volume}
  {159}},\ \bibinfo {pages} {634} (\bibinfo {year} {1989})}\BibitemShut
  {NoStop}%
\bibitem [{\citenamefont {Suzuki}\ \emph {et~al.}(1990)\citenamefont {Suzuki},
  \citenamefont {Kishio}, \citenamefont {Hasegawa},\ and\ \citenamefont
  {Kitazawa}}]{Suzuki_oxygen_nonstoichiometry_1990}%
  \BibitemOpen
  \bibfield  {author} {\bibinfo {author} {\bibfnamefont {K.}~\bibnamefont
  {Suzuki}}, \bibinfo {author} {\bibfnamefont {K.}~\bibnamefont {Kishio}},
  \bibinfo {author} {\bibfnamefont {T.}~\bibnamefont {Hasegawa}}, \ and\
  \bibinfo {author} {\bibfnamefont {K.}~\bibnamefont {Kitazawa}},\ }\href
  {\doibase https://doi.org/10.1016/0921-4534(90)90416-C} {\bibfield  {journal}
  {\bibinfo  {journal} {Physica C: Superconductivity}\ }\textbf {\bibinfo
  {volume} {166}},\ \bibinfo {pages} {357} (\bibinfo {year}
  {1990})}\BibitemShut {NoStop}%
\bibitem [{\citenamefont {Kim}\ and\ \citenamefont
  {Gaskell}(1993)}]{Kim_phase_stability_diagram_1993}%
  \BibitemOpen
  \bibfield  {author} {\bibinfo {author} {\bibfnamefont {J.}~\bibnamefont
  {Kim}}\ and\ \bibinfo {author} {\bibfnamefont {D.}~\bibnamefont {Gaskell}},\
  }\href {\doibase https://doi.org/10.1016/0921-4534(93)90549-6} {\bibfield
  {journal} {\bibinfo  {journal} {Physica C: Superconductivity}\ }\textbf
  {\bibinfo {volume} {209}},\ \bibinfo {pages} {381} (\bibinfo {year}
  {1993})}\BibitemShut {NoStop}%
\bibitem [{\citenamefont {Armitage}\ \emph {et~al.}(2010)\citenamefont
  {Armitage}, \citenamefont {Fournier},\ and\ \citenamefont
  {Greene}}]{armitage_progress_2010}%
  \BibitemOpen
  \bibfield  {author} {\bibinfo {author} {\bibfnamefont {N.~P.}\ \bibnamefont
  {Armitage}}, \bibinfo {author} {\bibfnamefont {P.}~\bibnamefont {Fournier}},
  \ and\ \bibinfo {author} {\bibfnamefont {R.~L.}\ \bibnamefont {Greene}},\
  }\href {\doibase 10.1103/RevModPhys.82.2421} {\bibfield  {journal} {\bibinfo
  {journal} {Reviews of Modern Physics}\ }\textbf {\bibinfo {volume} {82}},\
  \bibinfo {pages} {2421} (\bibinfo {year} {2010})}\BibitemShut {NoStop}%
\bibitem [{\citenamefont {Riou}\ \emph {et~al.}(2004)\citenamefont {Riou},
  \citenamefont {Richard}, \citenamefont {Jandl}, \citenamefont {Poirier},
  \citenamefont {Fournier}, \citenamefont {Nekvasil}, \citenamefont {Barilo},\
  and\ \citenamefont {Kurnevich}}]{Riou_2004_raman}%
  \BibitemOpen
  \bibfield  {author} {\bibinfo {author} {\bibfnamefont {G.}~\bibnamefont
  {Riou}}, \bibinfo {author} {\bibfnamefont {P.}~\bibnamefont {Richard}},
  \bibinfo {author} {\bibfnamefont {S.}~\bibnamefont {Jandl}}, \bibinfo
  {author} {\bibfnamefont {M.}~\bibnamefont {Poirier}}, \bibinfo {author}
  {\bibfnamefont {P.}~\bibnamefont {Fournier}}, \bibinfo {author}
  {\bibfnamefont {V.}~\bibnamefont {Nekvasil}}, \bibinfo {author}
  {\bibfnamefont {S.~N.}\ \bibnamefont {Barilo}}, \ and\ \bibinfo {author}
  {\bibfnamefont {L.~A.}\ \bibnamefont {Kurnevich}},\ }\href {\doibase
  10.1103/PhysRevB.69.024511} {\bibfield  {journal} {\bibinfo  {journal} {Phys.
  Rev. B}\ }\textbf {\bibinfo {volume} {69}},\ \bibinfo {pages} {024511}
  (\bibinfo {year} {2004})}\BibitemShut {NoStop}%
\bibitem [{\citenamefont {Richard}\ \emph {et~al.}(2004)\citenamefont
  {Richard}, \citenamefont {Riou}, \citenamefont {Hetel}, \citenamefont
  {Jandl}, \citenamefont {Poirier},\ and\ \citenamefont
  {Fournier}}]{Richard_2004_raman}%
  \BibitemOpen
  \bibfield  {author} {\bibinfo {author} {\bibfnamefont {P.}~\bibnamefont
  {Richard}}, \bibinfo {author} {\bibfnamefont {G.}~\bibnamefont {Riou}},
  \bibinfo {author} {\bibfnamefont {I.}~\bibnamefont {Hetel}}, \bibinfo
  {author} {\bibfnamefont {S.}~\bibnamefont {Jandl}}, \bibinfo {author}
  {\bibfnamefont {M.}~\bibnamefont {Poirier}}, \ and\ \bibinfo {author}
  {\bibfnamefont {P.}~\bibnamefont {Fournier}},\ }\href {\doibase
  10.1103/PhysRevB.70.064513} {\bibfield  {journal} {\bibinfo  {journal} {Phys.
  Rev. B}\ }\textbf {\bibinfo {volume} {70}},\ \bibinfo {pages} {064513}
  (\bibinfo {year} {2004})}\BibitemShut {NoStop}%
\bibitem [{\citenamefont {Xu}\ \emph {et~al.}(1996)\citenamefont {Xu},
  \citenamefont {Mao}, \citenamefont {Jiang}, \citenamefont {Peng},\ and\
  \citenamefont {Greene}}]{Xu_1996_oxygen_transport_properties}%
  \BibitemOpen
  \bibfield  {author} {\bibinfo {author} {\bibfnamefont {X.~Q.}\ \bibnamefont
  {Xu}}, \bibinfo {author} {\bibfnamefont {S.~N.}\ \bibnamefont {Mao}},
  \bibinfo {author} {\bibfnamefont {W.}~\bibnamefont {Jiang}}, \bibinfo
  {author} {\bibfnamefont {J.~L.}\ \bibnamefont {Peng}}, \ and\ \bibinfo
  {author} {\bibfnamefont {R.~L.}\ \bibnamefont {Greene}},\ }\href {\doibase
  10.1103/PhysRevB.53.871} {\bibfield  {journal} {\bibinfo  {journal} {Phys.
  Rev. B}\ }\textbf {\bibinfo {volume} {53}},\ \bibinfo {pages} {871} (\bibinfo
  {year} {1996})}\BibitemShut {NoStop}%
\bibitem [{\citenamefont {Kurahashi}\ \emph {et~al.}(2002)\citenamefont
  {Kurahashi}, \citenamefont {Matsushita}, \citenamefont {Fujita},\ and\
  \citenamefont {Yamada}}]{Kurahashi_2002_heat_treatment}%
  \BibitemOpen
  \bibfield  {author} {\bibinfo {author} {\bibfnamefont {K.}~\bibnamefont
  {Kurahashi}}, \bibinfo {author} {\bibfnamefont {H.}~\bibnamefont
  {Matsushita}}, \bibinfo {author} {\bibfnamefont {M.}~\bibnamefont {Fujita}},
  \ and\ \bibinfo {author} {\bibfnamefont {K.}~\bibnamefont {Yamada}},\ }\href
  {\doibase 10.1143/JPSJ.71.910} {\bibfield  {journal} {\bibinfo  {journal}
  {Journal of the Physical Society of Japan}\ }\textbf {\bibinfo {volume}
  {71}},\ \bibinfo {pages} {910} (\bibinfo {year} {2002})}\BibitemShut
  {NoStop}%
\bibitem [{\citenamefont {Mang}\ \emph
  {et~al.}(2004{\natexlab{a}})\citenamefont {Mang}, \citenamefont {Larochelle},
  \citenamefont {Mehta}, \citenamefont {Vajk}, \citenamefont {Erickson},
  \citenamefont {Lu}, \citenamefont {Buyers}, \citenamefont {Marshall},
  \citenamefont {Prokes},\ and\ \citenamefont
  {Greven}}]{Mang_2004_chemical_inhomogeneity}%
  \BibitemOpen
  \bibfield  {author} {\bibinfo {author} {\bibfnamefont {P.~K.}\ \bibnamefont
  {Mang}}, \bibinfo {author} {\bibfnamefont {S.}~\bibnamefont {Larochelle}},
  \bibinfo {author} {\bibfnamefont {A.}~\bibnamefont {Mehta}}, \bibinfo
  {author} {\bibfnamefont {O.~P.}\ \bibnamefont {Vajk}}, \bibinfo {author}
  {\bibfnamefont {A.~S.}\ \bibnamefont {Erickson}}, \bibinfo {author}
  {\bibfnamefont {L.}~\bibnamefont {Lu}}, \bibinfo {author} {\bibfnamefont
  {W.~J.~L.}\ \bibnamefont {Buyers}}, \bibinfo {author} {\bibfnamefont {A.~F.}\
  \bibnamefont {Marshall}}, \bibinfo {author} {\bibfnamefont {K.}~\bibnamefont
  {Prokes}}, \ and\ \bibinfo {author} {\bibfnamefont {M.}~\bibnamefont
  {Greven}},\ }\href {\doibase 10.1103/PhysRevB.70.094507} {\bibfield
  {journal} {\bibinfo  {journal} {Phys. Rev. B}\ }\textbf {\bibinfo {volume}
  {70}},\ \bibinfo {pages} {094507} (\bibinfo {year}
  {2004}{\natexlab{a}})}\BibitemShut {NoStop}%
\bibitem [{\citenamefont {Kang}\ \emph {et~al.}(2007)\citenamefont {Kang},
  \citenamefont {Dai}, \citenamefont {Campbell}, \citenamefont {Chupas},
  \citenamefont {Rosenkranz}, \citenamefont {Lee}, \citenamefont {Huang},
  \citenamefont {Li}, \citenamefont {Komiya},\ and\ \citenamefont
  {Ando}}]{Kang_2007_microscopic_annealing}%
  \BibitemOpen
  \bibfield  {author} {\bibinfo {author} {\bibfnamefont {H.~J.}\ \bibnamefont
  {Kang}}, \bibinfo {author} {\bibfnamefont {P.}~\bibnamefont {Dai}}, \bibinfo
  {author} {\bibfnamefont {B.~J.}\ \bibnamefont {Campbell}}, \bibinfo {author}
  {\bibfnamefont {P.~J.}\ \bibnamefont {Chupas}}, \bibinfo {author}
  {\bibfnamefont {S.}~\bibnamefont {Rosenkranz}}, \bibinfo {author}
  {\bibfnamefont {P.~L.}\ \bibnamefont {Lee}}, \bibinfo {author} {\bibfnamefont
  {Q.}~\bibnamefont {Huang}}, \bibinfo {author} {\bibfnamefont
  {S.}~\bibnamefont {Li}}, \bibinfo {author} {\bibfnamefont {S.}~\bibnamefont
  {Komiya}}, \ and\ \bibinfo {author} {\bibfnamefont {Y.}~\bibnamefont
  {Ando}},\ }\href {\doibase 10.1038/nmat1847} {\bibfield  {journal} {\bibinfo
  {journal} {Nature Materials}\ }\textbf {\bibinfo {volume} {6}},\ \bibinfo
  {pages} {224} (\bibinfo {year} {2007})}\BibitemShut {NoStop}%
\bibitem [{\citenamefont {Motoyama}\ \emph {et~al.}(1996)\citenamefont
  {Motoyama}, \citenamefont {Eisaki},\ and\ \citenamefont
  {Uchida}}]{motoyama_magnetic_1996}%
  \BibitemOpen
  \bibfield  {author} {\bibinfo {author} {\bibfnamefont {N.}~\bibnamefont
  {Motoyama}}, \bibinfo {author} {\bibfnamefont {H.}~\bibnamefont {Eisaki}}, \
  and\ \bibinfo {author} {\bibfnamefont {S.}~\bibnamefont {Uchida}},\ }\href
  {\doibase 10.1103/PhysRevLett.76.3212} {\bibfield  {journal} {\bibinfo
  {journal} {Phys. Rev. Lett.}\ }\textbf {\bibinfo {volume} {76}},\ \bibinfo
  {pages} {3212} (\bibinfo {year} {1996})}\BibitemShut {NoStop}%
\bibitem [{\citenamefont {Rice}(1997)}]{rice_rich_1997}%
  \BibitemOpen
  \bibfield  {author} {\bibinfo {author} {\bibfnamefont {T.}~\bibnamefont
  {Rice}},\ }\href {\doibase https://doi.org/10.1016/S0921-4526(97)00501-2}
  {\bibfield  {journal} {\bibinfo  {journal} {Physica B: Condensed Matter}\
  }\textbf {\bibinfo {volume} {241-243}},\ \bibinfo {pages} {5} (\bibinfo
  {year} {1997})},\ \bibinfo {note} {proceedings of the International
  Conference on Neutron Scattering}\BibitemShut {NoStop}%
\bibitem [{\citenamefont {Zaliznyak}\ \emph {et~al.}(1999)\citenamefont
  {Zaliznyak}, \citenamefont {Broholm}, \citenamefont {Kibune}, \citenamefont
  {Nohara},\ and\ \citenamefont {Takagi}}]{zaliznyak_anisotropic_1999}%
  \BibitemOpen
  \bibfield  {author} {\bibinfo {author} {\bibfnamefont {I.~A.}\ \bibnamefont
  {Zaliznyak}}, \bibinfo {author} {\bibfnamefont {C.}~\bibnamefont {Broholm}},
  \bibinfo {author} {\bibfnamefont {M.}~\bibnamefont {Kibune}}, \bibinfo
  {author} {\bibfnamefont {M.}~\bibnamefont {Nohara}}, \ and\ \bibinfo {author}
  {\bibfnamefont {H.}~\bibnamefont {Takagi}},\ }\href {\doibase
  10.1103/PhysRevLett.83.5370} {\bibfield  {journal} {\bibinfo  {journal}
  {Phys. Rev. Lett.}\ }\textbf {\bibinfo {volume} {83}},\ \bibinfo {pages}
  {5370} (\bibinfo {year} {1999})}\BibitemShut {NoStop}%
\bibitem [{\citenamefont {Hammerath}\ \emph {et~al.}(2011)\citenamefont
  {Hammerath}, \citenamefont {Nishimoto}, \citenamefont {Grafe}, \citenamefont
  {Wolter}, \citenamefont {Kataev}, \citenamefont {Ribeiro}, \citenamefont
  {Hess}, \citenamefont {Drechsler},\ and\ \citenamefont
  {Büchner}}]{hammerath_spin_2011}%
  \BibitemOpen
  \bibfield  {author} {\bibinfo {author} {\bibfnamefont {F.}~\bibnamefont
  {Hammerath}}, \bibinfo {author} {\bibfnamefont {S.}~\bibnamefont
  {Nishimoto}}, \bibinfo {author} {\bibfnamefont {H.-J.}\ \bibnamefont
  {Grafe}}, \bibinfo {author} {\bibfnamefont {A.~U.~B.}\ \bibnamefont
  {Wolter}}, \bibinfo {author} {\bibfnamefont {V.}~\bibnamefont {Kataev}},
  \bibinfo {author} {\bibfnamefont {P.}~\bibnamefont {Ribeiro}}, \bibinfo
  {author} {\bibfnamefont {C.}~\bibnamefont {Hess}}, \bibinfo {author}
  {\bibfnamefont {S.-L.}\ \bibnamefont {Drechsler}}, \ and\ \bibinfo {author}
  {\bibfnamefont {B.}~\bibnamefont {Büchner}},\ }\href {\doibase
  10.1103/PhysRevLett.107.017203} {\bibfield  {journal} {\bibinfo  {journal}
  {Physical Review Letters}\ }\textbf {\bibinfo {volume} {107}},\ \bibinfo
  {pages} {017203} (\bibinfo {year} {2011})}\BibitemShut {NoStop}%
\bibitem [{\citenamefont {Simutis}\ \emph {et~al.}(2013)\citenamefont
  {Simutis}, \citenamefont {Gvasaliya}, \citenamefont {Månsson}, \citenamefont
  {Chernyshev}, \citenamefont {Mohan}, \citenamefont {Singh}, \citenamefont
  {Hess}, \citenamefont {Savici}, \citenamefont {Kolesnikov}, \citenamefont
  {Piovano}, \citenamefont {Perring}, \citenamefont {Zaliznyak}, \citenamefont
  {Büchner},\ and\ \citenamefont {Zheludev}}]{simutis_spin_2013}%
  \BibitemOpen
  \bibfield  {author} {\bibinfo {author} {\bibfnamefont {G.}~\bibnamefont
  {Simutis}}, \bibinfo {author} {\bibfnamefont {S.}~\bibnamefont {Gvasaliya}},
  \bibinfo {author} {\bibfnamefont {M.}~\bibnamefont {Månsson}}, \bibinfo
  {author} {\bibfnamefont {A.~L.}\ \bibnamefont {Chernyshev}}, \bibinfo
  {author} {\bibfnamefont {A.}~\bibnamefont {Mohan}}, \bibinfo {author}
  {\bibfnamefont {S.}~\bibnamefont {Singh}}, \bibinfo {author} {\bibfnamefont
  {C.}~\bibnamefont {Hess}}, \bibinfo {author} {\bibfnamefont {A.~T.}\
  \bibnamefont {Savici}}, \bibinfo {author} {\bibfnamefont {A.~I.}\
  \bibnamefont {Kolesnikov}}, \bibinfo {author} {\bibfnamefont
  {A.}~\bibnamefont {Piovano}}, \bibinfo {author} {\bibfnamefont
  {T.}~\bibnamefont {Perring}}, \bibinfo {author} {\bibfnamefont
  {I.}~\bibnamefont {Zaliznyak}}, \bibinfo {author} {\bibfnamefont
  {B.}~\bibnamefont {Büchner}}, \ and\ \bibinfo {author} {\bibfnamefont
  {A.}~\bibnamefont {Zheludev}},\ }\href {\doibase
  10.1103/PhysRevLett.111.067204} {\bibfield  {journal} {\bibinfo  {journal}
  {Physical Review Letters}\ }\textbf {\bibinfo {volume} {111}},\ \bibinfo
  {pages} {067204} (\bibinfo {year} {2013})}\BibitemShut {NoStop}%
\bibitem [{\citenamefont {Simutis}\ \emph {et~al.}(2017)\citenamefont
  {Simutis}, \citenamefont {Gvasaliya}, \citenamefont {Beesetty}, \citenamefont
  {Yoshida}, \citenamefont {Robert}, \citenamefont {Petit}, \citenamefont
  {Kolesnikov}, \citenamefont {Stone}, \citenamefont {Bourdarot}, \citenamefont
  {Walker}, \citenamefont {Adroja}, \citenamefont {Sobolev}, \citenamefont
  {Hess}, \citenamefont {Masuda}, \citenamefont {Revcolevschi}, \citenamefont
  {Büchner},\ and\ \citenamefont {Zheludev}}]{simutis_spin_2017}%
  \BibitemOpen
  \bibfield  {author} {\bibinfo {author} {\bibfnamefont {G.}~\bibnamefont
  {Simutis}}, \bibinfo {author} {\bibfnamefont {S.}~\bibnamefont {Gvasaliya}},
  \bibinfo {author} {\bibfnamefont {N.~S.}\ \bibnamefont {Beesetty}}, \bibinfo
  {author} {\bibfnamefont {T.}~\bibnamefont {Yoshida}}, \bibinfo {author}
  {\bibfnamefont {J.}~\bibnamefont {Robert}}, \bibinfo {author} {\bibfnamefont
  {S.}~\bibnamefont {Petit}}, \bibinfo {author} {\bibfnamefont {A.~I.}\
  \bibnamefont {Kolesnikov}}, \bibinfo {author} {\bibfnamefont {M.~B.}\
  \bibnamefont {Stone}}, \bibinfo {author} {\bibfnamefont {F.}~\bibnamefont
  {Bourdarot}}, \bibinfo {author} {\bibfnamefont {H.~C.}\ \bibnamefont
  {Walker}}, \bibinfo {author} {\bibfnamefont {D.~T.}\ \bibnamefont {Adroja}},
  \bibinfo {author} {\bibfnamefont {O.}~\bibnamefont {Sobolev}}, \bibinfo
  {author} {\bibfnamefont {C.}~\bibnamefont {Hess}}, \bibinfo {author}
  {\bibfnamefont {T.}~\bibnamefont {Masuda}}, \bibinfo {author} {\bibfnamefont
  {A.}~\bibnamefont {Revcolevschi}}, \bibinfo {author} {\bibfnamefont
  {B.}~\bibnamefont {Büchner}}, \ and\ \bibinfo {author} {\bibfnamefont
  {A.}~\bibnamefont {Zheludev}},\ }\href {\doibase 10.1103/PhysRevB.95.054409}
  {\bibfield  {journal} {\bibinfo  {journal} {Physical Review B}\ }\textbf
  {\bibinfo {volume} {95}},\ \bibinfo {pages} {054409} (\bibinfo {year}
  {2017})}\BibitemShut {NoStop}%
\bibitem [{\citenamefont {Mang}\ \emph
  {et~al.}(2004{\natexlab{b}})\citenamefont {Mang}, \citenamefont {Vajk},
  \citenamefont {Arvanitaki}, \citenamefont {Lynn},\ and\ \citenamefont
  {Greven}}]{mang_spin_2004}%
  \BibitemOpen
  \bibfield  {author} {\bibinfo {author} {\bibfnamefont {P.~K.}\ \bibnamefont
  {Mang}}, \bibinfo {author} {\bibfnamefont {O.~P.}\ \bibnamefont {Vajk}},
  \bibinfo {author} {\bibfnamefont {A.}~\bibnamefont {Arvanitaki}}, \bibinfo
  {author} {\bibfnamefont {J.~W.}\ \bibnamefont {Lynn}}, \ and\ \bibinfo
  {author} {\bibfnamefont {M.}~\bibnamefont {Greven}},\ }\href {\doibase
  10.1103/PhysRevLett.93.027002} {\bibfield  {journal} {\bibinfo  {journal}
  {Physical Review Letters}\ }\textbf {\bibinfo {volume} {93}},\ \bibinfo
  {pages} {027002} (\bibinfo {year} {2004}{\natexlab{b}})}\BibitemShut
  {NoStop}%
\bibitem [{\citenamefont {Motoyama}\ \emph {et~al.}(2007)\citenamefont
  {Motoyama}, \citenamefont {Yu}, \citenamefont {Vishik}, \citenamefont {Vajk},
  \citenamefont {Mang},\ and\ \citenamefont {Greven}}]{motoyama_spin_2007}%
  \BibitemOpen
  \bibfield  {author} {\bibinfo {author} {\bibfnamefont {E.~M.}\ \bibnamefont
  {Motoyama}}, \bibinfo {author} {\bibfnamefont {G.}~\bibnamefont {Yu}},
  \bibinfo {author} {\bibfnamefont {I.~M.}\ \bibnamefont {Vishik}}, \bibinfo
  {author} {\bibfnamefont {O.~P.}\ \bibnamefont {Vajk}}, \bibinfo {author}
  {\bibfnamefont {P.~K.}\ \bibnamefont {Mang}}, \ and\ \bibinfo {author}
  {\bibfnamefont {M.}~\bibnamefont {Greven}},\ }\href {\doibase
  10.1038/nature05437} {\bibfield  {journal} {\bibinfo  {journal} {Nature}\
  }\textbf {\bibinfo {volume} {445}},\ \bibinfo {pages} {186} (\bibinfo {year}
  {2007})}\BibitemShut {NoStop}%
\bibitem [{\citenamefont {Hendriksen}\ \emph {et~al.}(1993)\citenamefont
  {Hendriksen}, \citenamefont {Linderoth},\ and\ \citenamefont
  {Lindgård}}]{hendriksen_finite-size_1993}%
  \BibitemOpen
  \bibfield  {author} {\bibinfo {author} {\bibfnamefont {P.~V.}\ \bibnamefont
  {Hendriksen}}, \bibinfo {author} {\bibfnamefont {S.}~\bibnamefont
  {Linderoth}}, \ and\ \bibinfo {author} {\bibfnamefont {P.-A.}\ \bibnamefont
  {Lindgård}},\ }\href {\doibase 10.1103/PhysRevB.48.7259} {\bibfield
  {journal} {\bibinfo  {journal} {Physical Review B}\ }\textbf {\bibinfo
  {volume} {48}},\ \bibinfo {pages} {7259} (\bibinfo {year}
  {1993})}\BibitemShut {NoStop}%
\bibitem [{\citenamefont {Fujita}\ \emph {et~al.}(2003)\citenamefont {Fujita},
  \citenamefont {Kuroshima}, \citenamefont {Matsuda},\ and\ \citenamefont
  {Yamada}}]{fujita_neutron-scattering_2003}%
  \BibitemOpen
  \bibfield  {author} {\bibinfo {author} {\bibfnamefont {M.}~\bibnamefont
  {Fujita}}, \bibinfo {author} {\bibfnamefont {S.}~\bibnamefont {Kuroshima}},
  \bibinfo {author} {\bibfnamefont {M.}~\bibnamefont {Matsuda}}, \ and\
  \bibinfo {author} {\bibfnamefont {K.}~\bibnamefont {Yamada}},\ }\href
  {\doibase 10.1016/S0921-4534(03)00750-0} {\bibfield  {journal} {\bibinfo
  {journal} {Physica C: Superconductivity}\ }\textbf {\bibinfo {volume}
  {392-396}},\ \bibinfo {pages} {130} (\bibinfo {year} {2003})}\BibitemShut
  {NoStop}%
\bibitem [{\citenamefont {Ishii}\ \emph {et~al.}(2021)\citenamefont {Ishii},
  \citenamefont {Asano}, \citenamefont {Ashida}, \citenamefont {Fujita},
  \citenamefont {Yu}, \citenamefont {Greven}, \citenamefont {Okamoto},
  \citenamefont {Huang},\ and\ \citenamefont
  {Mizuki}}]{ishii_post-growth_2021}%
  \BibitemOpen
  \bibfield  {author} {\bibinfo {author} {\bibfnamefont {K.}~\bibnamefont
  {Ishii}}, \bibinfo {author} {\bibfnamefont {S.}~\bibnamefont {Asano}},
  \bibinfo {author} {\bibfnamefont {M.}~\bibnamefont {Ashida}}, \bibinfo
  {author} {\bibfnamefont {M.}~\bibnamefont {Fujita}}, \bibinfo {author}
  {\bibfnamefont {B.}~\bibnamefont {Yu}}, \bibinfo {author} {\bibfnamefont
  {M.}~\bibnamefont {Greven}}, \bibinfo {author} {\bibfnamefont
  {J.}~\bibnamefont {Okamoto}}, \bibinfo {author} {\bibfnamefont {D.-J.}\
  \bibnamefont {Huang}}, \ and\ \bibinfo {author} {\bibfnamefont
  {J.}~\bibnamefont {Mizuki}},\ }\href {\doibase
  10.1103/PhysRevMaterials.5.024803} {\bibfield  {journal} {\bibinfo  {journal}
  {Physical Review Materials}\ }\textbf {\bibinfo {volume} {5}},\ \bibinfo
  {pages} {024803} (\bibinfo {year} {2021})}\BibitemShut {NoStop}%
\bibitem [{\citenamefont {Keimer}\ \emph {et~al.}(1992)\citenamefont {Keimer},
  \citenamefont {Belk}, \citenamefont {Birgeneau}, \citenamefont {Cassanho},
  \citenamefont {Chen}, \citenamefont {Greven}, \citenamefont {Kastner},
  \citenamefont {Aharony}, \citenamefont {Endoh}, \citenamefont {Erwin},\ and\
  \citenamefont {Shirane}}]{keimer_magnetic_1992}%
  \BibitemOpen
  \bibfield  {author} {\bibinfo {author} {\bibfnamefont {B.}~\bibnamefont
  {Keimer}}, \bibinfo {author} {\bibfnamefont {N.}~\bibnamefont {Belk}},
  \bibinfo {author} {\bibfnamefont {R.~J.}\ \bibnamefont {Birgeneau}}, \bibinfo
  {author} {\bibfnamefont {A.}~\bibnamefont {Cassanho}}, \bibinfo {author}
  {\bibfnamefont {C.~Y.}\ \bibnamefont {Chen}}, \bibinfo {author}
  {\bibfnamefont {M.}~\bibnamefont {Greven}}, \bibinfo {author} {\bibfnamefont
  {M.~A.}\ \bibnamefont {Kastner}}, \bibinfo {author} {\bibfnamefont
  {A.}~\bibnamefont {Aharony}}, \bibinfo {author} {\bibfnamefont
  {Y.}~\bibnamefont {Endoh}}, \bibinfo {author} {\bibfnamefont {R.~W.}\
  \bibnamefont {Erwin}}, \ and\ \bibinfo {author} {\bibfnamefont
  {G.}~\bibnamefont {Shirane}},\ }\href {\doibase 10.1103/PhysRevB.46.14034}
  {\bibfield  {journal} {\bibinfo  {journal} {Physical Review B}\ }\textbf
  {\bibinfo {volume} {46}},\ \bibinfo {pages} {14034} (\bibinfo {year}
  {1992})}\BibitemShut {NoStop}%
\bibitem [{\citenamefont {Rømer}\ \emph {et~al.}(2013)\citenamefont {Rømer},
  \citenamefont {Chang}, \citenamefont {Christensen}, \citenamefont {Andersen},
  \citenamefont {Lefmann}, \citenamefont {Mähler}, \citenamefont {Gavilano},
  \citenamefont {Gilardi}, \citenamefont {Niedermayer}, \citenamefont
  {Rønnow}, \citenamefont {Schneidewind}, \citenamefont {Link}, \citenamefont
  {Oda}, \citenamefont {Ido}, \citenamefont {Momono},\ and\ \citenamefont
  {Mesot}}]{romer_glassy_2013}%
  \BibitemOpen
  \bibfield  {author} {\bibinfo {author} {\bibfnamefont {A.~T.}\ \bibnamefont
  {Rømer}}, \bibinfo {author} {\bibfnamefont {J.}~\bibnamefont {Chang}},
  \bibinfo {author} {\bibfnamefont {N.~B.}\ \bibnamefont {Christensen}},
  \bibinfo {author} {\bibfnamefont {B.~M.}\ \bibnamefont {Andersen}}, \bibinfo
  {author} {\bibfnamefont {K.}~\bibnamefont {Lefmann}}, \bibinfo {author}
  {\bibfnamefont {L.}~\bibnamefont {Mähler}}, \bibinfo {author} {\bibfnamefont
  {J.}~\bibnamefont {Gavilano}}, \bibinfo {author} {\bibfnamefont
  {R.}~\bibnamefont {Gilardi}}, \bibinfo {author} {\bibfnamefont
  {C.}~\bibnamefont {Niedermayer}}, \bibinfo {author} {\bibfnamefont {H.~M.}\
  \bibnamefont {Rønnow}}, \bibinfo {author} {\bibfnamefont {A.}~\bibnamefont
  {Schneidewind}}, \bibinfo {author} {\bibfnamefont {P.}~\bibnamefont {Link}},
  \bibinfo {author} {\bibfnamefont {M.}~\bibnamefont {Oda}}, \bibinfo {author}
  {\bibfnamefont {M.}~\bibnamefont {Ido}}, \bibinfo {author} {\bibfnamefont
  {N.}~\bibnamefont {Momono}}, \ and\ \bibinfo {author} {\bibfnamefont
  {J.}~\bibnamefont {Mesot}},\ }\href {\doibase 10.1103/PhysRevB.87.144513}
  {\bibfield  {journal} {\bibinfo  {journal} {Physical Review B}\ }\textbf
  {\bibinfo {volume} {87}},\ \bibinfo {pages} {144513} (\bibinfo {year}
  {2013})}\BibitemShut {NoStop}%
\bibitem [{\citenamefont {Li}\ \emph {et~al.}(2018)\citenamefont {Li},
  \citenamefont {Zhong}, \citenamefont {Stone}, \citenamefont {Kolesnikov},
  \citenamefont {Gu}, \citenamefont {Zaliznyak},\ and\ \citenamefont
  {Tranquada}}]{li_low-energy_2018}%
  \BibitemOpen
  \bibfield  {author} {\bibinfo {author} {\bibfnamefont {Y.}~\bibnamefont
  {Li}}, \bibinfo {author} {\bibfnamefont {R.}~\bibnamefont {Zhong}}, \bibinfo
  {author} {\bibfnamefont {M.~B.}\ \bibnamefont {Stone}}, \bibinfo {author}
  {\bibfnamefont {A.~I.}\ \bibnamefont {Kolesnikov}}, \bibinfo {author}
  {\bibfnamefont {G.~D.}\ \bibnamefont {Gu}}, \bibinfo {author} {\bibfnamefont
  {I.~A.}\ \bibnamefont {Zaliznyak}}, \ and\ \bibinfo {author} {\bibfnamefont
  {J.~M.}\ \bibnamefont {Tranquada}},\ }\href {\doibase
  10.1103/PhysRevB.98.224508} {\bibfield  {journal} {\bibinfo  {journal}
  {Physical Review B}\ }\textbf {\bibinfo {volume} {98}},\ \bibinfo {pages}
  {224508} (\bibinfo {year} {2018})}\BibitemShut {NoStop}%
\bibitem [{\citenamefont {Song}\ \emph {et~al.}(2012)\citenamefont {Song},
  \citenamefont {Park}, \citenamefont {Kim}, \citenamefont {Kim}, \citenamefont
  {Leem}, \citenamefont {Choi}, \citenamefont {Jung}, \citenamefont {Koh},
  \citenamefont {Han}, \citenamefont {Yoshida}, \citenamefont {Eisaki},
  \citenamefont {Lu}, \citenamefont {Shen},\ and\ \citenamefont
  {Kim}}]{song_oxygen-content-dependent_2012}%
  \BibitemOpen
  \bibfield  {author} {\bibinfo {author} {\bibfnamefont {D.}~\bibnamefont
  {Song}}, \bibinfo {author} {\bibfnamefont {S.~R.}\ \bibnamefont {Park}},
  \bibinfo {author} {\bibfnamefont {C.}~\bibnamefont {Kim}}, \bibinfo {author}
  {\bibfnamefont {Y.}~\bibnamefont {Kim}}, \bibinfo {author} {\bibfnamefont
  {C.}~\bibnamefont {Leem}}, \bibinfo {author} {\bibfnamefont {S.}~\bibnamefont
  {Choi}}, \bibinfo {author} {\bibfnamefont {W.}~\bibnamefont {Jung}}, \bibinfo
  {author} {\bibfnamefont {Y.}~\bibnamefont {Koh}}, \bibinfo {author}
  {\bibfnamefont {G.}~\bibnamefont {Han}}, \bibinfo {author} {\bibfnamefont
  {Y.}~\bibnamefont {Yoshida}}, \bibinfo {author} {\bibfnamefont
  {H.}~\bibnamefont {Eisaki}}, \bibinfo {author} {\bibfnamefont {D.~H.}\
  \bibnamefont {Lu}}, \bibinfo {author} {\bibfnamefont {Z.-X.}\ \bibnamefont
  {Shen}}, \ and\ \bibinfo {author} {\bibfnamefont {C.}~\bibnamefont {Kim}},\
  }\href {\doibase 10.1103/PhysRevB.86.144520} {\bibfield  {journal} {\bibinfo
  {journal} {Physical Review B}\ }\textbf {\bibinfo {volume} {86}},\ \bibinfo
  {pages} {144520} (\bibinfo {year} {2012})}\BibitemShut {NoStop}%
\bibitem [{\citenamefont {Dagotto}\ \emph {et~al.}(1992)\citenamefont
  {Dagotto}, \citenamefont {Riera},\ and\ \citenamefont
  {Scalapino}}]{dagotto_superconductivity_1992}%
  \BibitemOpen
  \bibfield  {author} {\bibinfo {author} {\bibfnamefont {E.}~\bibnamefont
  {Dagotto}}, \bibinfo {author} {\bibfnamefont {J.}~\bibnamefont {Riera}}, \
  and\ \bibinfo {author} {\bibfnamefont {D.}~\bibnamefont {Scalapino}},\ }\href
  {\doibase 10.1103/PhysRevB.45.5744} {\bibfield  {journal} {\bibinfo
  {journal} {Physical Review B}\ }\textbf {\bibinfo {volume} {45}},\ \bibinfo
  {pages} {5744} (\bibinfo {year} {1992})}\BibitemShut {NoStop}%
\bibitem [{\citenamefont {Dahm}\ \emph {et~al.}(2009)\citenamefont {Dahm},
  \citenamefont {Hinkov}, \citenamefont {Borisenko}, \citenamefont {Kordyuk},
  \citenamefont {Zabolotnyy}, \citenamefont {Fink}, \citenamefont {Büchner},
  \citenamefont {Scalapino}, \citenamefont {Hanke},\ and\ \citenamefont
  {Keimer}}]{dahm_strength_2009}%
  \BibitemOpen
  \bibfield  {author} {\bibinfo {author} {\bibfnamefont {T.}~\bibnamefont
  {Dahm}}, \bibinfo {author} {\bibfnamefont {V.}~\bibnamefont {Hinkov}},
  \bibinfo {author} {\bibfnamefont {S.~V.}\ \bibnamefont {Borisenko}}, \bibinfo
  {author} {\bibfnamefont {A.~A.}\ \bibnamefont {Kordyuk}}, \bibinfo {author}
  {\bibfnamefont {V.~B.}\ \bibnamefont {Zabolotnyy}}, \bibinfo {author}
  {\bibfnamefont {J.}~\bibnamefont {Fink}}, \bibinfo {author} {\bibfnamefont
  {B.}~\bibnamefont {Büchner}}, \bibinfo {author} {\bibfnamefont {D.~J.}\
  \bibnamefont {Scalapino}}, \bibinfo {author} {\bibfnamefont {W.}~\bibnamefont
  {Hanke}}, \ and\ \bibinfo {author} {\bibfnamefont {B.}~\bibnamefont
  {Keimer}},\ }\href {\doibase 10.1038/nphys1180} {\bibfield  {journal}
  {\bibinfo  {journal} {Nature Physics}\ }\textbf {\bibinfo {volume} {5}},\
  \bibinfo {pages} {217} (\bibinfo {year} {2009})}\BibitemShut {NoStop}%
\bibitem [{\citenamefont {Yu}\ \emph {et~al.}(2009)\citenamefont {Yu},
  \citenamefont {Li}, \citenamefont {Motoyama},\ and\ \citenamefont
  {Greven}}]{yu_universal_2009}%
  \BibitemOpen
  \bibfield  {author} {\bibinfo {author} {\bibfnamefont {G.}~\bibnamefont
  {Yu}}, \bibinfo {author} {\bibfnamefont {Y.}~\bibnamefont {Li}}, \bibinfo
  {author} {\bibfnamefont {E.~M.}\ \bibnamefont {Motoyama}}, \ and\ \bibinfo
  {author} {\bibfnamefont {M.}~\bibnamefont {Greven}},\ }\href {\doibase
  10.1038/nphys1426} {\bibfield  {journal} {\bibinfo  {journal} {Nature
  Physics}\ }\textbf {\bibinfo {volume} {5}},\ \bibinfo {pages} {873} (\bibinfo
  {year} {2009})}\BibitemShut {NoStop}%
\bibitem [{\citenamefont {Wilks}(1938)}]{wilks_large-sample_1938}%
  \BibitemOpen
  \bibfield  {author} {\bibinfo {author} {\bibfnamefont {S.~S.}\ \bibnamefont
  {Wilks}},\ }\href {\doibase 10.1214/aoms/1177732360} {\bibfield  {journal}
  {\bibinfo  {journal} {The Annals of Mathematical Statistics}\ }\textbf
  {\bibinfo {volume} {9}},\ \bibinfo {pages} {60} (\bibinfo {year}
  {1938})}\BibitemShut {NoStop}%
\bibitem [{\citenamefont {Krighaar}(2026)}]{NBI_Magnetism_NCCOgithub_2026}%
  \BibitemOpen
  \bibfield  {author} {\bibinfo {author} {\bibfnamefont {K.~M.~L.}\
  \bibnamefont {Krighaar}},\ }\href@noop {} {\enquote {\bibinfo {title} {Ncco\_
  analysis\_ code\_ 2026: Data analysis for - emergence of low-energy spin
  waves in superconducting electron-doped cuprates},}\ }\bibinfo {howpublished}
  {\url{https://github.com/NBI-Magnetism-Group/NCCO_Analysis_Code_2026}}
  (\bibinfo {year} {2026}),\ \bibinfo {note} {gitHub repository}\BibitemShut
  {NoStop}%
\end{thebibliography}%

\end{document}


\title{Supplemental Material: \\ Emergence of low-energy spin waves in superconducting electron-doped cuprates}

\author{Kristine M. L. Krighaar}
 \email{kristine.krighaar@nbi.ku.dk}
 \affiliation{Nanoscience Center, Niels Bohr Institute, University of Copenhagen, 2100 Copenhagen, Denmark} 

\author{Jeppe J. Cederholm}
\affiliation{Nanoscience Center, Niels Bohr Institute, University of Copenhagen, 2100 Copenhagen, Denmark}
\affiliation{Institute Laue-Langevin (ILL), 71 Avenue des Martyrs, CS20156, 38042 Grenoble, France}
\affiliation{Laboratory for Quantum Magnetism, Institute of Physics, \'{E}cole Polytechnique F\'{e}d\'{e}rale de Lausanne (EPFL), CH-1015 Lausanne, Switzerland}

\author{Ellen M. S. Schriver}
\affiliation{Nanoscience Center, Niels Bohr Institute, University of Copenhagen, 2100 Copenhagen, Denmark}

\author{Henrik Jacobsen}
\affiliation{Nanoscience Center, Niels Bohr Institute, University of Copenhagen, 2100 Copenhagen, Denmark}
\affiliation{Data Management and Software Centre, Asmussens Allé 305, 2800 Kongens Lyngby, Denmark}

\author{Christine P. Lauritzen}
\affiliation{Nanoscience Center, Niels Bohr Institute, University of Copenhagen, 2100 Copenhagen, Denmark}
\affiliation{Condensed Matter and Interfaces, Debye Institute for Nanomaterials Science, Utrecht University, 3508 TA Utrecht, The Netherlands}

\author{Igor Zaliznyak}
\affiliation{Condensed Matter Physics and Materials Science Division, Brookhaven National Laboratory, Upton, NY 11973, USA}

\author{C\'edric H. Qvistgaard}
\affiliation{Nanoscience Center, Niels Bohr Institute, University of Copenhagen, 2100 Copenhagen, Denmark}
\affiliation{Department of Energy Conversion and Storage, Technical University of Denmark, 2800 Kgs. Lyngby, Denmark}

\author{Ursula Hansen}
\affiliation{Institute Laue-Langevin (ILL), 71 Avenue des Martyrs, CS20156, 38042 Grenoble, France}

\author{Ahmed Alshemi}
\affiliation{Division of Synchrotron Radiation Research, Department of Physics, Lund University, SE-22100 Lund, Sweden}

\author{Dongjoon Song}
\affiliation{Quantum Matter Institute, University of British Columbia, Vancouver, British Columbia, Canada, V6T 1Z4}

\author{Anton P. J. Stampfl}
\affiliation{Australian Nuclear Science and Technology Organisation, Lucas Heights, NSW 2234, Australia}%

\author{Jean-Claude Grivel}
\affiliation{Department of Energy Conversion and Storage, Technical University of Denmark, 2800 Kgs. Lyngby, Denmark}

\author{Kim Lefmann}%
\affiliation{%
Nanoscience Center, Niels Bohr Institute, University of Copenhagen, 2100 Copenhagen, Denmark}%

\author{Machteld E. Kamminga }
 \email{m.e.kamminga@uu.nl}
 \affiliation{Condensed Matter and Interfaces, Debye Institute for Nanomaterials Science, Utrecht University, 3508 TA Utrecht, The Netherlands}

\maketitle
\tableofcontents

\section{Sample Details}

\subsection{Annealing procedures}

A 39\,mm single crystal of optimally doped Nd\(_{2-x}\)Ce\(_x\)CuO\(_4\) (NCCO) was synthesized using the traveling-solvent floating-zone method. The crystal was subsequently divided using a diamond wire saw into three pieces:
\begin{itemize}
  \item \textbf{Sample A} (l=15\,mm, d = 6mm, m=2.5718 g): retained as the as-grown reference sample.
  \item \textbf{Sample B} (l=15\,mm, d = 6mm, m= 2.5711 g): reductively annealed, superconducting - not used in the data presented in the main paper (measured at TAIPAN, ANSTO). This sample was accidentally damaged prior to the IN20 measurements.
  \item \textbf{Sample C} (l=9\,mm, d = 6mm, 1.7521 g): reductively annealed, superconducting (measured at IN20, ILL).
\end{itemize}

\textbf{Sample B} was annealed in a tube furnace under a continuous nitrogen gas flow of 4\,L/min. The temperature was ramped at 2\,\textdegree{}C/min up to 900\,\textdegree{}C, held at that temperature for 20\,hours, and then cooled down at the same rate. The total annealing time was approximately 35\,hours.

\textbf{Sample C} underwent a two-step annealing process. The first step was performed at 900\,\textdegree{}C under nitrogen gas flow with an oxygen partial pressure below 15\,ppm for 24\,hours. This was followed by a second annealing step at 500\,\textdegree{}C under flowing oxygen for an additional 24\,hours.

\subsection{SQUID data}

Field-cooled (FC) and zero-field-cooled (ZFC) magnetic susceptibility measurements were performed using a SQUID magnetometer under an applied magnetic field of approximately $10$\,Oe. These measurements were carried out to confirm the superconducting transition in the annealed samples and a lack of transition in the as-grown sample.

\begin{figure}[H]
    \centering
    \includegraphics[width=\linewidth]{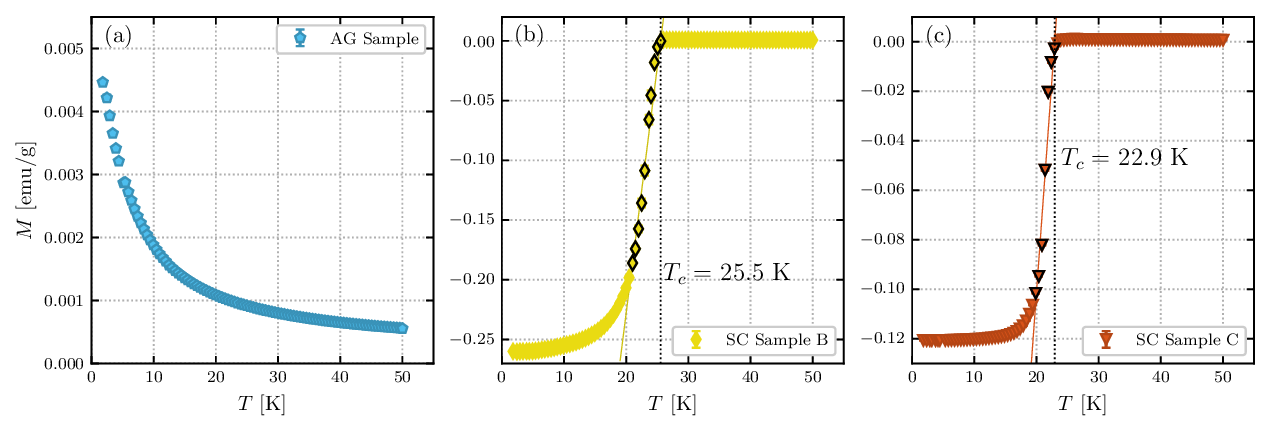}
    \caption{Susceptibility (ZFC) for (a) as-grown sample, (b) annealed sample B and (c) annealed sample C. Note that the data are not normalized to $\chi$, because the crystal sizes were too large to properly fit between the coils of the magnetometer, making comparison between samples impossible. The black outlined points are the data points contributing to the linear fit.}
    \label{fig:SQUID_data}
\end{figure}

\ref{fig:SQUID_data}(a) shows how the annealed sample only possess a small paramagnetic signal while the onset of diamagnetic screening, seen in Fig. \ref{fig:SQUID_data}(b) and (c) provides clear evidence of superconductivity in the annealed samples. For the SQUID measurement of sample B, a small pellet 0.0337\,g of the powder from the damaged sample was used. The magnetic susceptibility was measured shortly prior to each of the two neutron scattering experiments to ensure that the superconducting properties of the samples had not degraded over time or during transport. The data shown in the figure corresponds to the measurement performed before the final experiment on the IN20 spectrometer.

The critical temperatures are determined by the fits shown in Fig. \ref{fig:SQUID_data}. Here the slope of the transition was determined by a linear fit, with the black outlined points being the points included in the fit. The onset critical temperature was determined to be where this line crosses $M=0$. Both critical temperatures yield similar onset temperatures of $T_c = 25.5$ K for sample B, used for the TAIPAN experiment and $T_c = 22.9$ K for sample C used for the IN20 experiment. This shows that both annealing procedures yield very similar results with little difference in $T_c$. 

\subsection{Laue patterns}

\begin{figure}[H]
    \centering
    \includegraphics[width=0.32\textwidth]{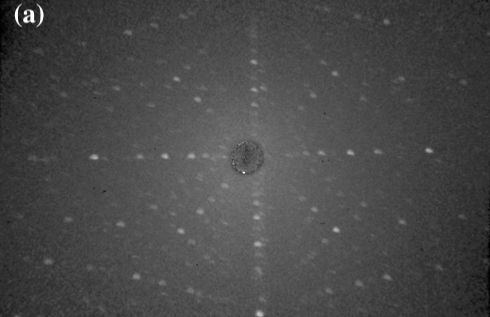}
    \includegraphics[width=0.32\textwidth]{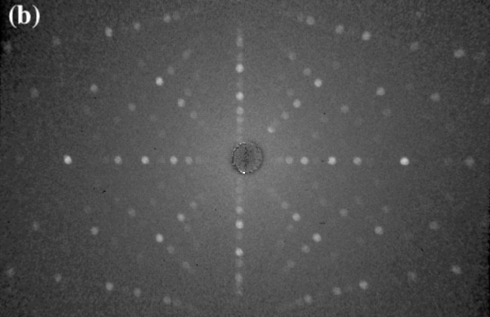}
    \includegraphics[width=0.32\textwidth]{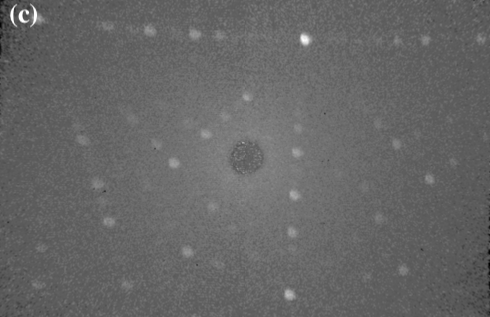}
    \caption{X-ray Laue pictures, (a) The as-grown sample A, (b) the annealed and superconducting sample B and (c) the annealed and superconducting sample C.} 
    \label{fig:Laue}
\end{figure}

Fig.~\ref{fig:Laue} shows X-ray Laue pictures of the three samples. All crystals show high crystallinity.

\subsection{Sample Quality}

We consider the quality of our samples. A high quality is reflected both in the magnetization measurements , the x-ray Laue pictures, and in the $q$-scans shown on Fig. 2 and Fig.~S6-S11 in the Supplemental Material. However, we observe a twin in one of the annealed samples rotated $45^{\circ}$ (no twin was observed in the as-grown sample; see also section II.B.) This is a common feature of the traveling floating zone method, to occur at the first part of the crystal growth. To investigate any effect of the twin, we are able to compare the results of the integrated intensities from IN20 to the data from TAIPAN, as shown in Fig.~S18 in the Supplemental Material. When converting $I$ to $S(\omega)$, by normalizing to the phonon, we see that the results exhibit consistent behavior of the gap sizes, across both instruments. Furthermore, we see a strong agreement between our results and previous results on the annealed NCCO by Zhao et al.\cite{zhao_neutron-spin_2007}, giving credibility to our study.

\section{Experimental details - IN20 @ILL}

The IN20 spectrometer was configured with a double-focusing PG(002) monochromator and analyser, operating at a fixed final wavevector of 
$k_f = 2.662$ Å$^{-1}$ The incident beam ($k_i$) included a neutron velocity selector before the monochromator, while the scattered beam ($k_f$) contained a PG filter between the sample and analyser. A radial collimator was positioned between the analyser and detector.

\subsection{Overview of the measurements performed at IN20}
\begin{figure}[H]
    \centering
    \includegraphics[width=0.85\linewidth]{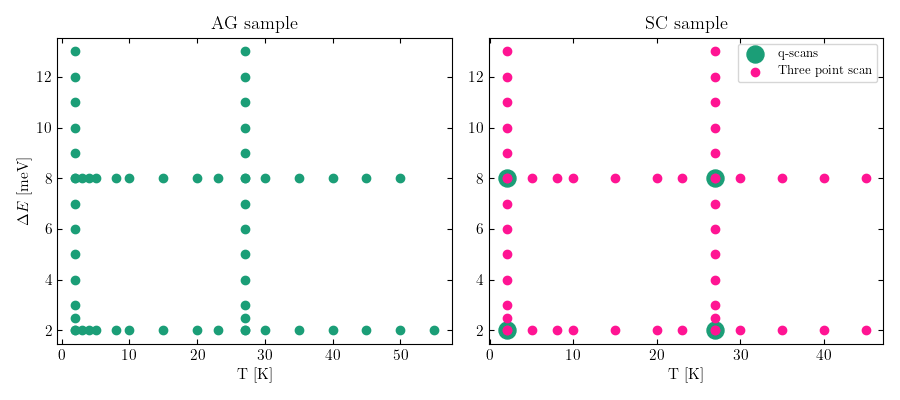}
    \caption{Map of the measurements performed at IN20. While all measurements on the AG sample (left) were performed using $q$-scans, most measurements on the SC sample (right) were performed using 3-point scans, to save measurement time.}
    \label{fig:IN20_map}
\end{figure}

Fig.~\ref{fig:IN20_map} presents a visual overview of the combinations of temperatures and energy transfers, where the magnetic signal at (0.5 0.5 0) was measured during the IN20 experiment. 

\subsection{Observation of a crystal twin for sample C}

\begin{figure}[H]
    \centering
    \includegraphics[width=0.5\linewidth]{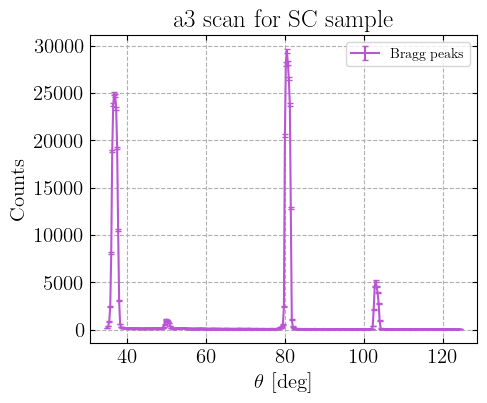}
    \caption{Sample rotation scan shows the twin of the superconducting sample C.}
    \label{fig:Twin_of_SC_sample}
\end{figure}

A sample rotation scan of the (0 2 0) peak, performed on IN20 and shown in Fig.~\ref{fig:Twin_of_SC_sample}, reveals the presence of a secondary crystallographic domain, indicated by the pronounced peak below 40$^{\circ}$. This domain is rotated by approximately 45° relative to the main orientation and accounts for roughly half of the sample volume. This is evident from the relative intensities of the two peaks: comparing 25,000 to 30,000 counts suggests that the larger peak originates from 54\% of the sample, corresponding to a mass of 0.95 g. A consistency check using the phonon intensities further down shows relative peak intensities of 0.00018 for the superconducting sample and 0.00052 for the 2.57 g as-grown sample. From these ratios, the superconducting sample mass is estimated as 1.028 g, which agrees well with the diffraction-based estimate, within the expected uncertainty from rounding when reading the y-axis by eye. Furthermore, two smaller domains can be observed from the peaks at approx. 55$^{\circ}$ and 105$^{\circ}$.

While this indicates a significant degree of twinning, it does not present a problem for the presented analysis. The rotated domain gives rise to scattering at symmetry-related positions in reciprocal space, which are well separated from those of the primary domain and can be readily identified or excluded. Moreover, due to the high symmetry of the crystal structure and the nature of the excitations studied, the spectral features originating from both domains are expected to be equivalent or related by symmetry. All integrated intensities were extracted from regions in reciprocal space where the contribution from the rotated domain is minimal or absent. Therefore, the presence of the twin does not affect the validity of the extracted quantities or the conclusions drawn from them.

\subsection{Data Analysis}

This section outlines the analysis performed on the IN20 data set to extract the integrated intensities of the excitations. All data were normalized using phonon measurements to correct for instrumental factors and allow quantitative comparison.

We begin by describing the phonon normalization procedure used to convert raw counts into units of $S(\mathbf{Q}, \omega)$. Following this, the fitting procedure applied to the normalized peak profiles is detailed, explaining how the spectral weight was extracted and tracked across different conditions.

\subsubsection{Phonon Normalization}
To convert the measured magnetic neutron scattering intensity given in counts per monitor, \( \tilde{I}_{\rm M}(\mathbf{Q}, \omega) \) into the dynamical structure factor \( S(\mathbf{Q}, \omega) \) and the dynamic suseptibility \( \chi''(\omega) \), we apply the phonon normalization method as outlined by Xu \textit{et al.} \cite{xu_absolute_2013} This procedure accounts for factors, such as the magnetic form factor, the Debye-Waller factor, and instrumental constants:

\begin{equation}
S(\mathbf{Q}, \omega) = \frac{13.77\ {\rm barns}^{-1} \tilde{I}_{\rm M}(\mathbf{Q}, \omega)}{g^2|f(\mathbf{Q})|^2 e^{-2W} N k_{\rm f} R_0} 
\end{equation}
and

\begin{equation}
\chi'' (\mathbf{Q}, \omega) = (10^3 m)\frac{\pi}{2}(1-e^{-\omega/k_B T} )\frac{13.77barns^{-1} \tilde I_M(\mathbf{Q}, \omega)}{|f(\mathbf{Q})|^2 e^{-2w} N k_f R_0} .
\end{equation}

where the used parameters and symbols are defined as:
\begin{description}
    \item[{\bf Q}] The neutron scattering vector.
    \item[$\omega$] The neutron energy transfer.
    \item[$ \tilde{I_{\rm M}}(\mathbf{Q}, \omega)$] The measured magnetic intensity.
    \item[$g$] The $g$-factor of the magnetic ion.
    \item [$k_B$] is the Boltzmann constant 0.08617 meV/K.
    \item[$f(\mathbf{Q})$] The magnetic form factor; the calculation method is described below.
    \item[$e^{-2W}$] The Debye-Waller factor, at low temperatures, $\exp(-2W) \approx 1$.
    \item[$N$] The number of unit cells in the sample.
    \item[$k_{\rm f}$] The magnitude of final neutron wavevector.
    \item[$R_0$] The volume of the spectrometer resolution function. 
\end{description}

The product \( N k_{\rm f} R_0 \) is determined by measurements of a phonon with a known scattering cross section:

\begin{equation}
    Nk_fR_0 = \frac{\int \tilde I_{\rm Ph} (\mathbf{Q},\omega)d\mathbf{q}}{e^{-2W} |F_{\rm N} (\mathbf{G})|^2 \cos^2(\beta) \frac{m_{\rm n}}{M} \frac{(\hbar \mathbf{Q})^2}{2m_n} \frac{n_q}{E}\frac{1}{d\omega/dq}}
\end{equation}

The resolution volume can be obtained from a phonon measured using a constant-energy scan. For clarity, a list of parameters and symbols used in the normalization procedure is given below:

\begin{description}
    \item[$\int \tilde{I_{\rm Ph}}(\mathbf{Q}, \omega)\, d\mathbf{q}$] The average integrated intensity of the two phonon branches.
    \item[$\beta$] Angle between the momentum transfer $\mathbf{Q}$ and the phonon polarization direction. Here, we assume $\beta \approx 0$.
    \item[$\frac{(\hbar \mathbf{Q})^2}{2m_n}$] Phonon-specific quantity is the neutron energy at wave-vector.
    \item[$F_N(\mathbf{G})$] Nuclear structure factor of the phonon, we here calculate it using VESTA \cite{Momma_VESTA_software,belokoneva1991preparation} (in barns).
    \item[$m_{\rm n} = 1.675 \times 10^{-27}$ \text{kg} $= 1.009\,u$] Neutron mass.
    \item[$M$] Total mass of atoms in the unit cell.
    \item[$n_q = 1/(1 - e^{-E/k_B T})$] Bose factor for the phonon.
    \item[$d\omega/dq$] The phonon velocity, measured at the energy $E$. 
\end{description}

\begin{figure}[H]
    \centering
    \includegraphics[width=0.7\linewidth]{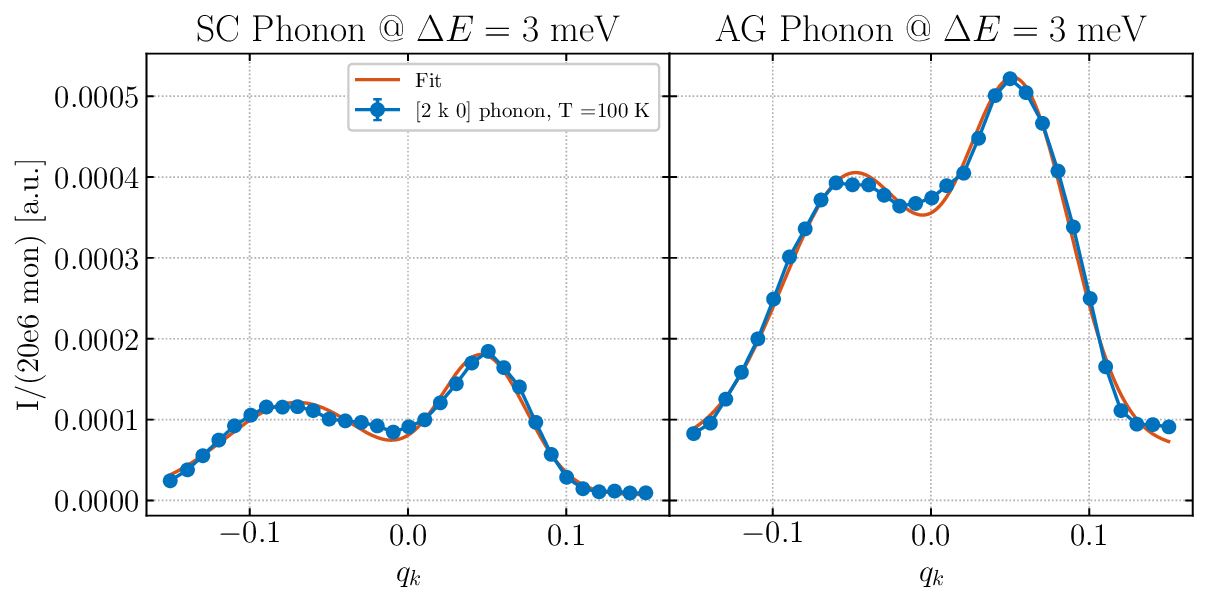}
    \caption{Measurements of transverse acoustic phonons, as constant-energy scans around (2 0 0) from IN20. (left) the SC sample, (right) the AG sample. The solid, red curves are a fits to a double-gaussian line shape from the two phonon branches.}
    \label{fig:IN20_Phonons}
\end{figure}

The integrated phonon intensities are determined from the measurements of a transverse acoustic phonon around (2 0 0), as shown for the two samples in Fig.~\ref{fig:IN20_Phonons}.
All values used for the normalization procedure for both samples are tabulated in Table \ref{tab:IN20_norm}. It is seen that the effective volume of the SC sample is around 30\% of that of the AG sample. This is explained by the smaller mass and the formation of a second domain in the AG sample.

\begin{table}[H]
\centering
\setlength{\extrarowheight}{3pt} 
\resizebox{\textwidth}{!}{%
\begin{tabular}{c|cccccccc}
\hline
Sample Name & $f(\mathbf{Q})$ & $Nk_fR_0$ & $\int \tilde{I}_{\rm Ph}\, d\mathbf{q}$ 
& $F_N(2\,0\,0)^2$ & $m/M$ & $(\hbar \mathbf{Q})^2/2m$ 
& $n_q/E$ & $d\omega/dq$ \\
\hline\hline
Units & -- & meV/barn & r.l.u. & barns & -- & meV & meV$^{-1}$ & meV/r.l.u \\
\hline
As-grown Sample A 
& 0.975 & 0.00094 & 3.67e-05 & 89.794 
& 0.00121 & 18.096 & 1.134 & 57.4 \\
\hline
Annealed Sample C 
& 0.975 & 0.00028 & 1.24e-05 & 89.794 
& 0.00121 & 18.099 & 1.134 & 51.1 \\
\hline
\end{tabular}%
}
\caption{Table of phonon normalization parameters for the samples measured at IN20}
\label{tab:IN20_norm}
\end{table}

\subsubsection{$q$-scan fits}

Our neutron data consist of a series of constant-E scans. We fit each of these scans to two models: one which is just a linear dependence on $q$ to model the background signal, and one where a Gaussian profile is added to model a magnetic contribution. To determine whether the experimental $q$-scan data show significant presence of a magnetic signal or not, we perform a statistical comparison using the likelihood ratio test, based on a result known as Wilks' theorem. This method states that the difference in likelihood, between nested expressions, goes as a $\chi^2$-distribution with the $n$ degrees of freedom being the difference between the degrees of freedom in the two hypotheses.\cite{wilks_large-sample_1938} In other words, we test the statistical significance of including a Gaussian peak in the fitting function.

We calculate the reduced chi-squared (\( \chi^2_{\text{red}} \)) for both models, for all scans. We then apply Wilks' theorem to compute the statistical significance of the chi-squared difference (\( \Delta\chi^2 \)). In this particular case, the confidence level was set to within $p = 0.05$.

A $p$-value is derived from this comparison:
\begin{itemize}
    \item If \( p < 0.05 \), the Gaussian model is preferred.
    \item If \( p > 0.05 \), the constant model is considered sufficient.
\end{itemize}

The errorbars of the integrated magnetic intensity, $A$, for the scans that were fitted with a Gaussian model is taken as the diagonal element of the covariance matrix of the fit. For the scans where the straight line model is preferred by Wilks' theorem, the value of $A$ is set to zero, while the errorbar of $A$ is chosen as the fitting error of ($A$) from the alternative Gaussian fit.

\subsubsection{3-point scans}
Because of limited measurement time at IN20, we decided to perform 3-point scan instead of $q$-scan to map out the excitations in the, previously studied, superconducting sample. Since similar measurements have been reported elsewhere, data quality only had to be to a verification standard. Here we outline the data treatment procedure used in order to estimate an integrated intensity for the 3-point scans, in order to compare them to the q-scans integrated intensities. 

\begin{figure}[ht]
    \centering
    \includegraphics[width=0.32\linewidth]{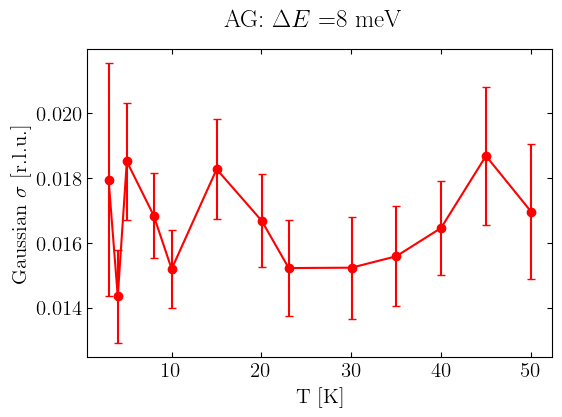}
    \includegraphics[width=0.32\linewidth]{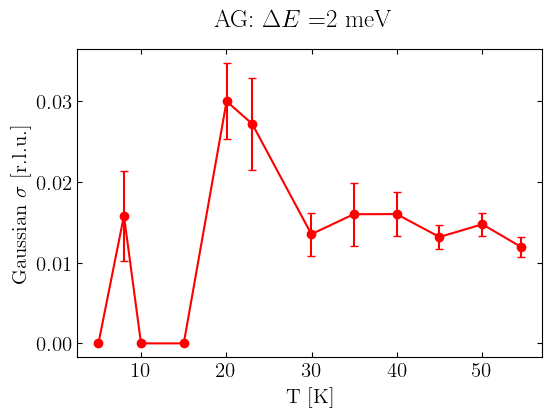}
    \includegraphics[width=0.32\linewidth]{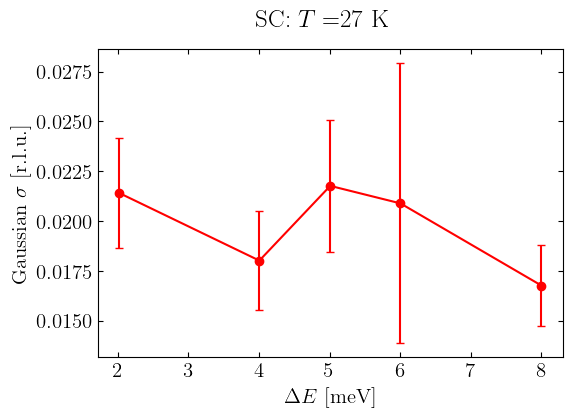}
    \caption{Fitting value $\sigma$ of different q-scans from IN20. The x-axis displays the $\Delta E$ of temperature for the particular scan. The points which indicate $\sigma = 0$ are scans which have been chosen via Wilk's Theorem to be a slope without a peak. (left) The As-grown sample as a function of temperature at $\Delta E = 8$ meV (middle) The As-grown sample as a function of temperature at $\Delta E = 2$ meV (right) Superconducting sample as a function of energy transfer.}
    \label{fig:Sigma_distribution}
\end{figure}

At both IN20 and TAIPAN, the magnetic signal is resolution limited in $q$ over the energy range we investigate. This can be seen from Fig.\ref{fig:Sigma_distribution} where we show a selection of the fitted widths ($\sigma$) from the Gaussian peaks. We see that across different scans and samples, we have almost identical widths, showing that we are resolution limited in $q$.
The zero widths are from scan, which have been chosen via Wilk's theorem not to contain a Gaussian peak. 

In order to estimate an area for the 3-point scans, we use the fact that we are resolution limited to determine an average Gaussian width, $\sigma$, from all the $q$-scans. Assuming that the 3-point scans follow a Gaussian line shape, the area could be determined by: 
\begin{equation}
    A_{\rm 3p} = a \sqrt{2\pi}\sigma,
\end{equation}
where the amplitude is the difference between the count rate at the centre point and the average of the two side points, 
\begin{equation}
    a = c_0 - (c_1 + c_{-1})/2,
\end{equation}

and the error on the amplitude is found via error propagation.
\begin{equation}
    \sigma_a = \sqrt{\sigma_{c_0}^{2} + (\sigma_{c_{-1}}^{2} + \sigma_{c_1}^{2})/4}
\end{equation}
 
The uncertainty of $A_{\rm 3p}$ was determined through error propagation.

\begin{equation}
    \sigma_{A_{\rm 3p}} = \sqrt{2\pi\,\bar{\sigma}^{2} a_{\mathrm{err}}^{2} + 2\pi\,a^{2} \bar{\sigma}_{\mathrm{err}}^{2}}
\end{equation}

Here $\bar{\sigma}$ is the average sigma determined from an average of the fitted gaussian $\sigma$'s. $\bar{\sigma}_{err}$ is the error on $\bar{\sigma}$.

\subsubsection{Results: scans with fits}

\textbf{As-grown sample - A}

\begin{figure}[H]
    \centering
    \includegraphics[width=\linewidth]{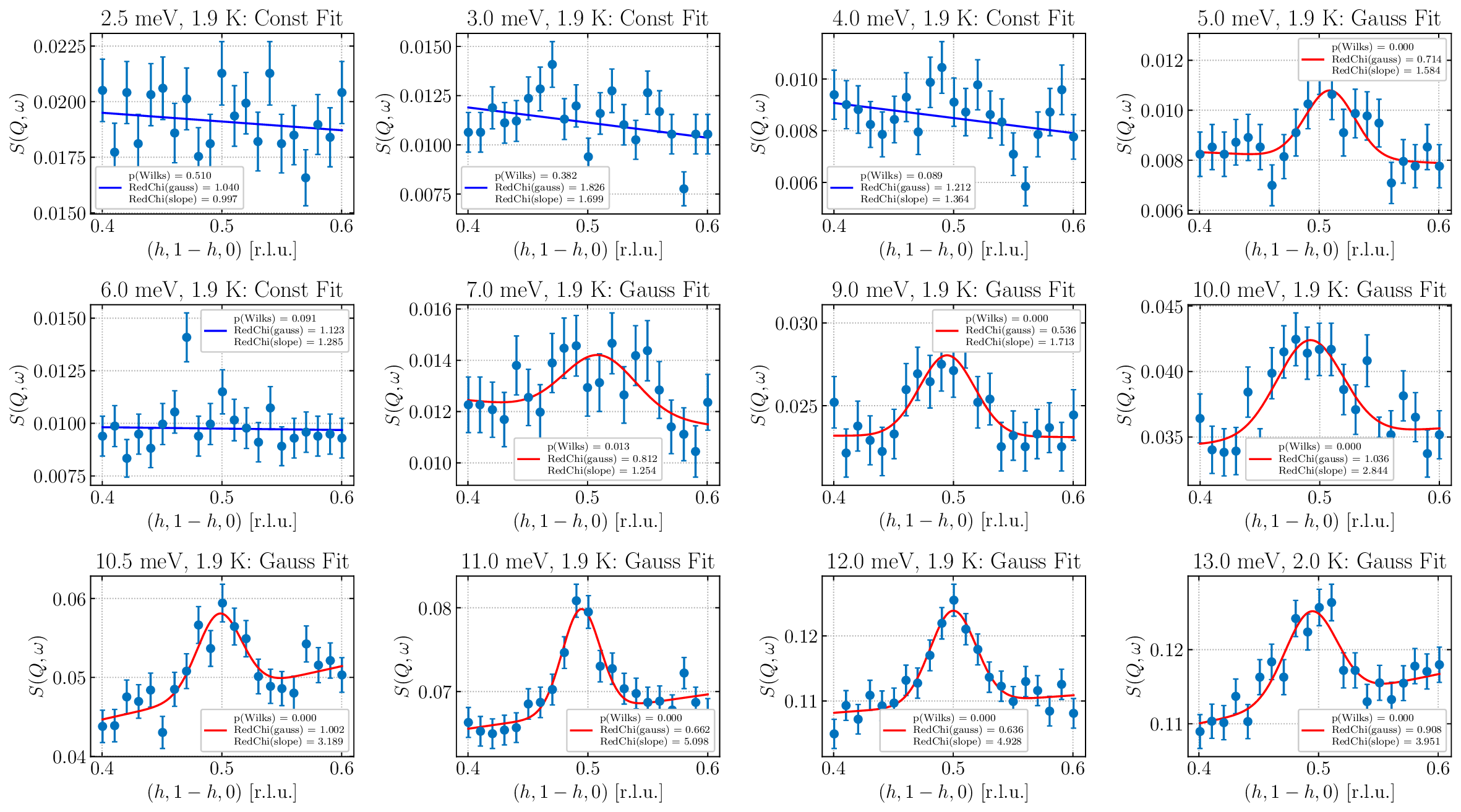}
    \caption{Scans performed for the A sample at 1.9 K. The energy for each scan is given in the title of each figure.  Blue markers show the data, while the red or blue line is a fit of the data as described in the text.}
    \label{fig:AG_A_2K}
\end{figure}

\begin{figure}[H]
    \centering
    \includegraphics[width=\linewidth]{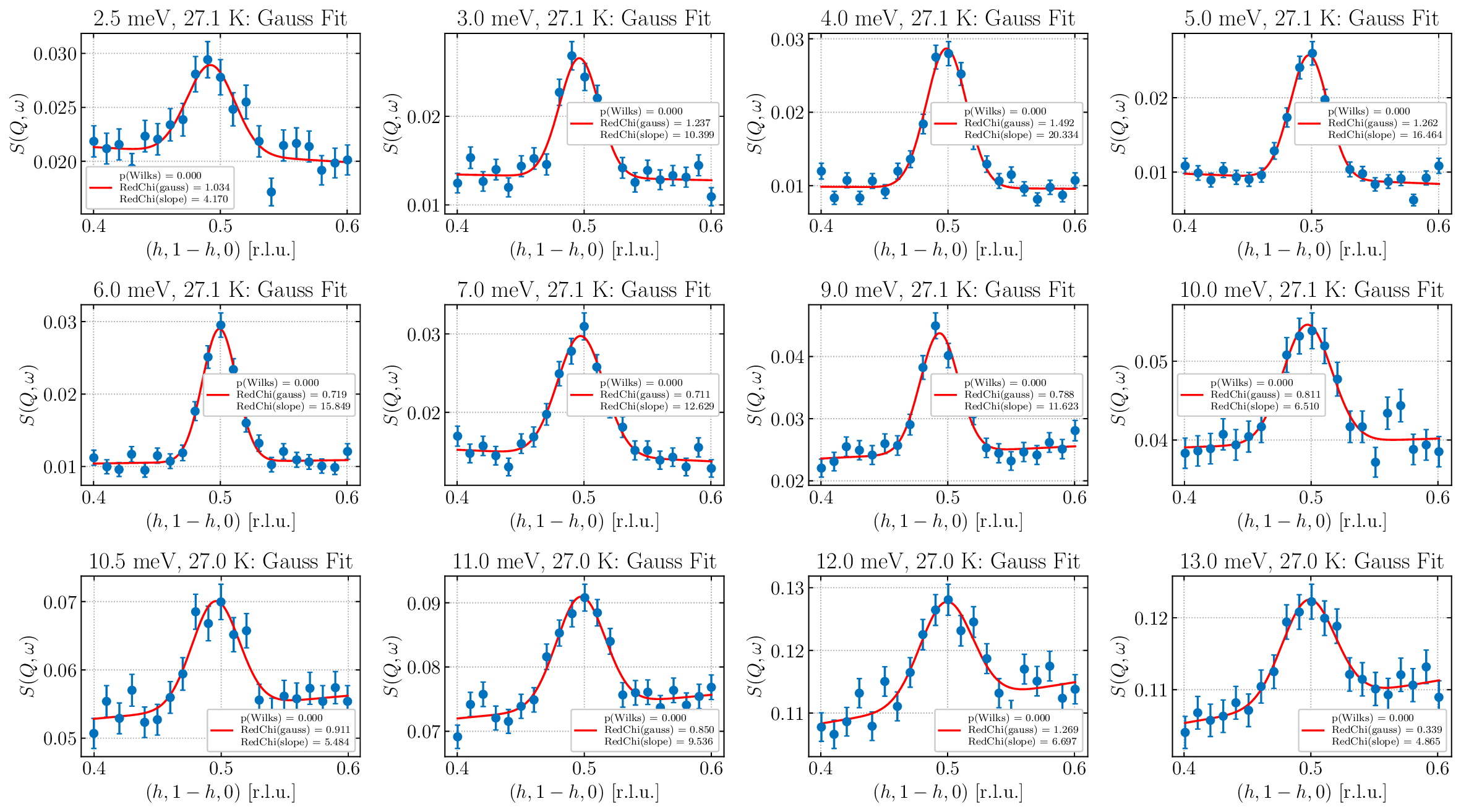}
    \caption{Scans performed for the A sample at 27 K. The energy for each scan is given in the title of each figure.  Blue markers show the data, while the red or blue line is a fit of the data as described in the text.}
    \label{fig:AG_A_27K}
\end{figure}

\begin{figure}[H]
    \centering
    \includegraphics[width=\linewidth]{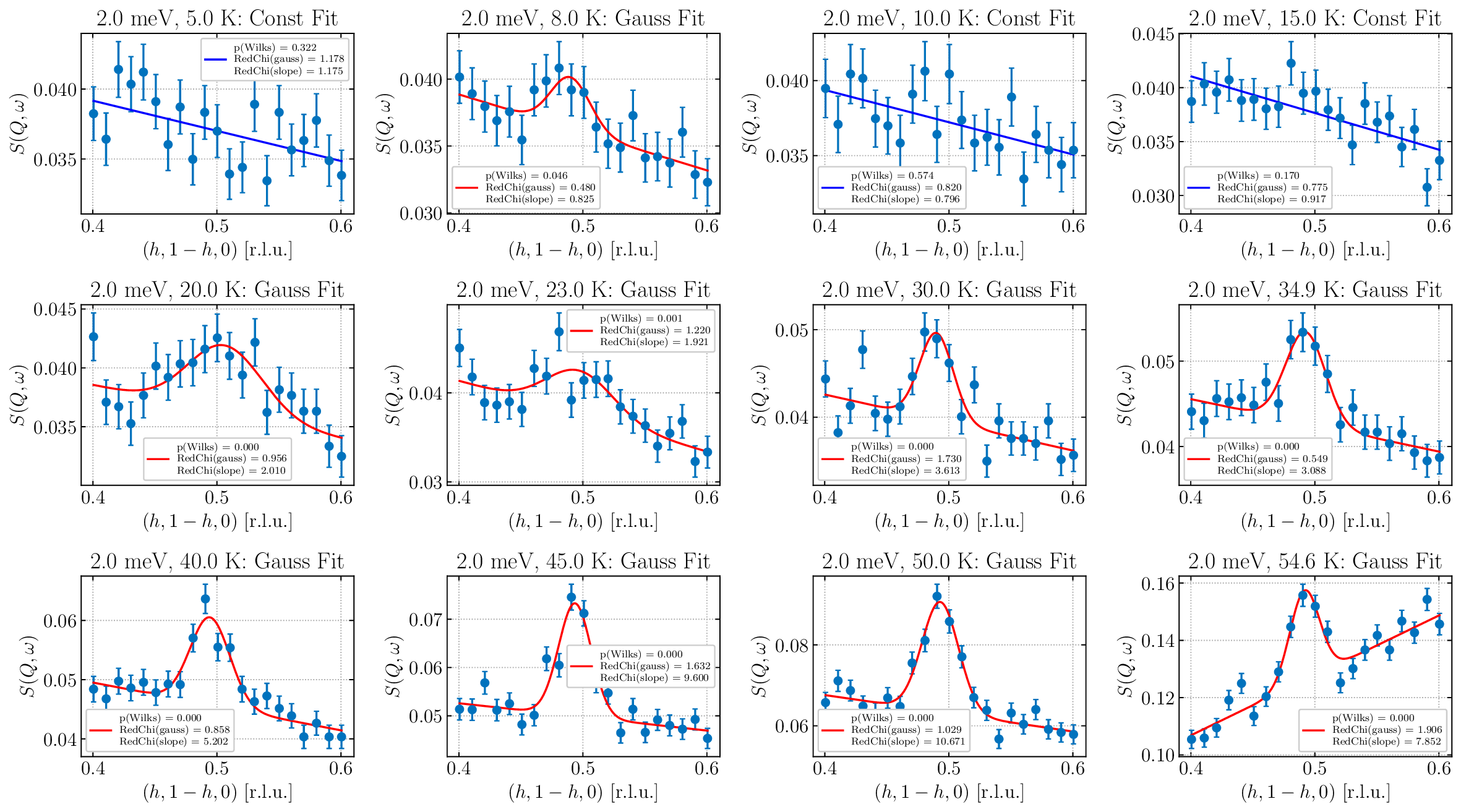}
    \caption{Scans performed for the A sample at 2 meV. The energy for each scan is given in the title of each figure.  Blue markers show the data, while the red or blue line is a fit of the data as described in the text.}
    \label{fig:AG_A_2meV}
\end{figure}

\begin{figure}[H]
    \centering
    \includegraphics[width=\linewidth]{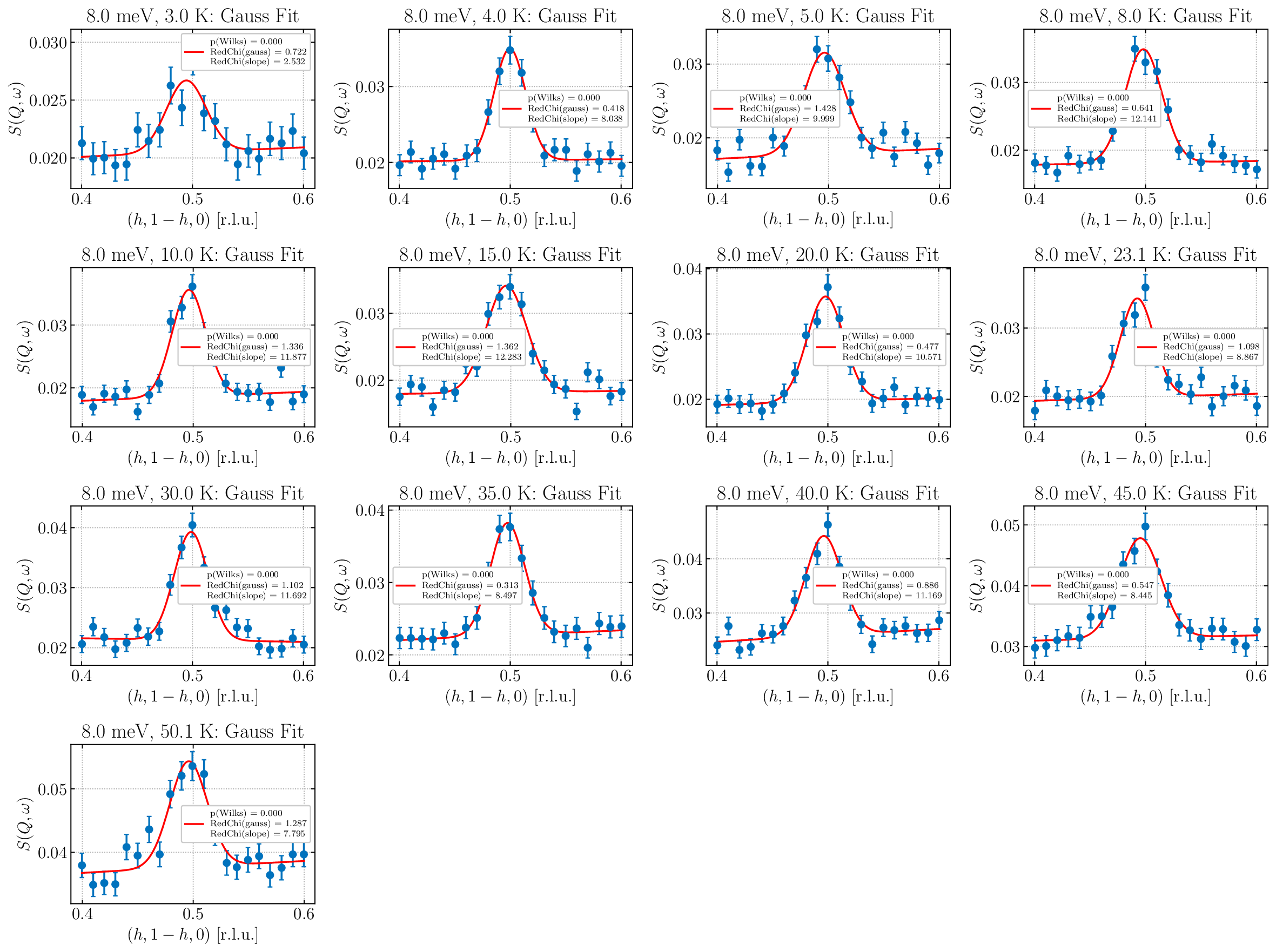}
    \caption{Scans performed for the A sample at 8 meV. The energy for each scan is given in the title of each figure.  Blue markers show the data, while the red or blue line is a fit of the data as described in the text.}
    \label{fig:AG_A_8meV}
\end{figure}

In Figs.~\ref{fig:AG_A_2K} - \ref{fig:AG_A_8meV} all the normalized and fitted data from the $q$-scans are presented. The lines are the fits and the color shows the corresponding model selected by Wilk's theorem: Blue is a linear model and red is linear + Gauss model. In the legend, the $p$-values for both models are printed, including the result of the Wilk test for every dataset. 

\textbf{SC Sample - Sample C}

\begin{figure}[H]
    \centering
    \includegraphics[width=0.6\linewidth]{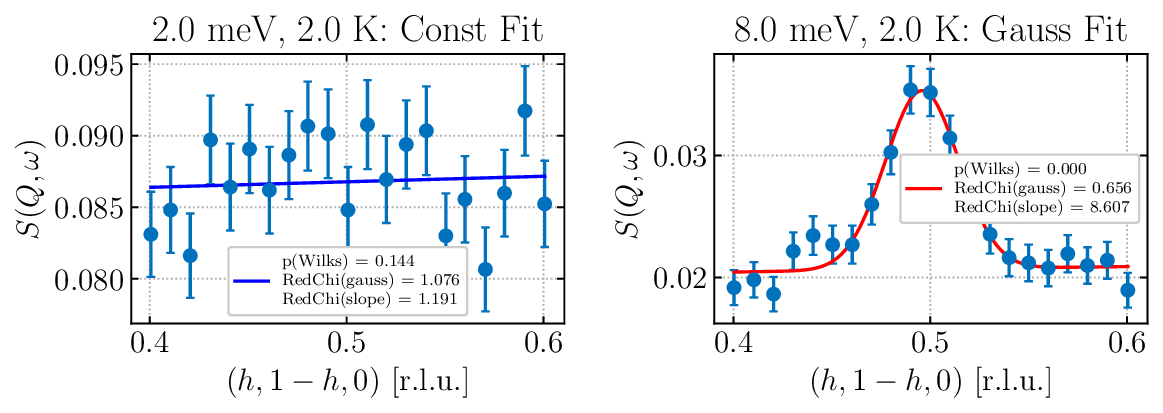}
    \caption{Scans performed for the C sample at 2 K. The energy for each scan is given in the title of each figure.  Blue markers show the data, while the red or blue line is a fit of the data as described in the text.}
    \label{fig:SC_C_2K}
\end{figure}

\begin{figure}[H]
    \centering
    \includegraphics[width=0.9\linewidth]{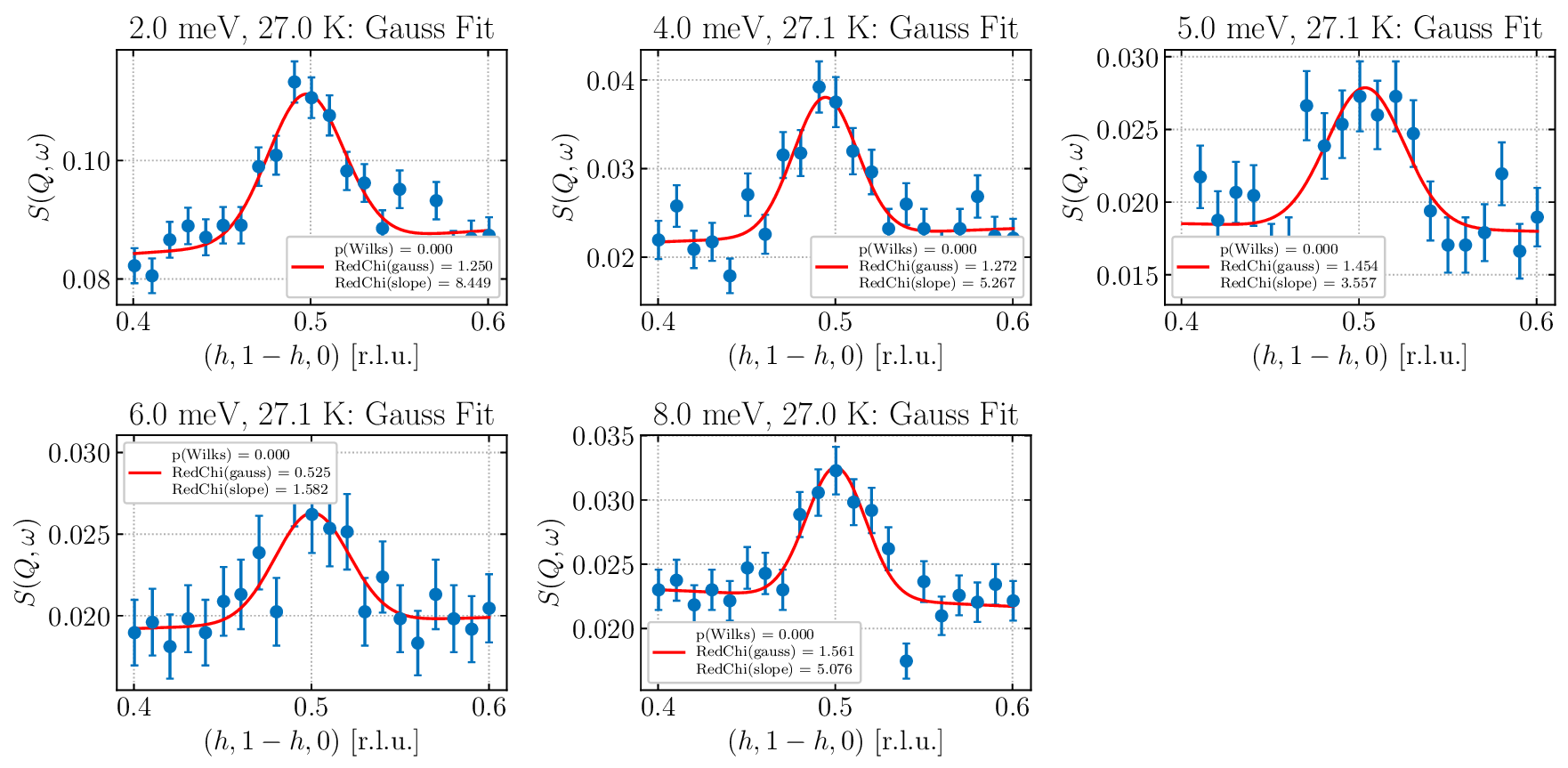}
    \caption{Scans performed for the C sample at 27 K. The energy for each scan is given in the title of each figure.  Blue markers show the data, while the red or blue line is a fit of the data as described in the text.}
    \label{fig:SC_C_27K}
\end{figure}

This section presents the data from sample C in Fig.\ref{fig:SC_C_2K} and \ref{fig:SC_C_27K}. Since this crystal contains a large twin, the effective mass of the sample is roughly three times smaller than that of the other two samples (roughly half of 1.75~g as compared to more than 2.5~g for the other two samples). Hence, $q$-scans were more time consuming and we only performed a few. In order to cover the needed range in energy transfer and temperature, the rest of the data was taken by 3-point measurements, as shown in Fig.~\ref{fig:IN20_map}.

\clearpage
\subsection{Integrated intensities in $S(\omega)$}

Here we present the results of the integrated intensities, presented in the main text, in units of $S(\omega)$. 

\begin{figure}[H]
    \centering
    \includegraphics[width=0.48\linewidth]{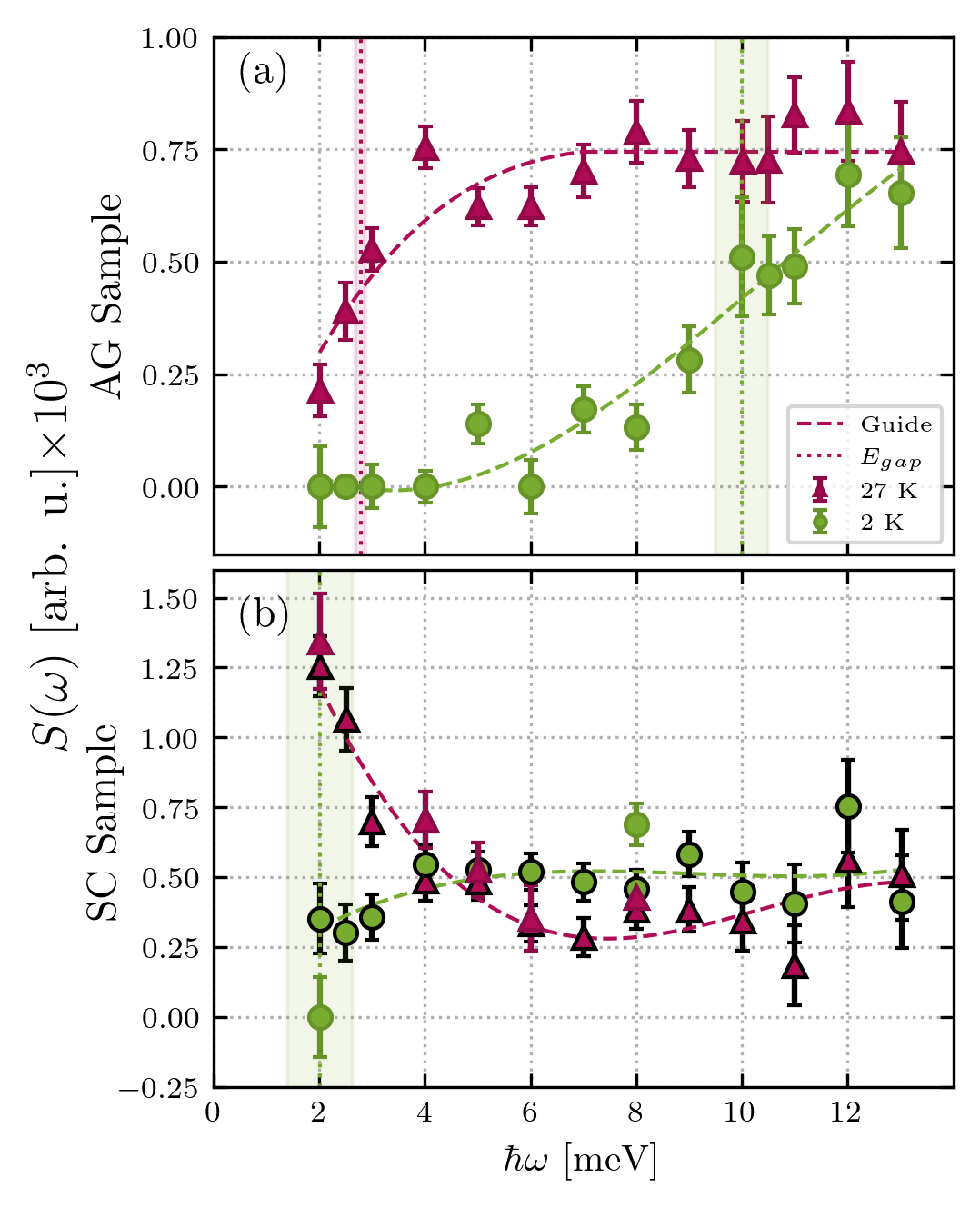}
    \includegraphics[width=0.48\linewidth]{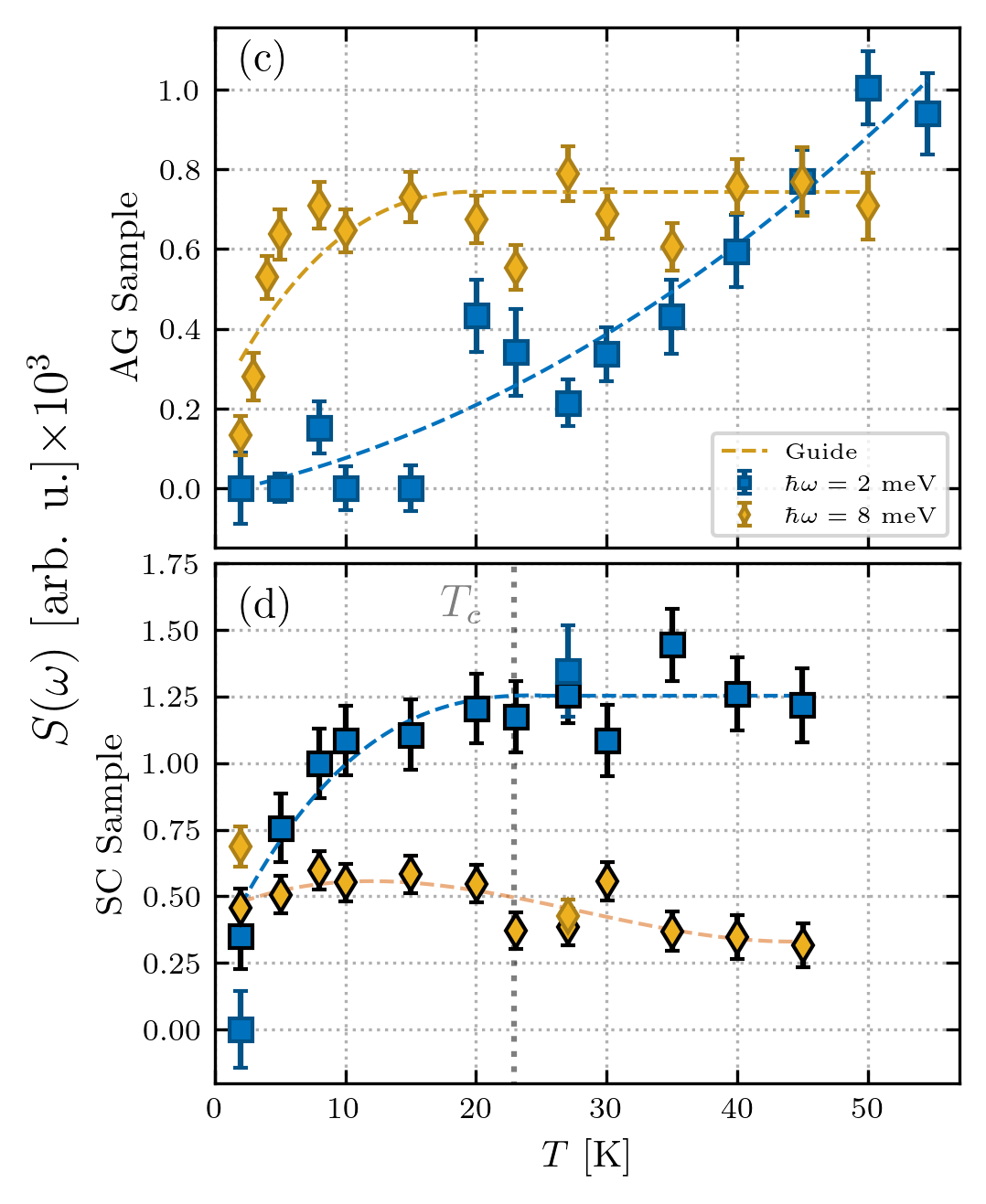}
    \caption{(a) and (b) show the integrated intensity, in units of $S(\omega)$, as a function of energy transfer. (c) and (d) likewise show the temperature dependence. Dashed lines are guide to the eye, while vertical lines indicate the onset of the magnetic spin pseudogap $E_{gap}$. The gap values are found by fits to $\chi''(\omega)$ indicated in the main text and section II.E.}
    \label{fig:Main_plots_IN20_S}
\end{figure}

\begin{figure}[H]
    \centering
    \includegraphics[width=0.49\linewidth]{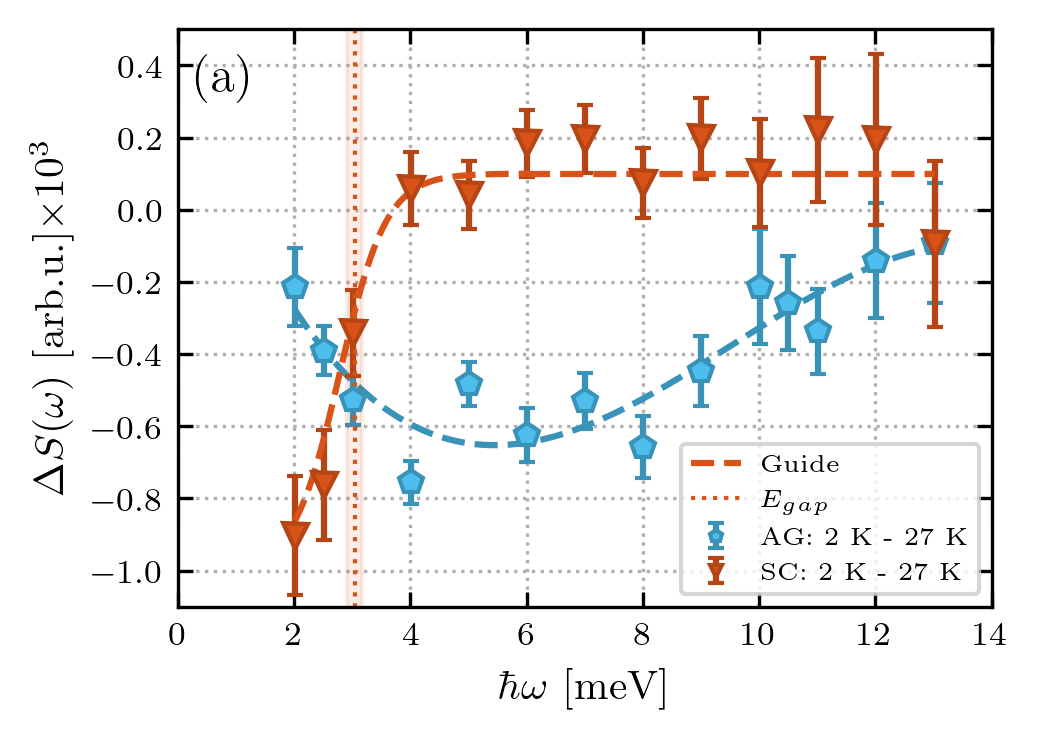}
    \includegraphics[width=0.49\linewidth]{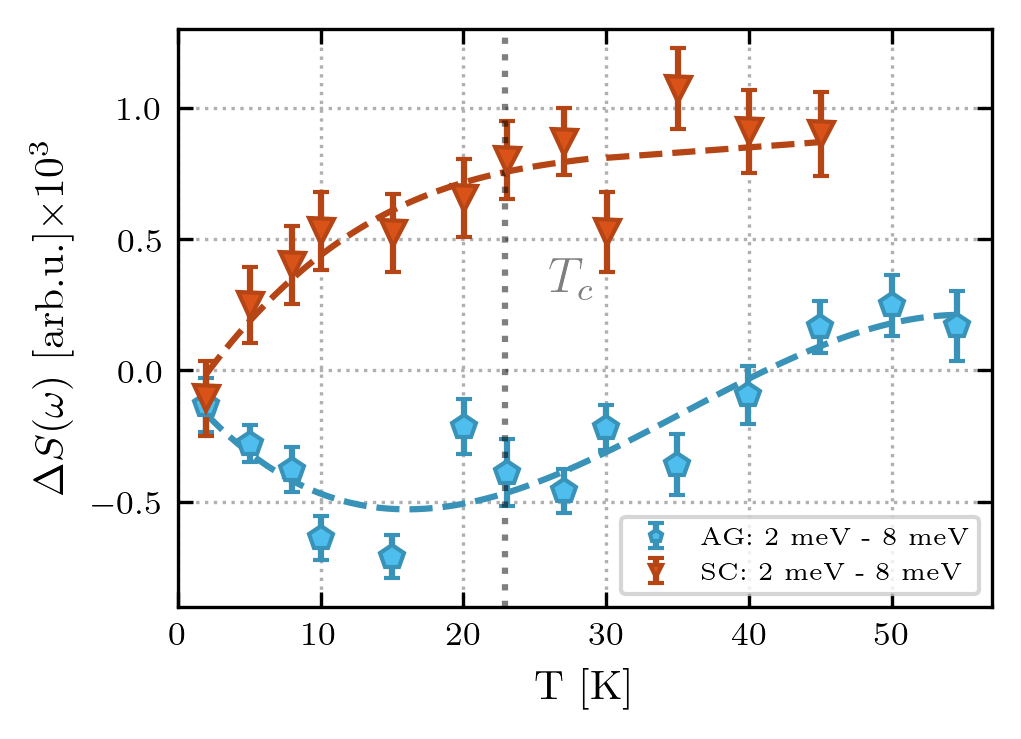}
    \caption{Replications of Fig. 4 and Fig. 5 from the main text, with the intensities converted into units of $S(\omega)$. Dashed lines are guide to the eye, while vertical lines indicate the onset of the magnetic spin pseudogap $E_{gap}$. The gap values are found by fits to $\chi''(\omega)$ indicated in the main text and section II.E.}
    \label{fig:Main_plots_diffrence_IN20_S}
\end{figure}

\clearpage

\subsection{Methods for determining the gap size.}

The in plane spin wave dispersion is very steep, governed by an exchange $J \sim 130$~meV. In the absence of a spin gap the wave-vector resolution of our triple-axis measurements therefore integrates over the dispersive spin-wave response, resulting in an energy-independent response in the investigated energy range. The suppression of the spectral weight at low energy that we observe can be described by an error function representing an onset of magnetic scattering above a threshold energy, $E_{gap}$,
\begin{equation}
\label{Suppl_Eq:Chi_gap}
\chi''(E) = \frac{\chi_{0}}{2} \left[ Erf{\left(\frac{E-E_{gap}}{2\sqrt{\ln{2}}\,\Gamma_{FW}} \right)} + Erf{\left(\frac{E+E_{gap}}{2\sqrt{\ln{2}}\,\Gamma_{FW}} \right)} \right],
\end{equation}
where $\mathrm{Erf}(x)$ is the error function, $\Gamma_{FW}$ is the Gaussian full width at half maximum (FWHM) of the onset as a function of energy $\bigl(2\sqrt{\ln{2}}\,\Gamma_{FW} = \sigma \sqrt{2}\bigr)$, with $\sigma$ being the Gaussian standard deviation, and $\chi_{0}$ is the normalization prefactor determined by the asymptotic high-energy value. The function for $\chi''(E)$ in Eq.~\eqref{Suppl_Eq:Chi_gap} is odd in energy and thus provides a physically valid model consistent with causality. The autocorrelation function $S(E)$ describing the neutron-scattering intensity is then obtained by applying the detailed-balance factor imposed by the fluctuation--dissipation theorem, 
\begin{equation}
\label{Suppl_Eq:S_gap}
S(E) = \frac{\chi''(E)}{\pi \left(1-e^{-\frac{E}{k_B T}} \right)} ,
\end{equation}
where $T$ is the temperature and $k_B$ is the Boltzmann constant. 

This fitting procedure was applied to the as-grown sample at 27 K and the superconducting sample at 2 K, as shown in Fig.~3, in the main text. The superconducting data at 27 K were not fitted, as they exhibit an increase toward zero energy transfer. Likewise, the as-grown data at 2 K were not fitted because no well-defined upper limit is observed, and the signal may continue to rise at higher energies. Consequently, the onset of the gap was estimated as the lowest energy at which the 27 K and 2 K data sets exhibit overlapping error bars. The associated uncertainty was taken as the energy difference to the nearest neighbouring data point.
The resulting uncertainty on the gap for the superconducting sample at 2 K is relatively large. To obtain a more reliable estimate, the subtracted data shown in Fig.~4 were also fitted using a similar procedure described above. The only change being that a constant $(b)$ was added to the error function, as the bottom of the subtracted data is lower than zero. This constant was fixed to the minimum value of the data points.

Overview of the gap values and the uncertainties: 
\begin{equation*}
    E_{\mathrm{gap}} (\mathrm{AG}, 2K) = 10.0 \pm 0.5 \,\, \,\,\, [\mathrm{meV}]
\end{equation*}
\begin{equation*}
    E_{\mathrm{gap}} (\mathrm{AG}, 27K) = 2.8 \pm 0.1\,\, \,\,\, [\mathrm{meV}]
\end{equation*}
\begin{equation*}
    E_{\mathrm{gap}} (\mathrm{SC}, 2K) = 2.0 \pm 0.6\,\, \,\,\, [\mathrm{meV}]
\end{equation*}
\begin{equation*}
    E_{\mathrm{gap}} (\mathrm{SC}, \mathrm{T-difference}) = 3.0 \pm 0.1 \,\, \,\,\, [\mathrm{meV}]
\end{equation*}

\section{Experimental Details - TAIPAN @ANSTO}

We conducted measurements on the TAIPAN triple-axis spectrometer at ANSTO. We aimed to examine the temperature dependence of both the magnetic diffraction signal and the inelastic signal. Due to the difference in neutron flux at the two instruments, we performed fewer $q$-scans at TAIPAN than performed later at IN20. 

In the following sections, we present an overview of the results. We begin with the magnetic diffraction data, followed by the normalization data obtained from the phonon measurements, and then the fitted magnetic inelastic $q$-scans. We conclude with a comparison of the integrated intensities similar to the one for the IN20 data; Fig.~3 in the main text. The inelastic results are not included in the main text, since it would only make the plots more crowded and difficult to read.

\subsection{Magnetic diffraction data}
\begin{figure}[H]
    \centering
    \includegraphics[width=0.49\linewidth]{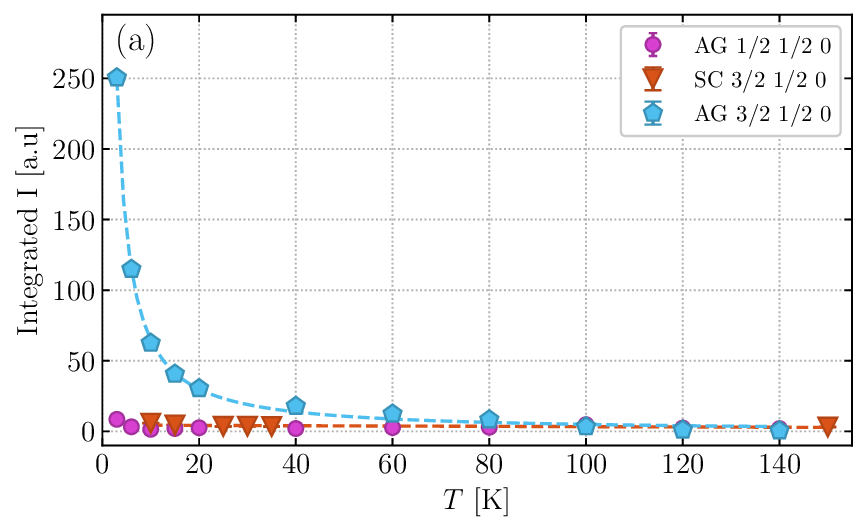}
    \includegraphics[width=0.49\linewidth]{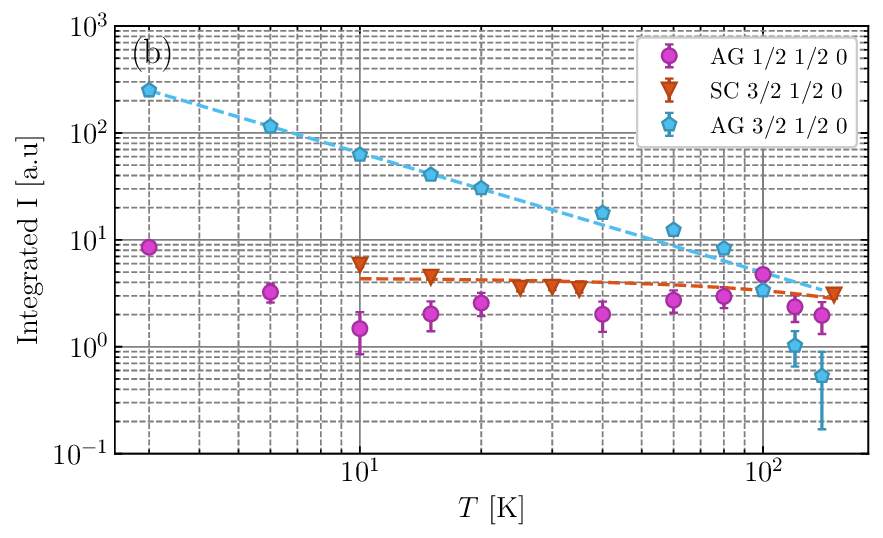}
    \caption{Integrated area of diffraction peaks as a function of temperature. (a) on a linear scale and (b) on log-log scale. The lines are fit to a power law as described in the text.}
    \label{fig:Taipan diffraction}
\end{figure}

In addition to the inelastic measurements, we also measured the temperature dependence of the antiferromagnetic diffraction peak in samples A and B. Fig.~\ref{fig:Taipan diffraction} shows the integrated intensity of the magnetic diffraction data for the AG crystal, sample A, and one of the SC crystals, sample B. The data was measured at the antiferromagnetic reflection (3/2 1/2 0), in analogy to Ref.~\cite{yamada_commensurate_2003}. We tried also measuring at the presumed antiferromagnetic reflection (1/2 1/2 0), but found almost no signal (As seen by pink circles in Fig. ~\ref{fig:Taipan diffraction}) at any temperature. We did not investigate this any further. 

In Fig. \ref{fig:Taipan diffraction} we have fitted the (3/2, 1/2, 0) signal from thee as-grown sample to a power law $I(T) = aT^{-k}$ and the similar superconducting data has been fitted to a linear slope. Furthermore, we see that the diffraction signal in the AG sample falls off fast with increasing $T$, and only 15\% of the signal remains at 15~K. In contrast to the low temperatures, the signal disappears completely above 120~K. For the SC sample, the low-temperature signal is around an order of magnitude smaller. These diffraction results are roughly consistent with earlier studies \cite{yamada_commensurate_2003} despite earlier measurment showing broad peak in the (1/2, 1/2, 0) intensity of the as-grown sample around 50 K.\cite{Matsuda1992}

\subsection{Spectroscopy data}
For the experiment, the superconducting sample was measured first and the as-grown sample was measured second.

During phonon measurements of the as-grown sample we observed that the phonon intensity did not seem to follow Bose statistics. Initially, all measurements of the superconducting sample relied on a single thermometer at the bottom of the cryostat near the sample, while a second thermometer at the top, was fixed at 3 K. This configuration created a temperature gradient at elevated temperatures e.g., 100 K, which only became apparent during measurements of the as-grown sample. To normalize the phonon signal in the superconducting sample that we had already measured, measurements of the as-grown sample were made at both the correct temperature setting and the wrong setting. This intensity ratio was in turn used to normalize the intensity of the phonon intensity in the superconducting sample such that a temperature reading of 100 K corresponded to a sample temperature of 46 K. However, the systematic error introduced by this method is difficult to estimate.

To check the effect on the magnetic signal at 27 K, control measurements with corrected and incorrect temperature settings show consistent intensities, suggesting that the 27 K data of the superconducting sample remain valid despite the earlier oversight.

The conversion and fitting procedure used for the TAIPAN data follows the same described above for the IN20 experiment.
\clearpage

\subsubsection{Phonon Normalization Parameters}

\begin{figure}[H]
    \centering
    \includegraphics[width=0.7\linewidth]{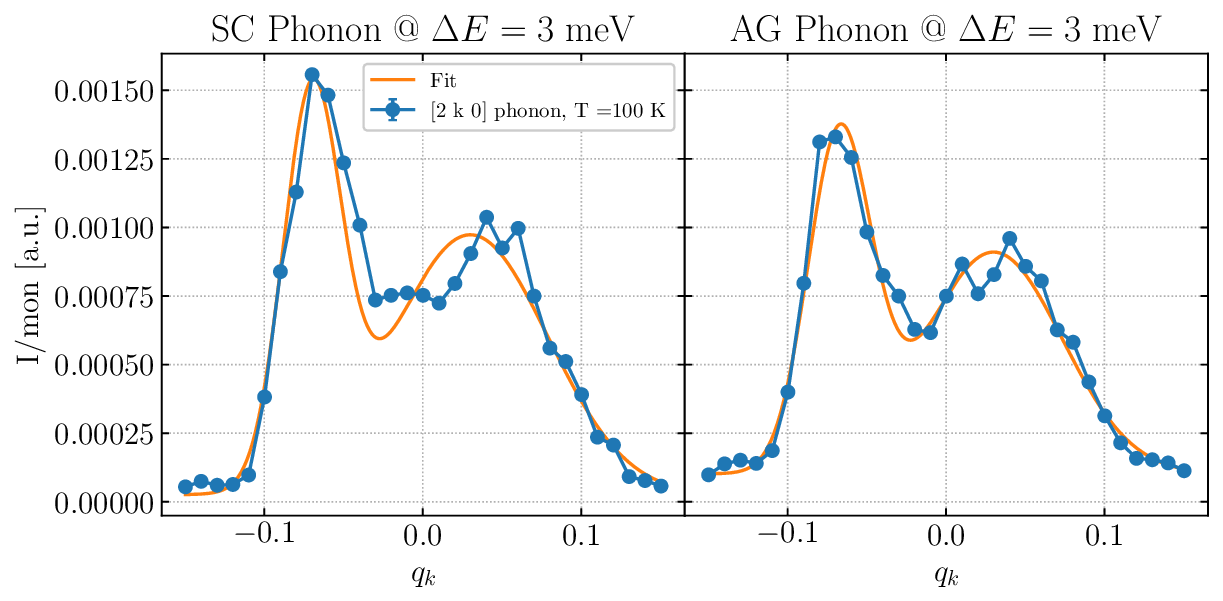}
    \caption{Measurements of transverse acoustic phonons, as constant-energy scans around (2 0 0) from TAIPAN. (left) the SC sample, (right) the AG sample. The solid, red curves are a fits to a double-gaussian line shape from the two phonon branches.}
    \label{fig:Taipan_Phonons}
\end{figure}

Fig. \ref{fig:Taipan_Phonons} show a constant energy scan of the phonon and the corresponding fitted values needed for the normalization is listed in Tables \ref{tab:Taipan_norm}.

\begin{table}[H]
\centering
\setlength{\extrarowheight}{3pt}
\resizebox{\textwidth}{!}{%
\begin{tabular}{c|cccccccc}
\hline
Sample Name & $f(\mathbf{Q})$ & $Nk_fR_0$ & $\int \tilde{I}_{\rm Ph}\, d\mathbf{q}$ 
& $F_N(2\,0\,0)^2$ & $m/M$ & $(\hbar \mathbf{Q})^2/2m$ 
& $n_q/E$ & $d\omega/dq$ \\
\hline\hline
Units & -- & meV/barn & r.l.u. & barns & -- & meV & meV$^{-1}$ & meV/r.l.u \\
\hline
As-grown Sample A 
& 0.976 & 0.00205 & 7.43e-05 & 89.794 
& 0.00121 & 18.094 & 1.134 & 61.6 \\
\hline
Annealed Sample B 
& 0.976 & 0.00242 & 9.02e-05 & 89.794 
& 0.00121 & 18.095 & 1.134 & 60.0 \\
\hline
\end{tabular}%
}
\caption{Table of parameters used for normalization for the samples measured at TAIPAN}
\label{tab:Taipan_norm}
\end{table}
\clearpage

\subsubsection{Results: scans with fits}

\textbf{As-grown Sample - A} \newline

\begin{figure}[H]
    \centering
    \includegraphics[width=\linewidth]{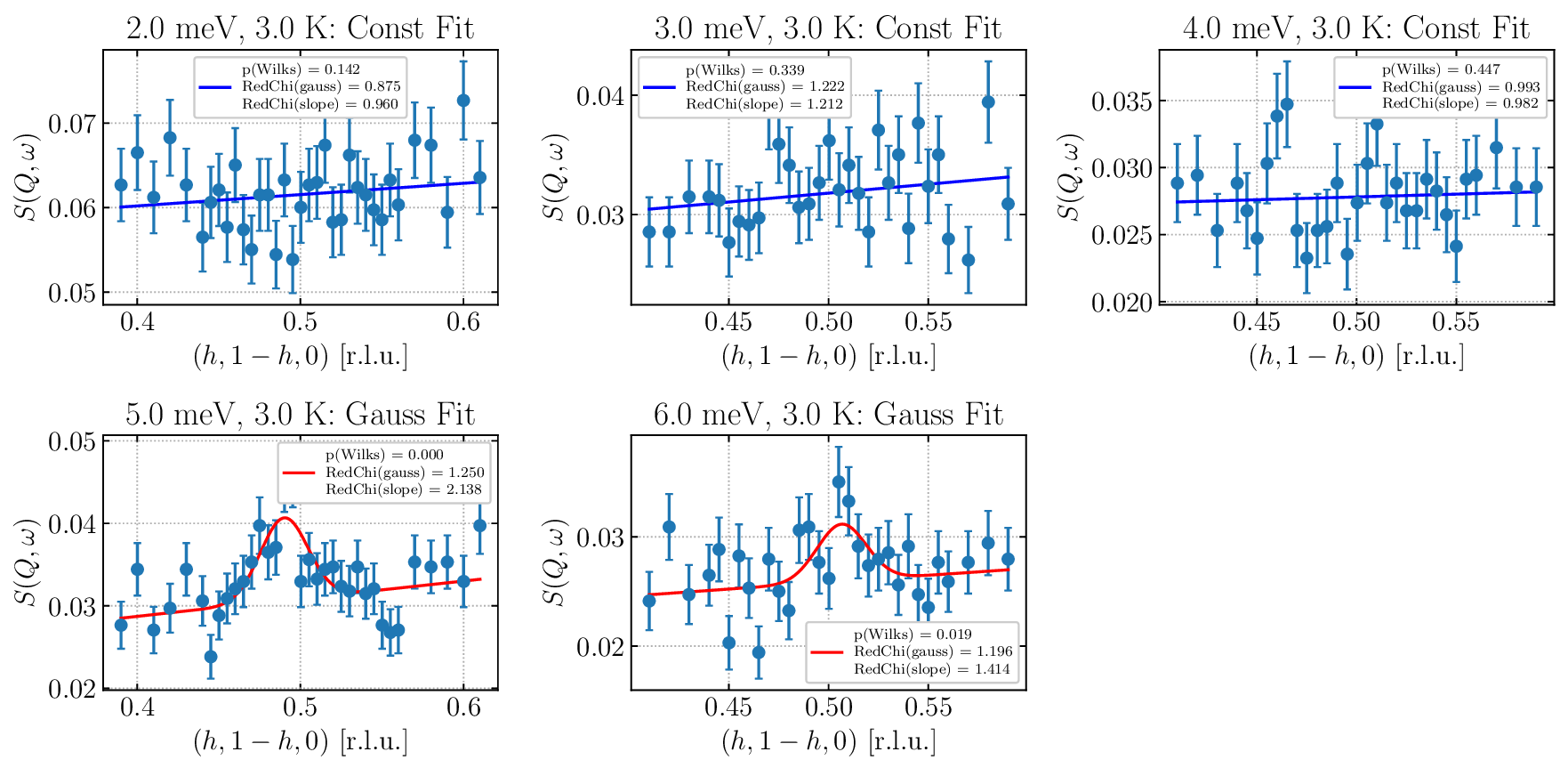}
    \caption{Scans performed for the A sample at 3 K. The energy for each scan is given in the title of each figure.  Blue markers show the data, while the red or blue line is a fit of the data as described in the text.}
    \label{fig:Taipan_AG_A_3K}
\end{figure}

\begin{figure}[H]
    \centering
    \includegraphics[width=\linewidth]{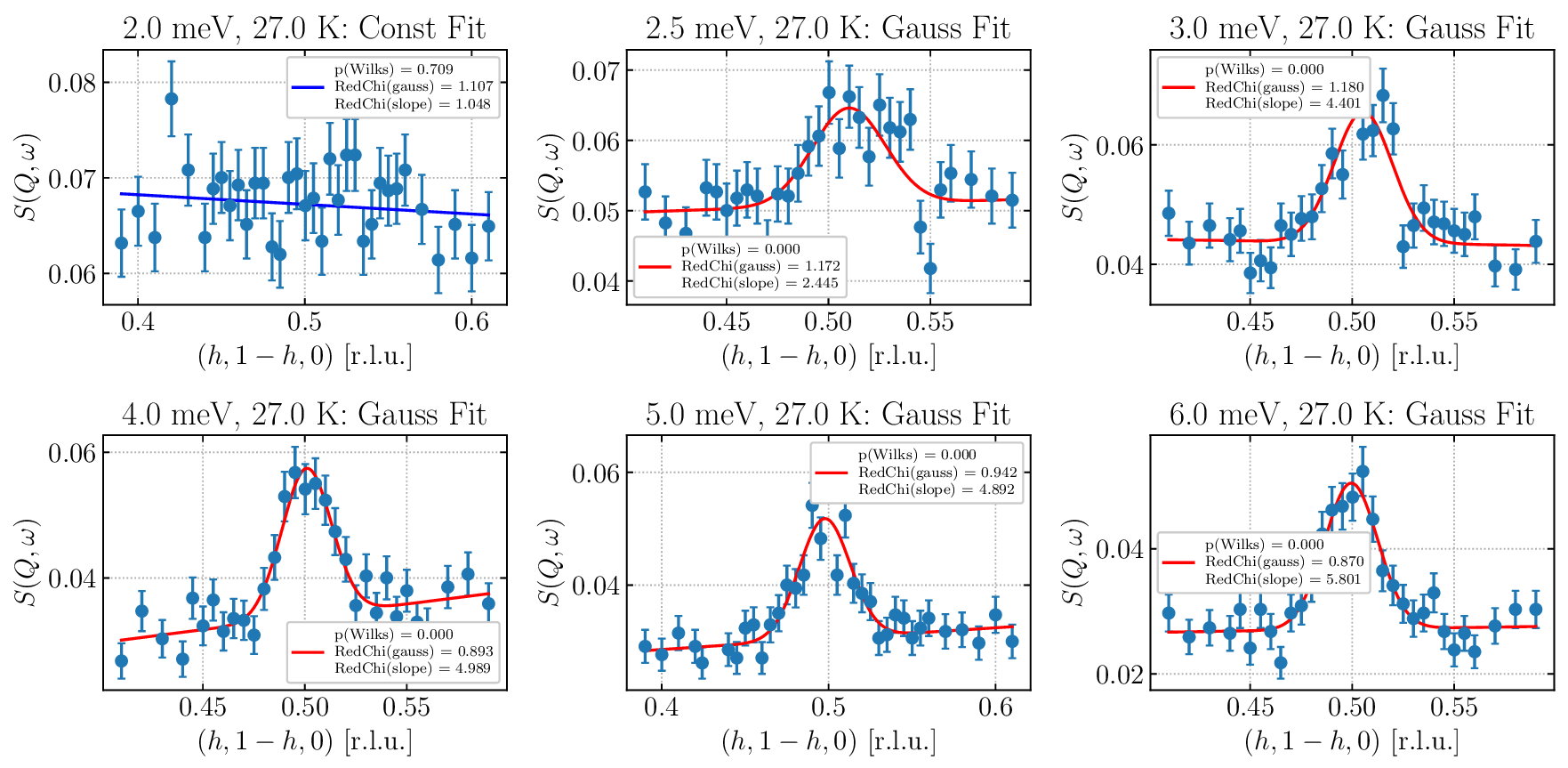}
    \caption{Scans performed for the A sample at 27 K. The energy for each scan is given in the title of each figure.  Blue markers show the data, while the red or blue line is a fit of the data as described in the text.}
    \label{fig:Taipan_AG_A_27K}
\end{figure}

\textbf{Superconducting Sample - B}

\begin{figure}[H]
    \centering
    \includegraphics[width=\linewidth]{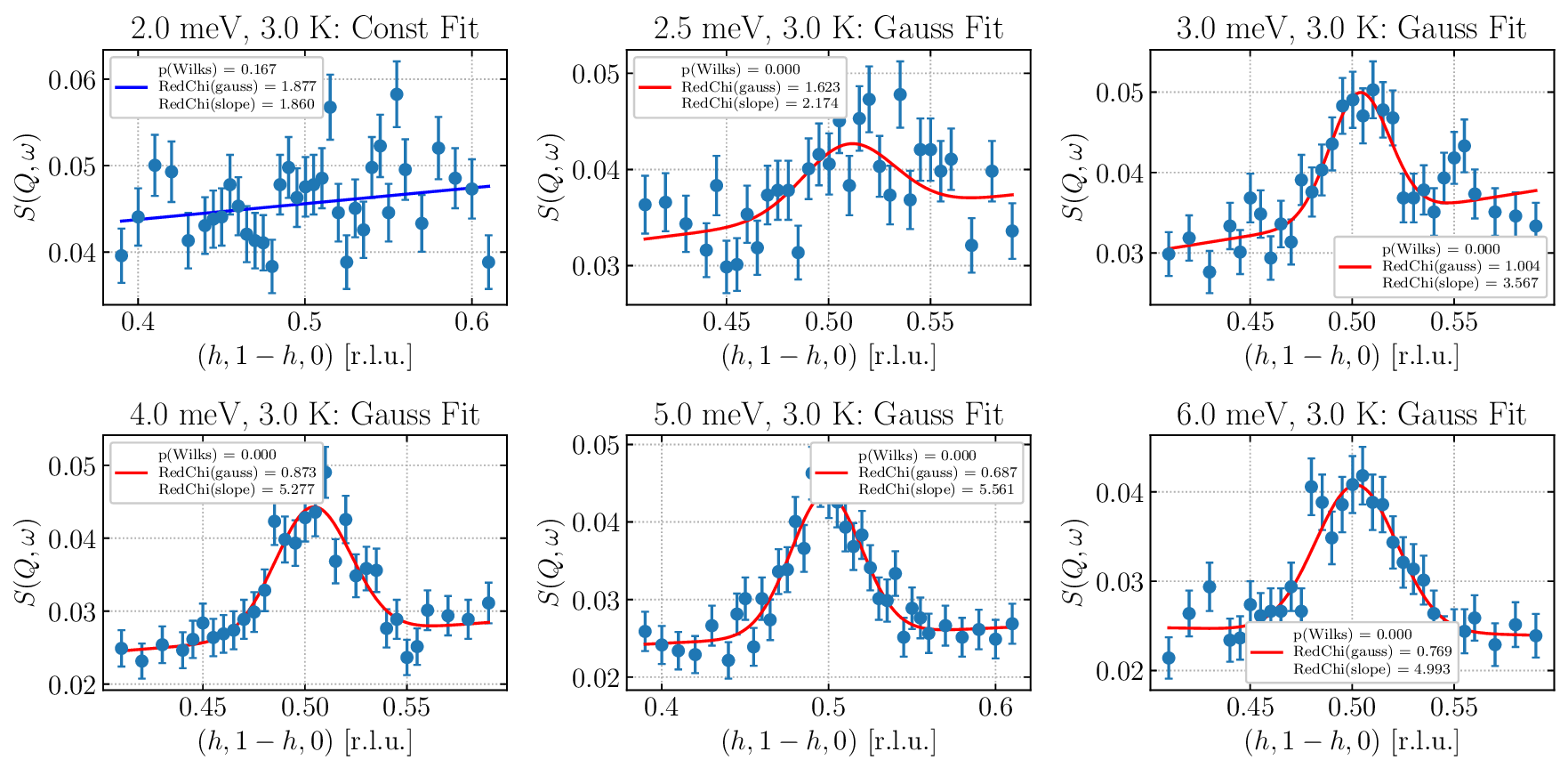}
    \caption{Scans performed for the B sample at 3 K. The energy for each scan is given in the title of each figure.  Blue markers show the data, while the red or blue line is a fit of the data as described in the text.}
    \label{fig:Taipan_SC_B_3K}
\end{figure}

\begin{figure}[H]
    \centering
    \includegraphics[width=0.7\linewidth]{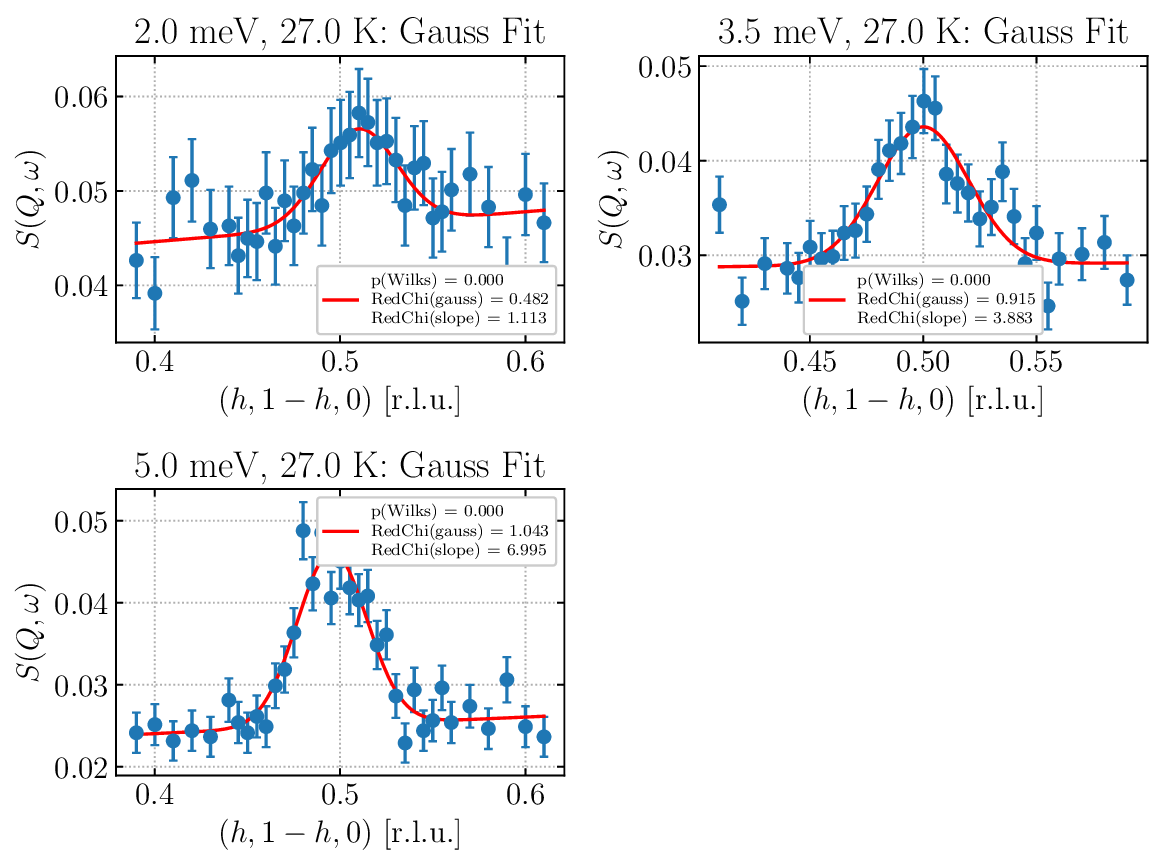}
    \caption{Scans performed for the B sample at 27 K. The energy for each scan is given in the title of each figure.  Blue markers show the data, while the red or blue line is a fit of the data as described in the text.}
    \label{fig:Taipan_SC_B_27K}
\end{figure} 

\subsubsection{Spectroscopy results}

\begin{figure}[H]
    \centering
    \includegraphics[width=0.49\linewidth]{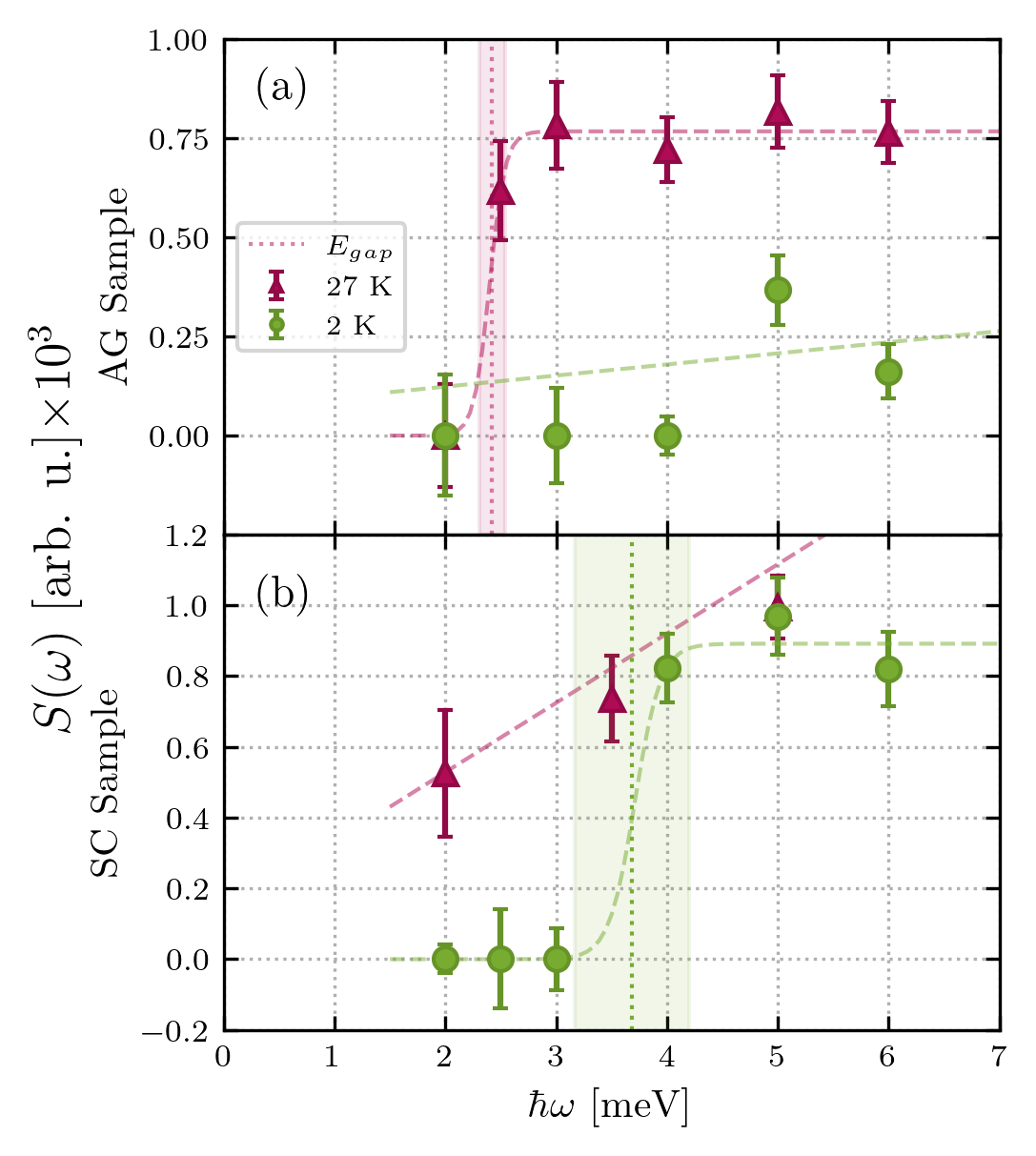}
    \includegraphics[width=0.49\linewidth]{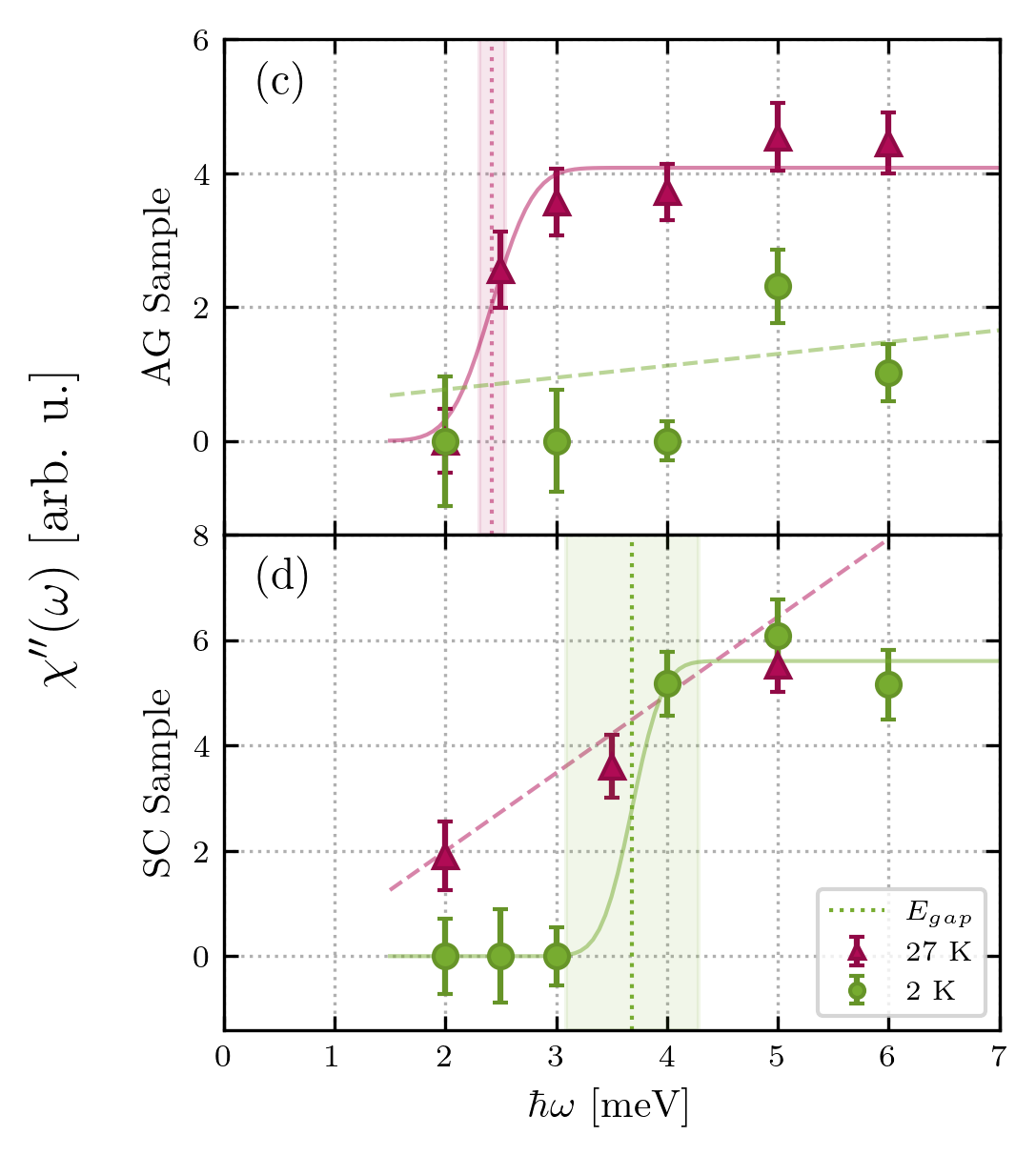}
    \caption{Integrated intensities as a function of energy transfer from the TAIPAN data. (a) and (b) in units of $S(\omega)$ and (c) and (d) in units of $\chi^{\prime \prime}(\omega)$. Top panel: as-grown (AG) sample. Bottom panel: annealed, superconducting (SC) sample. Dashed eyes are guide to the eye.}
    \label{fig:Taipan_temperature_dependence}
\end{figure}

The total overview of the energy dependence of the magnetic response, similar to that show in Fig.~3 in the main text, is seen here in Fig. \ref{fig:Taipan_temperature_dependence}. The gaps are determined in the same way as for the IN20 gaps described in section II.E, and the determined gaps are,  
\begin{equation*}
    E_{\mathrm{gap}} (\mathrm{AG}, 27K) = 2.4 \pm 0.1\,\, \,\,\, [\mathrm{meV}]
\end{equation*}
\begin{equation*}
    E_{\mathrm{gap}} (\mathrm{SC}, 2K) = 3.7 \pm 0.5 \,\, \,\,\, [\mathrm{meV}]
\end{equation*}

This result from TAIPAN shows excellent agreement to the IN20 data, both in the relative behavior between the superconducting and as-grown sample, but also for at the temperatures. The only slight difference between Taipan and IN20 is the 27 K point at 2 meV in the superconducting sample. Here the point is slightly lower on the Taipan than the IN20. However, it is still significantly higher than the corresponding 2 K point and the slight decrease might be due to the temperature control issue in the cryostat explained previously. This would result in the sample having a slightly lower temperature than 27 K which, in turn, should lead to a reduced intensity, consistent with our observations.


\section{Calculating the cut-off wavelength from the size of the spin gap}
We here calculate the cut-off wavelength from the size of the spin gap. We start by making a simple spin wave model of perfect-lattice non-SC NCCO. Here, we assume the all spins reside on the Cu$^{2+}$ ions ($S=1/2$) and that each spin interact with only it nearest neigbours with a Heisenberg interaction of strength $J$.  In this textbook model, the spin wave energies are given by
\begin{equation}
    E_q = S \sqrt{J^2({\bf 0})-J^2({\bf q})} ,
\end{equation}
where the Fourier transformed interaction is given by
\begin{equation}
    J({\bf q}) = 2 J \left( \cos(a q_x) + \cos(a q_y) \right) .   
\end{equation}
and $q_x$ and $q_y$ are the components of the wave vector {\bf q} along the direction of the neigbour bonds. This leads to a spin wave energy of
\begin{equation}
    E_q = 2 S J \sqrt{\sin^2(a q_x) + \sin^2(a q_y)} .
\end{equation}
The top of the spin wave band is found for ${\bf q} = (\pm \pi/(2a)\, , \, \pm \pi/(2a))$ with an energy of $E_{\rm max} = 2 \sqrt{2} S J$.
For NCCO, this is measured to be $E_{\rm max,NCCO} = 450$~meV.\cite{ishii_post-growth_2021}

In a low-$q$ approximation, the spin wave energy can be approximated by its linear term
\begin{equation}
    E_{\rm lin} = 2 S J a |q| .
\end{equation}
If a spin pseudo-gap, $\Delta$ opens up due to the suppression of long wavelength (or small-$q$) spin waves, we will find the minimal $q$-value through
\begin{equation}
    \Delta = 2 J a S q_{\rm min} .
\end{equation}
This will, in turn, lead to a maximal wavelength of
\begin{equation}
    \lambda_{\rm max} = \frac{2 \pi}{q_{\rm min}} = \frac{4 \pi J a S}{\Delta} = \sqrt{2} \pi a \frac{E_{\rm max,NCCO}}{\Delta} .
\end{equation}
Since the maximal wavelength is approximately twice the patch size, we get
\begin{equation}
      d_{\rm max} =   \frac{\sqrt{2} \pi a E_{\rm max,NCCO}}{2 \Delta} .  
\end{equation}

A tabulated list over the different spin pseudogaps and the corresponding patch sizes are seen below in \ref{tab:patches}.

\begin{table}[H]
    \centering
    \begin{tabular}{c|c|c}
        \hline
         & Gap size [meV] & $d_{\rm max}$  [nm] \\
        \hline \hline
         IN20 AG sample (A) & 10 & 39 \\
         IN20 SC sample (B) &  3.0 & 138  \\
         TAIPAN SC Sample (C) & 3.7  & 107  \\
         \hline
    \end{tabular}
    \caption{Effective sizes of the spin patches calculated from the simple spin wave model of the gap sizes. The gap cannot be estimated from the as-grown sample on the TAIPAN experiment due to the limited number of energy points and is therefore not included.}
    \label{tab:patches}
\end{table}

\bibliographystyle{apsrev4-1}
\bibliography{Main_NCCO}